\documentclass{imsart}% \documentclass[11pt]{article}
\usepackage[psamsfonts]{amsfonts}
\usepackage{theorem}
\usepackage{version}
\usepackage{pstricks,pst-node,pst-text,pst-3d}
\usepackage{latexsym,psfig}
\usepackage{hyperref}

\usepackage{amsmath}%,natbib}
\usepackage{graphicx}
\usepackage{amssymb}
\usepackage[utf8]{inputenc}
\usepackage[mathscr]{eucal}
\usepackage{fancyhdr}
\usepackage{fancybox}
\usepackage{color}
\usepackage{longtable}
\usepackage{xcolor}
\usepackage{caption}

\usepackage{stackengine}
\stackMath
\newcommand\tenq[2][1]{%
\def\useanchorwidth{T}%
\ifnum#1>1%
\stackunder[0pt]{\tenq[\numexpr#1-1\relax]{#2}}{\scriptscriptstyle\thicksim}%
\else%
\stackunder[1pt]{#2}{\scriptscriptstyle\thicksim}%
\fi%
}

\RequirePackage[colorlinks,citecolor=blue,urlcolor=blue]{hyperref}

\newtheorem{lem}{Lemma}[section]

\newtheorem{Prop}{Proposition}[section]

\newtheorem{prop}{Proposition}[section]

\newtheorem{defin}{Definition}[section]

\newtheorem{Thm}{Theorem}[section]

\newtheorem{cor}{Corollary}[section]
\newtheorem{rem}{Remark}[section]

\newcommand{\pms}{{\scriptscriptstyle \pm}}
\newcommand{\cqfd}{\hfill $\square$}

\newcommand{\R}{\mathbb R}

\newcommand{\n}{^{(n)}}

\newcommand{\Xb}{\mathbf{X}}

\newcommand{\thetab}{{\pmb \theta}}

\newcommand{\Thetab}{{\pmb \Theta}}

\newcommand{\pr}{^{\prime}}

\begin{document}

\begin{frontmatter}

\title{%Center-Outward
 Distribution~and~Quantile  Functions, Ranks, and Signs in dimension $d$:%\vspace{-1.5mm}
\\  a measure transportation approach% \protect\thanksref{T1}
}
\runtitle{Distribution and Quantile  Functions %,  Quantiles, Ranks, and Signs
 in $\mathbb{R}^{\MakeLowercase{d}}$
}

\begin{aug}
% indicate corresponding author with \corref{}
% \author{\fnms{John} \snm{Smith}\thanksref{a}\corref{}\ead[label=e1]{smith@foo.com}\ead[label=e2,url]{www.foo.com}}
% \address[a]{\printead{e1};\printead{e2}}
\author{\fnms{Marc} \snm{Hallin,}\corref{}\ead[label=e1]{mhallin@ulb.ac.be }}
\author{\fnms{Eustasio} \snm{del Barrio,}\ead[label=e2]{tasio@eio.uva.es }}
 \author{\fnms{Juan} \snm{Cuesta Albertos,  }\ead[label=e3]{juan.cuesta@unican.es }\\ }
\and
\author{\fnms{Carlos} \snm{Matr\' an}\ead[label=e4]{carlos.matran@uva.es }}
\address{\printead{e1}\\ ECARES and  D\' epartement de Math\' ematique\\ Universit\' e libre de Bruxelles,  Brussels,  Belgium} 
\address{{\printead{e2}}\\ Departamento de Estad\'\i stica
%\\ \hspace{4mm}
 e Investigaci\`on Operativa\\  Facultad de Ciencias, Universidad de Valladolid, Spain}
\address{\printead{e3}\\ Departamento de Matem\' aticas\\ Facultad de Ciencias,  Universidad de Cantabria, Santander,   Spain}
\address{\printead{e4}\\ Departamento de Estad\'\i stica %\\ \hspace{4mm}
 e Investigaci\'on Operativa\\  Facultad de Ciencias,  Universidad de Valladolid, Spain}

\runauthor{Hallin et al. }

  \affiliation{Universidad de Valladolid,  Universidad de Cantabria, \\ Universit\' e libre de Bruxelles, and Universidad de Valladolid}

\end{aug}

\begin{abstract}
Unlike the real line, the %$d$-dimensional
 real space ${\mathbb R}^d$, for $d\geq~\!2$, is not canonically ordered. As a consequence, such fundamental 
 %and strongly order-related
  univariate concepts as quantile and distribution functions,  and their empirical counterparts, involving ranks and signs,    do not canonically extend to the multivariate context. Palliating that lack  of a canonical ordering has been an open problem for more than half a century,  generating an abundant literature and motivating, among others, the development of statistical depth and copula-based methods. We  show that, unlike the many definitions  % that have been
   proposed in the literature, the measure transportation-based ranks  introduced in Chernozhukov et al.~(2017) enjoy all the properties that make  univariate  ranks  a successful tool for semiparametric %statistical
 inference. Related with those ranks, we   propose a new {\it center-outward} definition of multivariate distribution and quantile functions, along with their empirical counterparts, for which we establish  a Glivenko-Cantelli result. %---the quintessential property of an empirical distribution function.
  Our approach is  based on McCann~(1995) % , is geometric rather than analytical
   and our results, unlike those of %the Monge-Kantorovich one   in
 Chernozhukov et al.~(2017),  do not require any moment assumptions. The  resulting ranks and signs   are shown to be strictly distribution-free and essentially maximal ancillary in the sense of Basu~(1959) which, in semiparametric models involving noise with unspecified density,  can be interpreted as a finite-sample form of semiparametric efficiency.  
%maximal invariant under the action of  a data-driven class of (order-preserving)  transformations generating the family of absolutely continuous distributions; that maximal invariance, in view of a general result by Hallin and Werker~(2003),  is the theoretical foundation of the  semiparametric efficiency preservation property of ranks.  The corresponding quantiles are equivariant under the same transformations.
 Although constituting a sufficient summary of the sample,  empirical center-outward distribution functions  are defined at observed values only. 
% , with   collections of isolated observations as empirical quantile contours and sign curves.
  A continuous extension to the entire $d$-dimensional space, yielding smooth empirical quantile contours and sign curves  while preserving the essential monotonicity and Glivenko-Cantelli features, is provided.  % Based on those results, a 
  A numerical study of the resulting empirical %center-outward
   quantile contours   is conducted.\vspace{-1mm} 
\end{abstract}

\begin{keyword}[class=MSC]
\kwd[Primary ]{62G30}
%\kwd{60K35}
\kwd[; secondary ]{62B05}
\end{keyword}%\vspace{-1mm}
\begin{keyword}
%\kwd{sample}
\kwd{Multivariate distribution function; multivariate quantiles, multivariate ranks; multivariate signs;   Glivenko-Cantelli theorem; Basu theorem; 
distri\-bu\-tion-freeness; cyclical monotonicity}\vspace{-2.5mm}
\end{keyword}

\end{frontmatter}

%%%%%%%%, 
%%%%%%%%
%%%%%%%%

\vspace{-2mm}

\section{Introduction\vspace{-1.5mm}}\label{introintro}
\setcounter{equation}
{0}

Unlike the real line, the real space ${\mathbb R}^d$, for $d\geq 2$, is not canonically ordered. As a consequence, such fundamental  concepts as quantile and distribution functions,  which are strongly related to the ordering of the observation space,  and their empirical counterparts---ranks and empirical quantiles---playing, in dimension~$d=1$,   a fundamental role in statistical inference,  do not canonically extend   to dimension~$d\geq 2$. 

Of course, a classical concept of distribution function---the familiar one, based on marginal orderings---does exist. That concept, from a probabilistic point of view,  does the job   of characterizing the underlying distribution. However, the corresponding quantile function does not mean much (see, e.g., Genest and Rivest~(2001)), and the corresponding empirical versions (related to their  population counterparts via a Glivenko-Cantelli result) do not possess any of the properties that make them successful inferential tools in dimension~$d=1$. 

That observation about traditional multivariate distribution functions  is not new:  palliating the  lack  of a ``natural'' ordering of ${\mathbb R}^d$---hence, defining  statistically sound concepts of distribution and quantile  functions---has been an open problem for more than half a century,   generating an abundant literature that includes, among others, the  theory of copulas and the theory of statistical depth. 

A number of most  ingenious solutions have been proposed,    each of them extending some chosen features    of the well-understood univariate concepts, with which they   coincide for $d=~\!1$. 
  Coinciding, for $d=1$, with the univariate concepts obviously is  important, but    hardly sufficient for qualifying  as a statistically pertinent multivariate extension. For statisticians, distribution and quantile functions  are not just probabilistic notions: above all, their empirical versions (empirical quantiles and ranks)  constitute fundamental  tools for inference. A multivariate extension yielding quantiles and ranks that do not enjoy, in dimension~$d\geq 2$, the properties  that make  traditional ranks    natural and successful  tools for inference for $d=1$ is not a statistically sound extension.  
  
  Those  inferential  concerns are at the heart of the 
  approach   adopted~here.\vspace{-2mm}% is placing those  at the heart of the problem. 
  
\subsection{Ranks and rank-based inference\vspace{-1.5mm}} 
\label{orderRd}

To facilitate the exposition, let us focus on ranks and   their role in testing problems.   Univariate rank-based methods naturally enter the picture in the context of semiparametric statistical models %or experiments
 under which the distribution~${\rm P}\n_{\thetab,f}$ of some real-valued observa\-tion $\Xb=(X_1,\ldots,X_n)\pr\!$, besides a  finite-dimensional parame\-ter of interest $\thetab\!\in\!\Thetab$, also depends  on the unspecified density~$f\in{\mathcal F}_1$ (${\mathcal F}_1$   the family of   Lebesgue densities over $\mathbb{R}$) of some unobserved  univariate   noise $Z_i(\thetab)$, say.  More precisely, %assume that~
 $\Xb\sim{\rm P}\n_{\thetab,f}$ iff the {\it $\thetab$-residuals}~$Z_1(\thetab),\ldots,Z_n(\thetab)=:{\bf Z}\n(\thetab)\vspace{-0.5mm}$  are i.i.d.\!\footnote{Although i.i.d.-{\it ness} can be relaxed into {\it exchangeability},  we are sticking to the former.}~\!with density~$f$.  In such  models---call them\vspace{-.6mm}  i.i.d.~noise models\footnote{Typical examples are linear models, with $Z_i(\thetab)=X_i-{\bf c}_i\pr\thetab$ (${\bf c}_i$ a $q$-vector of covariates and~$\thetab \in\mathbb{R}^q$), or first-order autoregressive models, with $Z_i(\theta)=X_i-\theta X_{i-1}$ (where $i$ denotes time and $\theta\in (-1,1)$; see, e.g., Hallin and Werker~(1998)), etc.}---testing the null hypothesis~$H\n_0\!:\thetab=\thetab_0\vspace{-0mm}$ (that is, ${\rm P}\n_{\thetab,f}\!\in\!{\mathcal P}\n_{\thetab_0}\!:=\!\{{\rm P}\n_{\thetab_0,f}\vert f\!\in\!{\mathcal F}_1\}$)  reduces to the problem of testing that~$Z_1(\thetab_0),\ldots,Z_n(\thetab_0)$ are i.i.d.\  with unspecified density~$f\in{\cal F}_1$.  %(distribution ${\rm P}^n_{\! f}$). 
% Denote by~${\cal P}_1$ the family of Lebesgue-absolutely continuous distributions over~$\mathbb{R}$, by ${\cal F}_1$ the corresponding family of densities.
   Invariance arguments suggest tests based on the ranks ${\bf R}\n(\thetab_0)$ of the residuals ${\bf Z}\n(\thetab_0)$\footnote{Those ranks indeed are maximal invariant under the group of continuous monotone increasing transformations of $Z_1(\thetab_0),\ldots,Z_n(\thetab_0)$; see, e.g., Example 7 in Lehmann and Scholz~(1992).}; such tests are {\it distribu\-tion-free} under $H_0\n$. %---a finite-sample property holding in all fixed-$\thetab_0$ submodels. 
 
Distribution-freeness~(DF) is often considered as the trademark and main virtue of %\vspace{-0.8mm}
    (univariate\vspace{-0.5mm})
     ranks; it guarantees the validity and similarity of rank-based   tests of~$H_0\n\!$.  Distribution-freeness  
 alone is not sufficient, though, for explaining the success of rank tests: %, and efficiency is no less important: 
 other classes of distribution-free methods indeed can be constructed, such as sign or runs tests, that do not perform as well 
% (in models driven by i.i.d.~noise  with unspecified density)
  as the rank-based ones. The reason is that, unlike the ranks, they do not fully exploit the information available once %the possible impact of
  the nuisance (the unknown~$f$) has been controlled for via some minimal sufficient statistic.  That %important
   feature of ranks originates in the fact that \vspace{1mm}
  
\noindent (DF$^+$)\!  (essential maximal ancillarity) {\it  the sub-$\sigma$-field generated by the residual  ranks~${\bf R}\n(\thetab)$ is {\em essentially maximal ancillary}  
    (hence distribution-free) for~${\mathcal P}\n_\thetab\!\!$ in the sense of Basu~(1959) (see, e.g., Example 7 in Lehmann and Scholz~(1992)).}\vspace{1.9mm}
 
 \noindent  while   the sub-$\sigma$-field generated by the residual order statistic~${\bf Z}\n_{{\scriptscriptstyle{(\, .\, )}}}(\thetab)$ is {\it minimal sufficient} and {\it complete} (still for~${\mathcal P}\n_\thetab$).
    
    In families satisfying the condition  (Koehn and Thomas~1975) 
 of non-existence  of a 
  {\it splitting set}---which is the case here whenever  $f$ ranges over ${\cal F}_1$---Theorems~1  and~2 in Basu~(1955)
    imply that essential maximal ancillarity 
    is equivalent to ``essential maximal independence with
     respect to the complete (hence minimal)  sufficient statistic ${\bf Z}\n_{{\scriptscriptstyle{(\, .\, )}}}(\thetab)$."\footnote{We refer to Appendix~\ref{Basuapp} for precise definitions, a  more general  and stronger version of this  property,   and a proof.}   
Intuitively, thus, and leaving aside the required mathematical precautions, the order statistic~${\bf Z}\n_{{\scriptscriptstyle{(\, .\, )}}}(\thetab)$, being minimal sufficient  for~${\mathcal P}\n_\thetab\!\!$, is carrying all the information about the nuisance $f$ and nothing but that information, while the ranks, being (essentially) ``maximal independent of~${\bf Z}\n_{{\scriptscriptstyle{(\, .\, )}}}(\thetab)$,'' are carrying whatever information is left for~$\thetab$. This  can be interpreted as  a finite-sample form of  semiparametric efficiency\footnote{Semiparametric efficiency indeed is characterized as  asymptotic orthogonality, with respect to the central sequences carrying information about parametric perturbations of the nuisance; asymptotic orthogonality here is replaced with finite-sample independence.}  
%the {\it Basu factorization property} of ranks.   See   Proposition~\ref{DFProp} and Appendix~\ref{Basuapp} for  a  more general formulation and a proof of this %important 
%property
 \color{black}

In the same vein, it also has been shown (Hallin and Werker~2003) that, under appropriate regularity conditions, univariate ranks  preserve semiparametric efficiency  in models where 
%under the %regularity
% assumptions required for
 that concept makes sense: \vspace{0.5mm}%vspace{-1.5mm} 
%\begin{enumerate}
%\item[(HW)]

\noindent (HW) (preservation of semiparametric efficiency) {\it the semiparametric efficiency  bound at arbitra\-ry~$(\thetab ,f)$   can be reached, under ${\rm P}\n_{\thetab, f}\vspace{0mm}$, via rank-based procedures (tests that are measurable functions of %with respect to
 the ranks of $\thetab$-residuals~$Z_i(\thetab)$).} \vspace{0.5mm}
%\end{enumerate}

The latter property, contrary to (DF) and (DF$^+$), is of a local and asymptotic nature; in Hallin and Werker~(2003), it follows from the maximal invariance property of ranks under a group of order-preserving transformations of~$\mathbb{R}^n$  {\it generating} the fixed-$\thetab$ submodel (that is, yielding a unique orbit in the fa\-mily~${\cal P}\n_\thetab$ 
%$\{{\rm P}\n_{\thetab,f}\vert f\in{\cal F}_1\}$
 of fixed-$\thetab$ model distributions). Being intimately related to the concept of order-preserving transformation, this invariance approach is much more delicate in dimension  $d>1$. For lack of space, we do not investigate  it any further here, leaving a formal multivariate extension of (HW) for further research.% for $d>1$. 

Properties  (DF$^+$) and (HW), which indicate, roughly, that the order statistic only carries  information about the nuisance $f$ while the ranks carry all the information available about $\thetab$,  are those  a statistician  definitely would like to see satisfied by any sensible multivariate extension of the concept.\vspace{-1mm}
%with ${\cal F}^d$ and~${\cal F}^d_*$ substituted for~${\cal F}^1$ and  ${\cal F}^1_*$
%in a $d$-dimensional setting, by
% the concept of ranks associated with the empirical counterpart of any sensible definition of a multivariate distribution function. 

\subsection{Multivariate ranks  and the ordering of  ${\mathbb R}^d$, $d\geq 2$\vspace{-1mm}}
\label{orderRd2}

 The problem of ordering  ${\mathbb R}^d$ for $d\geq 2$, thus defining multivariate concepts of ranks, signs, empirical distribution functions and quantiles, is not new, and has a rather long history in statistics. Many concepts have been proposed in the literature, a complete list of which cannot be given here. Focusing again on ranks,   four types of {\it multivariate ranks}, essentially, can be found:
 
 \noindent (a) {\it Componentwise  ranks.}  The idea of   componentwise  ranks  goes back as far as Hodges~(1955),  Bickel (1965) or  Puri and Sen (1966, 1967, 1969). It  culminates in the monograph by Puri and Sen (1971), where inference procedures based on componentwise ranks are proposed, basically, for all     classical problems of multivariate analysis. Time-series testing methods based on the same ranks have been considered in Hallin, Ingenbleek, and Puri~(1989).  That strand of literature is still alive: see   Chaudhuri and Sengupta~(1993), %Nordhausen, Oja, and Tyler~(2006),
  Segers, van den Akker, and Werker~(2015), ... to quote only a very few.   Componentwise  ranks actually are intimately related to copula transforms, of which they constitute the empirical version: rather than solving the tricky problem of ordering  $\mathbb{R}^d$,   they bypass    it  by considering~$d$ univariate marginal rankings. As a consequence, they crucially depend   on the choice of a coordinate system. Unless the underlying distribution has independent components (Nordhausen et al.~2009,  Ilmonen and Paindaveine~2011,   Hallin and Mehta~2015) coinciding with the chosen coordinates,  componentwise ranks in general are not even asymptotically distribution-free: neither (DF) nor (DF$^+$)~hold. 
% Nor are they invariant under any model-generating class of transformations; a transformation-retransformation approach has been proposed by Chakraborty and Chaudhuri (1996, 1998), which ensures affine-invariance---but  the group of affine transformations  is not a generating group in this context.
%  As a consequence,   (DF$^+$)  cannot hold. 
 
 \noindent (b) {\it Spatial ranks and signs.}  This class of multivariate ranks  (M\" ott\" onen and Oja 1995;  M\" ott\" onen et al.~1997; Chaudhuri 1996;  Koltchinskii 1997;     Oja and Randles~2004,  Oja~2010,   and many others) includes several very ingenious, elegant, and appealing   concepts.  Similar ideas also have been developed by   Choi and Marden~(1997) and, more recently,  in high dimension, by Biswas,  Mukho\-padhyay,  and  Ghosh (2014) and Chakraborthy and Chaudhuri~(1996, 2014, 2017). We refer to Marden~(1999), Oja~(1999) or %the monograph by
  Oja~(2010) for a systematic exposition and exhaustive list of  references.  All those concepts are  extending the traditional univariate ones but none of them enjoys (DF)\footnote{Biswas et al.~(2014) is an exception, but fails on~(DF$^+$)}, let alone~(DF$^+$). %, but fails to satisfy (HW)). 
% Their invariance properties at best extend to classes (actually, groups)  of  rotations, scale  or affine transformations, which are not generating groups: neither (HW$^*$) nor (HW) are satisfied. 

 \noindent (c) {\it Depth-based ranks.}  Those ranks have been considered in Liu~(1992), Liu and Singh~(1993), He and Wang~(1997),  Zuo and He~(2006), Zuo and Serfling~(2000), among others; see Serfling~(2002) for a general introduction on statistical depth, Hallin et al.~(2010) for the related concept of quantile, L\`opez-Pintado and   Romo~(2012) for functional extensions, Zuo~(2018) for a state-of-the art survey in a regression context.     
  Depth-based ranks,  in general, are distribution-free  but fail to satisfy (DF$^+$). 
%   At best (Monge-Kantorovich depth, as recently proposed by Chernozhukov et~al.~(2017), is an exception), they also are affine-invariant; affine transformations, however, fail to be a generating group: neither (HW$^*$) nor (HW)  hold. 

 \noindent (d) {\it Mahalanobis ranks and signs/interdirections. }   When considered jointly with interdirections (Randles 1989), lift interdirections (Oja and Paindaveine 2005),  Tyler angles or Mahalanobis signs (Hallin and Paindaveine~2002a,~c),    Mahalanobis ranks do satisfy     (DF$^+$), but in elliptical models only---when~$f$ is limited to  the family of elliptical densities.  %with unspecified radial densities. 
  There, they have been used, quite successfully, in a variety of multivariate models,  including one-sample location (Hallin and Paindaveine 2002a), $k$-sample location (Um and Randles~1998), serial dependence
(Hallin and Paindaveine 2002b), linear models with VARMA errors (Hallin and Paindaveine
2004a, 2005a, 2006a), VAR order identification (Hallin and Paindaveine 2004b), shape
(Hallin and Paindaveine~2006b; Hallin, Oja, and Painda\-veine~2006), homogeneity of scatter
(Hallin and Paindaveine 2008), principal and common principal components (Hallin, Paindaveine,
and Verdebout~2010, 2013, 2014). Unfortunately, the tests  developed in those references  cease to be valid, and the related R-estimators no longer are root-$n$ consistent,  under non-elliptical densities.

None of those multivariate rank concepts, thus,    enjoys   distribution-freeness {\it and} (DF$^+$)---except, but only over the class of elliptically symmetric distributions, % and along with the adequate concept of sign,
 for the Mahalanobis ranks and signs.  A few other concepts have been proposed as well, related to {\it cone orderings} (Belloni and Winkler~2011;   Hamel and Kostner~2018), which require some subjective (or problem-specific) preliminary choices, and similarly fail to achieve  distribution-freeness, hence~(DF$^+$). %\smallskip\vspace{-3.6mm}

The lack, % fact that, 
%contrary to the real line~$\mathbb R$,   the real space $\mathbb{R}^d$
 for $d\geq 2$,  
% the real space $\mathbb{R}^d$ does not admit
 of a canonical   ordering of $\mathbb{R}^d$ places an essential difference between dimensions~$d=1$ and~$d\geq~\!2$. Whereas the same ``exogenous"  left-to-right  ordering of $\mathbb R$ applies both in population and in the sample,     pertinent orderings  of $\mathbb{R}^d$  are bound to be ``endogenous",  that is, distribution-specific in populations,  and data-driven (hence, random) in samples. This is the case for the concepts developed under  (b)-(d) above; it also holds for the concept we are proposing in this paper.     Each distribution, each sample, thus, is to produce its own ordering, inducing  %(related forms of)
   quantile and distribution functions, and classes of order-preserving transformations. As a result,     datasets, at best,  can be expected to produce, via adequate concepts of multivariate ranks and signs, consistent empirical versions of the %unavailable  
   underlying population ordering. That consistency typically takes the form of a Glivenko-Cantelli~(GC)  result connecting an empirical {\it center-outward distribution function} to its population version.  A quintessential feature of Glivenko-Cantelli is its insensitivity to continuous order-preserving transformations of the data. That feature is not compatible with moment assumptions, since the existence of moments is not preserved under such transformations. Moment assumptions (as in Boeckel et~al.~(2018) or Chernozhukov et al.~(2017) where (weak) consistency is established  under compactly supported distributions), therefore, are  somewhat inappropriate  in this context.%the intrinsically  ordinal context of distribution and  quantile functions. 

No ordering of $\mathbb{R}^d$ for $d\geq 2$ moreover can be expected to be of the one-sided  ``left-to-right" type, since   ``left" and ``right" do not make sense anymore. A depth-type center-outward ordering is by far more sensible.  All this  calls for %a fundamental
 revisiting the traditional  univariate concepts from a center-outward perspective, while   disentangling (since they are to be based on distinct orderings)   the population  concepts from their sample counterparts. \vspace{-1mm}

\subsection{Outline of the paper\vspace{-1mm}} 

This paper consists of a main text and an online appendix. Except for the proofs, the main text is self-contained and the reader familiar with measure transportation and statistical decision can skip most of the appendix. For those who are less familiar with those topics, however, we recommend the following plan for fruitful reading. After the introduction (Section~\ref{introintro}), one may like to go to Appendix~\ref{transpsec1} for a brief  and elementary account  of some classical facts in measure transportation, then to Appendix~\ref{reviewsec}  for a short review of the (scarce)  literature on  relation of that theory to multivariate ranks and quantiles. Appendix~\ref{1dimsec} is describing how the traditional univariate case, where the concepts of distribution and quantile functions, ranks, and signs are familiar, naturally enters the realm of measure trasportation once the usual distribution function $F$ is replaced by the  so-called center-outward one $2F-1$. The paper then really starts with Section~\ref{MKsec}, where the main concepts---center-outward distribution and quantile functions, ranks,  signs, quantile contours and quantile regions---are defined and their main properties---regularity of distribution and quantile functions, nestedness and connectedness of quantile regions,  distribution-freeness of ranks and signs, their maximal ancillarity property and their Glivenko-Cantelli asymptotics---are stated. Proofs are given in Appendices~\ref{Proofsec2} and~\ref{Basuapp} and  the relation, under ellipticity, to Mahalanobis ranks and signs    is discussed in Appendix~\ref{ellsec}. Up to that point, empirical distribution and quantile functions are defined at the observations only. Section~\ref{sec3} shows how to extend them into smooth functions defined over the entire space $\mathbb{R}^d$ while preserving their   gradient of convex function nature,  without  which they no longer would  qualify as distribution and quantile functions. This smooth extension is shown (Proposition~\ref{PropGC}) to satisfy an extended Glivenko-Cantelli property; proofs are concentrated in Appendix~\ref{Proofsec3}. The tools we are using  throughout are  exploiting the concept of {\it cyclical monotonicity} and the approach initiated by McCann~(1995).\footnote{This fact is emphasized by a shift in the terminology: as our approach is no longer  based on  Monge-Kantorovich optimization ideas, we consistently adopt the terminology {\it center-outward} ranks and signs instead of {\it Monge-Kantorovich} ranks and signs.} Section~\ref{numsec} provides some numerical results. The algorithms we are using can handle samples of size as large as~$n=20000$ in dimension 2 (the complexity of the algorithms in $\mathbb{R}^d$ only depend on $n$, not on $d$); simulations demonstrate the  power of empirical center-outward  quantile functions as descriptive tools. Further numerical results, and a comparison with Tukey depth are given in Appendix~\ref{NUMAPP}. Section~\ref{Consec} concludes with a discussion of  some perspectives for further research.%\vspace{-1mm}

\color{black}
  \subsection{Notation%\vspace{-1mm}
  } 
  
  Throughout,  let $\mu_d$ stand for the Lebesgue measure over $\mathbb{R}^d$ equipped with its Borel $\sigma$-field ${\mathcal B}_d$. Denote by ${\cal P}_d$  the family of  Lebesgue-absolutely continuous distributions  %${\rm P}_{\! f}$
   over~$(\mathbb{R}^d, {\mathcal B}_d)$, %where~$f:=d{\rm P}_{\!f}/d\mu_d$,
    by ${\cal F}_d:=\{f:=d{\rm P}/d\mu_d,\,  {\rm P}\!\in\!~\!{\cal P}^d\}$ the   corresponding family of densities, by ${\mathcal B}_d^n$ the $n$-fold product ${\mathcal B}_d\times\ldots\times {\mathcal B}_d$,     by~${\rm P}\n$ or $ {\rm P}_{\! f}^{(n)}$  the %joint 
  distribution of an i.i.d.~$n$-tuple with margi\-nals~${\rm P} = {\rm P}_{\! f}\in{\cal P}_d$, by ${\mathcal P}\n_d$ the corresponding collection $\{ {\rm P}_{\! f}^{(n)}\!, f\in{\cal F}_d \}$; ${\cal P}_d\n$\!-a.s.\ means ${\rm P}\n$\!-a.s. for all ${\rm P}\in {\cal P}_d\n\!\!$. Write $\overline{\text{spt}}(\rm P)$  for the support of $\rm P$,   spt$(\rm P)$ for its interior,    
  %  of  all {\it nonvanishing} Lebesgue densities over~$\mathbb{R}^d$, $d\in\mathbb{N}$---to be precise,  the family
%such that $\log f$ is locally bounded, that is, such that for all~$D\in\mathbb{R}^+$ there exist~$\Lambda_{D;f}$ and~$\lambda_{D;f}$  in~$(0,\infty)$  such that $\lambda_{D;f}\leq f({\bf x})\leq \Lambda_{D;f}$ for~$\Vert{\bf x}\Vert\leq D$; for simplicity, we refer to it as the family of {\it nonvanishing} Lebesgue densities. Let~${\cal P}_d$ denote the corresponding family of distributions, ${\cal P}_d\n$ the family of joint distributions of i.i.d.~$n$-tuples with common distribution in ${\cal P}_d$. The probability measures and distribution functions asso\-ciated with  densities $f, g$, ...\  are denoted by~${\rm P}\!_f,{\rm P}_g$,~..., and~$F,G$,~...,  respectively;  
%%while~
%${\rm P}\!_f\n\!\!,\, {\rm P}\!_g\n\!\!$,\,\!\! ...  stand for the distributions of  i.i.d.~$n$-tuples with densities $f$, $g$, ...  
%The notation~
$\mathcal{S}_{d-1}$,  $\mathbb{S}_d$, and $\overline{\mathbb{S}}_d$  for  the unit  sphere, the open, and the closed   unit ball   in~$\mathbb{R}^d$, respectively.\vspace{-2mm}
%}

\setcounter{equation}{0}

\section{Distribution and quantile functions, ranks and signs in $\mathbb{R}^d$}\label{MKsec}\vspace{-1mm}

As announced in the introduction, our definitions of center-outward distribution and quantile functions are rooted in the main result of McCann~(1985). Those definitions in Hallin~(2017) are given under the assumption that $\rm P$ has a nonvanishing density with  support~$\mathbb{R}^d$. Under that assumption, one safely can define the center-outward distribution  function as the  unique gradient of a convex function~$\nabla\phi$ pushing $\rm P$ forward to the uniform distribution over the unit ball.  That gradient, moreover, is a homeomorphism between $\mathbb{R}^d\setminus\nabla\phi^{-1}(\{{\bf 0}\})$ and the punctured unit ball $\mathbb{S}_d\!\setminus\!\{{\bf 0}\}$  (Figalli~2019) and its inverse naturally qualifies as a %center-outward
 quantile  function---a very simple and intuitively clear characterization.

Things are more delicate when the support of $\rm P$ is a strict subset of~$\mathbb{R}^d$, as uniqueness of $\nabla\phi$ then  only holds $\rm P$-a.s., and requires the slightly more elaborate definitions developed here.   
%(where the quantile function is defined prior to the distribution function). 
The two approaches, however, coincide in  case ${\rm P}$ has a non vanishing density over~$\mathbb{R}^d$.\vspace{-1.5mm}
%\section{Monge-Kantorovich Ranks and Signs}\label{MKsec}
%We are now ready to propose our definition of distribution and quantile functions in $\mathbb{R}^d$, along with their empirical counterparts. %Throughout, we are considering  ${\rm P}\!_f\in{\cal P}_d$, with density  $f\in{\cal F}^d$. 
%To start with, observe that~${\rm U}_1$, which is the Lebesgue-uniform distribution over the unit ball $\mathbb{S}_1$, is also the product of the uniform measure over the unit sphere~${\cal S}_0=\{ -1,1\}$ with a uniform measure over the unit interval of distances from the origin. We similarly  
%, this spherical uniform ${\rm U}_d$ no longer coincides, for $d\geq 2$, with the Lebesgue-uniform measure over  $\mathbb{S}_d$. 

\subsection{Center-outward distribution and quantile functions in $\mathbb{R}^d$}\label{defsubsec}\vspace{-1.5mm}

 Recall that a convex function $\psi$ from $\mathbb{R}^d$ to~$\mathbb{R}\cup\{\infty\}$ (a) is continuous on the interior of~dom$(\psi):=\{{\bf x} : \psi({\bf x})<\infty\}$ and (b)  Lebesgue-a.e.\ differentiable, with gradient~$\nabla\psi$, on dom$(\psi)$.   By abuse of language and notation,  call gradient and denote as~$\nabla\psi$ any function    coinciding~$\mu_d$-a.e.~with that gradient. A  statement of McCann's main result  adapted to our needs is the following. \footnote{Below we are  borrowing from the measure transportation literature the convenient notation $T\#{\rm P}_1={\rm P}_2$ for the distribution ${\rm P}_2$  of $T(X)$ under ${X}\sim{\rm P}$---we say that $T$ is {\it pushing forward} ${\rm P}_1$ to ${\rm P}_2$. }\vspace{-1.5mm}
 
 \begin{Thm} [McCann~1985]\label{ThMcC} Let ${\rm P}_1$ and ${\rm P}_2$  denote two distributions in~${\cal P}_d$. Then, {\it (i)}  the class of functions\vspace{-1mm}   %(from~$\mathbb{R}^d$ to $\mathbb{R}^d$)
\begin{align}\label{nablaPsi}
%\nabla\Psi_{\rm P}:=\left\{
%\nabla\psi\,\left\vert \, \psi  : \mathbb{R}^d \to \mathbb{R}\text{ convex,   such that } \nabla\psi\#{\rm P}={\rm U}_d
%\right.\right\}
\nabla\Psi_{{\rm P}_1; {\rm P}_2}:=
\left\{
\nabla\psi\,\left\vert \, \psi  : \mathbb{R}^d \to \mathbb{R}\right.\right. \text{ convex, lower}& \text{  semi-continuous, and\vspace{-1.0mm}} \\ 
&\left.\text{ such that } \nabla\psi\#{\rm P}_1={\rm P}_2
\right\}\vspace{-1.5mm}
\nonumber\end{align}
is not empty; {\it (ii)} if $\nabla\psi^\prime$ and $\nabla\psi^{\prime\prime}$ are two elements of $\nabla\Psi_{{\rm P}_1; {\rm P}_2}$, they coincide~${\rm P}_1$-a.s.;\,\footnote{That is, ${\rm P}_1\!\left(\{{\bf x} : \nabla\psi^{\prime}({\bf x})\neq \nabla\psi^{\prime\prime}({\bf x})\}\right) = 0$; in particular, $\nabla\psi_1({\bf x})=\nabla\psi_2({\bf x})$ Lebesgue-a.e.\ for~${\bf x}\in\text{spt}({\rm P}_1)$.}   {\it (iii)} if ${\rm P}_1$ and ${\rm P}_2$ have finite moments of order two, any element of $\nabla\Psi_{{\rm P}_1; {\rm P}_2}$ is an optimal quadratic  transport pushing ${\rm P}_1$ forward to ${\rm P}_2$. %\vspace{-1.5mm}
\end{Thm}

Although  not mentioned in McCann's main result (p. 310 of McCann~(1995)), lower semi-continuity in \eqref{nablaPsi} can be imposed   without loss of generality (this follows, for instance, from his proof of uniqueness on p. 318).

Denoting by~${\rm U}_d$ the {\it spherical uniform} distribution over $\mathbb{S}_d$,\footnote{Namely,  the product of the uniform   over the unit sphere~${\cal S}_{d-1}$ with a uniform   over the unit interval of distances to the origin. While  ${\rm U}_d$  coincides,  for $d=1$, with the Lebesgue-uniform over $\mathbb{S}_1$, this is no longer the case for $d>1$;  we nevertheless  still call  it {\it uniform over the unit ball}.} consider Theorem~\ref{ThMcC} and \eqref{nablaPsi} for~${\rm P}_1={\rm U}_d$ and ${\rm P}_2={\rm P}\in {\cal P}_d$. Since the support of ${\rm U}_d$  is~$\overline{\mathbb{S}}_d$ (which is convex and compact), $\psi$ is uniquely determined over ${\mathbb{S}}_d$ if we impose, without loss of generality,~$\psi({\bf 0})=~\!0$.\footnote{Indeed, two convex functions with a.e. equal gradients on an open convex set are equal up  to an additive constant: see Lemma~2.1 in del Barrio and Loubes~(2019).} Outside ${\mathbb S}_d$ (that is, on a set with~${\rm U}_d$-probability zero), let us further impose again\vspace{-1mm}
 \begin{equation}\label{aeun}
\psi({\bf u}) = \infty\   \text{ for }  \ \Vert {\bf u}\Vert > 1\quad  \text{ and } \quad \psi({\bf u}) = \liminf_{\mathbb{S}_d \ni{\bf v}\to{\bf u}}\psi (\bf v)\   \text{ for }  \  \Vert {\bf u}\Vert  = 1.\vspace{-2mm}
\end{equation}
 The domain of $\psi$ is dom$(\psi):=\{{\bf u}\big\vert \psi({\bf u})<\infty \}=\overline{\mathbb{S}}_d$. A convex function is differentiable a.e.\ in the interior of its domain. Hence,   the gradient $\nabla\psi$ of~$\psi $ satisfying~\eqref{aeun}   exists, is unique a.e.\  in   $\mathbb{S}_d$, and    still belongs to  $\nabla\Psi_{{\rm U}_d; {\rm P}}$. %; being a.e. unique on ${\mathbb S}_d$.  

  Inspired by the univariate case as described in Section~\ref{meastranspsec},  we propose the following definitions of the center-outward  quantile function of ${\rm P} \in{\cal P}_d$. \vspace{-1mm}
  \begin{defin}\label{qfdef} {\em 
Call {\em center-outward quantile function} ${\bf Q} _{\pms}$ of ${\rm P} \in{\cal P}_d$  the a.e.\ unique element $\nabla\psi\in\nabla\Psi_{{\rm U}_d; {\rm P}}$ such that $\psi$ satisfies \eqref{aeun}. 
 }
 \end{defin}
   In general, thus,   ${\bf Q} _{\pms}$ is a {\it class} of Lebesgue-a.e.\ equal functions rather than a  function. Each element in that class   pushes ${\rm U}_d$ to $\rm P$, hence fully characterizes~$\rm P$. Such a.e. uniqueness, in probability and statistics, is not uncommon:   densities, conditional expectations, likelihoods, MLEs,  ... all are defined up to sets of probability zero. As we shall see, however, strict uniqueness  does hold for important families of distributions, for which $\psi$ is everywhere differentiable over $\mathbb{S}_d$. 

Next, let us proceed with the definition of the {\em center-outward distribution function} ${\bf F} _{\!{\pms}}$. Consider the Legendre transform\vspace{-1mm} 
\begin{equation}\label{Legendre} 
\phi({\bf x}):=\psi^*({\bf x}):=\sup_{{\bf u}\in{\mathbb{S}}_s}\left(\langle 
{\bf u},{\bf x}
\rangle -\psi({\bf u})
\right)\qquad {\bf x}\in\mathbb{R}^d\vspace{-2mm}
\end{equation}
of the a.e.-unique convex function $\psi$ (satisfying~$\psi({\bf 0})=0$ and~\eqref{aeun}) of which~${\bf Q} _{\pms}$ is the gradient. Being the $\sup$ of a 1-Lipschitz function,   $\phi$ also is 1-Lipschitz. It follows that $\phi$ is a.e.\ differentiable, with $\Vert \nabla\phi({\bf x})
\Vert \leq 1$, so that (Corollary~(A.27) in Figalli~(2017)), denoting by $\partial\phi ({\bf x})$ the subdifferential of $\phi$ at~$\bf x$,\footnote{Recall that the subdifferential of $\phi$ at ${\bf x}\in {\mathbb R}$ is the set  $\partial\phi ({\bf x})$ of all ${\bf z}\in {\mathbb R}^d$ such\linebreak  that~$\phi({\bf y}) - \phi({\bf x})\geq \langle {\bf z}, {\bf y} - {\bf x}
\rangle$ for all $\bf y$; $\phi$ is differentiable at $\bf x$ iff $\partial\phi ({\bf x})$ consists of a single point, $\nabla\phi({\bf x})$.} 
\begin{equation}\label{inball}
\partial\phi({\mathbb R}^d):=\bigcup _{{\bf x}\in{\mathbb R}^d}\partial\phi ({\bf x})\, \subseteq\overline{\mathbb{S}}_d.\vspace{-2mm}
\end{equation}
Moreover, since $\rm P$ has a density, Proposition 10 in McCann~(1995) implies that \vspace{-1mm}
\begin{equation}\label{inverse}
\nabla\psi\circ\nabla\phi ({\bf x}) = {\bf x}\ \text{ $\rm P$-a.s.\ and } \  \nabla\phi\circ\nabla\psi ({\bf u}) = {\bf u}\ \text{ ${\rm U}_d$-a.s. }\vspace{-1mm}
\end{equation}
In view of \eqref{inball} and the second statement in~\eqref{inverse}, ${\bf F}_\pms:=\nabla\phi$ takes  values in~$\overline{\mathbb{S}}_d$ and pushes $\rm P$ forward to ${\rm U}_d$. Moreover, there exist   subsets  $\breve{\text{spt}}({\rm P})$ and $\breve{\mathbb{S}}_d$ of~spt$({\rm P})$ and $\mathbb{S}_d$, respectively, such that (a) 
%${\text{spt}}({\rm P})\setminus
${\rm P}\big(\breve{\text{spt}}({\rm P})\big) =1=
%$ and $\mathbb{S}_d\setminus
{\rm U}_d\big(\breve{\mathbb{S}}_d\big)$,   %have Lebesgue measure zero,  
 (b)  the restriction to $\breve{\text{spt}}({\rm P})$ of $\nabla\phi=:{\bf F}_\pms$ and the restriction to~$\breve{\mathbb{S}}_d$ of $\nabla\psi=:{\bf Q}_\pms$    are bijections, and  (c) those restrictions are the  inverse of each other.  Accordingly,~${\bf F}_\pms$  qualifies as a  center-outward distribution function. \vspace{-1mm}
\color{black}

% In dimension $d$. %The following 
%The center-outward quantile function~${\bf Q}_\pms$ and $\nabla\phi$, thus,  are each other's  inverse;  in view of \eqref{inball} and the second statement in~\eqref{inverse}, $\nabla\phi$ has  values in $\overline{\mathbb{S}}_d$ and pushes $\rm P$ forward to ${\rm U}_d$. Accordingly, it qualifies as a  center-outward distribution function. % In dimension $d$. %The following definitions  coincide, for $d=1$, with the univariate ones given in Section~\ref{pmranksec}.\vspace{-1mm} % \smallskip
\begin{defin}\label{dfdef} {\em 
%{\rm 
%Let ${\rm P}_{\!f}\in{\cal P}_d$ and define %${\rm P}\in{\cal P}_d$
Call ${\bf F} _{\!{\pms}}\!:=\!\nabla\phi$ the {\em center-outward distribution function}   of~${\rm P}\! \in~\!\!{\cal P}_d$. %  any element of the nonempty class $\nabla\Psi_{\rm P}$ defined in \eqref{nablaPsi}. %the $\rm P$-a.s. unique  gradient  of  convex function $\nabla\psi$ %$\psi   }
}\vspace{-3mm}
 \end{defin}

The following propositions summarize the main properties of ${\bf F}_{\!\pms}$ and ${\bf Q}_{\pms}$, some of which already have been mentioned in previous comments. \vspace{-1mm}

%; the proofs of the others are elementary and left to the reader. 
\begin{prop}\label{Fpmprop} Let ${\bf Z}\sim{\rm P}\in{\cal P}_d$ and denote by ${\bf F}_{\!\pms}$ the center-outward distribution function of ${\rm P}$. Then,\vspace{-1mm}
\begin{enumerate}
\item[{\it (i)}] ${\bf F}_{\!\pms}$ takes values in $\overline{{\cal S}}_d$ and ${\bf F}_{\!\pms}\#{\rm P}={\rm U}_d$: ${\bf F}_{\!\pms}$, thus, is a probability-integral transformation;
\item[{\it (ii)}] $\Vert{\bf F}_{\!\pms}({\bf Z})\Vert $  is uniform over $[0,1]$, ${\bf S}({\bf Z}):={\bf F}_{\!\pms}({\bf Z})/ \Vert{\bf F}_{\!\pms}({\bf Z})\Vert$  uniform over~${\cal S}_{d-1}$, and they are mutually independent;  
\item[{\it (iii)}] ${\bf F}_{\!\pms}$ entirely characterizes $\rm P$;
\item[{\it (iv)}] for $d\!=\!1$, ${\bf F}_{\!\pms}\!$ coincides with $2F -1$    ($F$ the traditional distribution function).% where $F$ stands for the traditional distribution function.
\end{enumerate}\vspace{-2mm}
\end{prop}

For $q\in(0,1)$, define the {\em center-outward quantile region} and {\em center-outward quantile contour} of order  as \vspace{-1mm}
\begin{equation}\label{qcontdef}
\mathbb{C}(q)\!:=\!{\bf Q} _{\pms}(q\,\bar{\mathbb{S}}_d)\!=\!\{{\bf z}\big\vert \Vert{\bf F}_{\!\pms}({\bf Z})\Vert\leq q \} \text{ and }  \mathcal{C}(q)\!:=\!{\bf Q} _{\pms}(q\, {\mathcal{S}}_{d-1})\!=\!\{{\bf z}\big\vert \Vert{\bf F}_{\!\pms}({\bf Z})\Vert = q \},\vspace{-2mm}\end{equation}
respectively. \vspace{-2mm}

\begin{prop}\label{Qpmprop} %Let  ${\bf Q}_{\pms}$ denote the center-outward quantile  function of ${\rm P}\!\in~\!{\cal P}_d$. 
 Let ${\rm P}\!\in~\!{\cal P}_d$ have center-outward quantile  function ${\bf Q}_{\pms}$. Then,\vspace{-1mm}
\begin{enumerate}
\item[{\it (i)}] ${\bf Q}_{\pms}$ pushes ${\rm U}_d$ forward to ${\rm P}$, hence entirely characterizes $\rm P$;
\item[{\it (ii)}]  the {\em center-outward quantile region} $\mathbb{C}(q)$, $0< q <1$,  %:={\bf Q} _{\pms}(q\,\bar{\mathbb{S}}_d)=\{{\bf z}\big\vert \Vert{\bf F}_{\!\pms}({\bf Z})\Vert\leq q \}$
 has $\rm P$-probability content $q$;  
\item[{\it (iii)}]  ${\bf Q}_{\pms}(u)$ coincides, for $d=1$, with  $\inf\{x\big\vert F(x)\geq (1 +  u )/2)\}$, $u\in (-1,1)$, and~$\mathbb{C}(q)$, $q\in (0,1)$, with ($F$   the traditional   distribution function)\footnote{Since ${\bf Q}_{\pms}$ is only a.e.\  defined, one can as well use ${\text{spt}}({\rm P})$ in \eqref{univinteq}; this, however, no longer produces a closed region and may result in an empty set $\bigcap_{0<q<1}\mathbb{C}(q)$ of medians  in \eqref{mediandef}.}\vspace{-1mm}% the interval 
\begin{equation}\label{univinteq}
\left[\inf\{x\big\vert F(x)\geq (1-q)/2\},\, \inf\{x\big\vert F(x)\geq (1+q)/2\}\right]\bigcap\overline{\text{\rm spt}}({\rm P}).\vspace{-1mm}
\end{equation}
% where 
\end{enumerate}\vspace{-2mm}
\end{prop}
The modulus $\Vert{\bf F}_{\!\pms}({\bf x})\Vert $ thus is the order of the quantile contour %${\cal C}(\Vert{\bf F}_{\!\pms}({\bf x})\Vert)$
 and  the $\rm P$-probability content of the largest quantile region containing 
    ${\bf x}$; the unit vector~${\bf S}({\bf z}):={\bf F}_{\!\pms}({\bf z})/ \Vert{\bf F}_{\!\pms}({\bf z})\Vert$ has the interpretation of a multivariate sign. Note that the definition of~$\mathbb{C}(0)$ so far has been postponed.  %are called {\em center-outward medians}. 

% 
% \begin{defin}\label{dfdef} {\em 
%Call {\em center-outward quantile function} ${\bf Q} _{\!{\pms}}$ of ${\rm P} \in{\cal P}_d$  any element of the nonempty class $\left\{ {\bf F} _{\!{\pms}}^{-1} \big\vert {\bf F} _{\!{\pms}}\in\nabla\Psi_{\rm P}\right\}$.\footnote{This  alone does not imply that ${\bf Q}_\pms$ is a.e. the gradient of a convex function.}  
%  Denoting by~$q\,\bar{\mathbb{S}}_d$ and~$q\mathcal{S}_{d-1}$ the closed ball and the hypersphere centered at the origin with radius~$q\in~\![0,1)$, the quantile function~${\bf Q} _{\pms}$  characterizes center-outward  {\em quantile regions}   and {\em   contours}   $\mathbb{C}(q):={\bf Q} _{\pms}(q\,\bar{\mathbb{S}}_d)$ and~$\mathcal{C}(q):={\bf Q} _{\pms}(q\mathcal{S}_{d-1})$, respectively, of order $q$ (i.e., with probability contents $q$). The elements of  $\mathbb{C}(0)$  are called {\em center-outward medians}.
%  }
%   \end{defin}
%   
 %  Note that ${\rm P}(\mathbb{C}(0))= {\rm U}_d({\bf 0})=0$
    
  These properties are not entirely satisfactory, though,  and  a bijection between~$\breve{\text{spt}}({\rm P})$ and $\breve{\mathbb{S}}_d$ is not enough for meaningful quantile concepts to exist. \color{black}The terminology {\it quantile region} and {\it quantile contour}, indeed, calls for a collection of connected, closed, and strictly nested regions $\mathbb{C}(q)$---i.e., 
  such that~${\mathbb C}(q_1)\subsetneq{\mathbb C}(q)\subsetneq{\mathbb C}(q_2)$ %, %such that, 
   for   any~$0< q_1<q<q_2<1$---with continuous boundaries $\mathcal{C}(q)$ of Hausdorff dimension $d-1$;   a reasonable\footnote{By analogy with the definition of $\mathbb{C}(q)$ for $q>0$, one may be tempted to define ${\mathbb C}(0)$ as~${\bf Q}_\pm ({\bf 0})$. This yields for ${\mathbb C}(0)$ an arbitrary  point  in the subdifferential $\partial\psi({\bf 0})$ which, unless that subdifferential consists of a single point,     cannot satisfy~\eqref{mediandef}. } definition of a median set then is, with ${\mathbb C}(q)$ ($q\in (0,1)$) defined in \eqref{qcontdef},%\vspace{-1mm}  
  \begin{equation}\label{mediandef} {\mathbb C}(0):=\text{${\bigcap}_{0<q<1}$}{\mathbb C}(q).\vspace{-0mm}\end{equation} 
  
  Such attractive  properties do not hold, unfortunately, and the median set~${\mathbb C}(0)$, as defined in~\eqref{mediandef}  may be empty, unless ${\bf Q}_\pm$, hence  ${\bf F}_{\!\pm}$, enjoy some continuity properties, which require   regularity assumptions on~$\rm P$ and its support: see Appendix~\ref{NUMAPP} for examples.  A sufficient condition, as we shall see, is the continuity of ${\bf u}\mapsto{\bf Q}_\pm({\bf u})$, at least on $\mathbb{S}_d\!\setminus\!\{{\bf 0}\}$.

%  In dimension $d=1$, for instance, the regions  $\mathbb{C}(q)$ may be disconnected,  reducing to an   interval  for connected~spt${({\rm P})}$ only.  All those properties are satisfied, however, if~$\rm P$ is such that~${\bf u}\mapsto{\bf Q}_\pm({\bf u})$ is continuous, at least on $\mathbb{S}_d\!\setminus\!\{{\bf 0}\}$. 
  
To see this and understand the special role of ${\bf 0}$, recall that   ${\bf Q}_\pm$   is only a.e. defined. Hence, ${\bf Q}_\pm ({\bf 0})$ can take any value compatible with the convexity of~$\psi$---namely, any single point in the subdifferential~$\partial\psi({\bf 0})$ of the uniquely defined~$\psi$ satisfying~$\psi({\bf 0})=0$.  As a consequence, continuity of ${\bf Q}_\pm$   is impossible unless~$\partial\psi({\bf 0})$ (and all other subdifferentials---not just almost all of them) contains exactly one single point.  

Continuity of the restriction of ${\bf Q}_\pm$ to a closed spherical annulus $q^+ {\overline{\mathbb S}}_d \!\setminus\! q^- {\mathbb S}_d$   yields continuous contours ${\cal C}(q)$ and strictly nested closed regions ${\mathbb C}(q)$ for the orders~$q\in[q^-, q^+]$. Letting $q^+ = 1-\epsilon$ and $q^-=\epsilon$ with $\epsilon>0$ arbitrarily small, continuity of ${\bf Q}_\pm$ everywhere except possibly at $\bf 0$ thus yields continuous contours and strictly nested closed regions for the orders $q\in(0, 1)$. 

The definition of quantile regions implies that all possible values of ${\bf Q}_\pm({\bf 0})$ are contained in the intersection $\bigcap_{0<q<1}{\mathbb{C}}(q)$ of all regions of order $q>0$; hence,~$\partial\psi({\bf 0})\subseteq \bigcap_{0<q<1}{\mathbb{C}}(q)$.  Conversely,  any point  ${\bf u}\neq{\bf 0}\vspace{1mm}$ has a neighborhood~$V({\bf u})$ such that ${\bf 0}\notin V({\bf u})$.  Assuming that ${\bf Q}_\pm$ is continuous everywhere but at $\bf 0$, ${\bf Q}_\pm(V({\bf u}))\cap\bigcap_{0<q<1}{\mathbb{C}}(q)=\emptyset$. Hence, $\partial\psi({\bf 0})= \bigcap_{0<q<1}{\mathbb{C}}(q)\vspace{1mm}$.
 As the subdifferential of a convex function $\psi$, $\partial\psi({\bf 0})$, hence $\bigcap_{0<q<1}{\mathbb{C}}(q)$, is closed and convex. Because $\rm P$ has a density and $\bf 0$ is in the interior of $\psi$'s domain,  it also is compact and has Lebesgue measure zero (Lemma~A.22 in Figalli~(2017)).

It follows that by defining the median set as ${\mathbb C}(0):= \bigcap_{0<q<1}{\mathbb{C}}(q)=\partial\psi({\bf 0})\vspace{0.5mm}$ (instead of ${\mathbb{C}}(0):={\bf Q}_\pm({\bf 0})$, which is not uniquely determined), we do not need continuity at $\bf 0$ to obtain strict nestedness of all quantile contours and regions---now including   ${\mathbb C}(0)$---while \eqref{mediandef}, of course, is automatically satisfied.

This, which justifies giving up continuity at $\bf 0$ (and only there), is not an unimportant detail: Proposition~\ref{Figalliprop} below indeed shows that important classes of distributions yield quantile functions ${\bf Q}_\pm$ that are not continuous over  the ball $\mathbb{S}_d$ but nevertheless enjoy  continuity over the punctured ball~$\mathbb{S}_d\!\setminus\!\{{\bf 0}\}$. 

Denote by ${\cal P}_d^{\text{conv}}$ the class of  
 distributions~${\rm P}_{\!f}\in{\cal P}_d$ such that {\it (a)} $\overline{\text{spt}}({\rm P}_{\!f})$ is a  convex set\footnote{That convex set  %$\mathcal X$
  is not necessarily bounded.} %$\mathcal X$
   and, {\it (b)} for all~$D\in\mathbb{R}^+\!$, there exist constants~$\Lambda_{D;f}$ and~$\lambda_{D;f}$  in~$(0,\infty)$  such that~$\lambda_{D;f}\leq~\!f({\bf x})\leq~\!\Lambda_{D;f}$ for all~${\bf x}\in (D\,\mathbb{S}_d)\cap\overline{\text{spt}}({\rm P}_{\!f})$. That class includes the class ${\cal P}_d^+$ of distributions with support spt$({\rm P})=\mathbb{R}^d$ considered by Hallin~(2017) and Figalli~(2018).

The following result, which  establishes the continuity properties of   ${\bf F}_{\!\pms}$ and~${\bf Q} _{\pms}$ for ${\rm P}\in{\cal P}_d^{\text{conv}}$, extends the main result obtained for ${\cal P}_d^+$ by  Figalli (2018) and is borrowed, with some minor additions, from   del Barrio et al.~(2019).\vspace{-1mm} 

\setcounter{Prop}{2}

\begin{Prop}\label{Figalliprop}  Let ${\rm P}\in{\cal P}_d^{\text{\rm conv}}\!$ have density $f$ and support $\overline{\text{\rm spt}}({\rm P})$. %$\mathcal X$.
 Then,
its  center-outward  distribution function ${\bf F}_\pm\!=\!\nabla\phi$ is continuous and single-valued on $\mathbb{R}^d$ and
$\|{\bf F}_\pm(\mathbf{x})\|=1$ for   $\mathbf{x}\notin{\text{\rm spt}}({\rm P})$. Furthermore, 
there exists a compact convex set $K\subset \overline{\text{\rm spt}}({\rm P})$ with Lebesgue measure zero such that\vspace{-1mm} 
\begin{enumerate}
\item[{\it (i)}] ${\bf F}_\pm$ and  the center-outward quantile function ${\bf Q}_\pm\!=\!\nabla\psi$  are homeomorphisms between $\mathbb{S}_d\!\setminus\!\{{\bf 0}\}$ and~${\text{\rm spt}}({\rm P})\!\setminus\! K$, on which they are inverse of each other; for $d=1,\, 2$, however, $K$ contains a single point and the homeomorphisms are between $\mathbb{S}_d$ and~$ {\text{\rm spt}}({\rm P})$;
%(i)
%\footnote{Hence, $\nabla\Psi_{{\rm U}_d;{\rm P}}$ contains a unique element which is a gradient, not just  an a.e.\ gradient, and~$\phi$ is unique up to an additive constant} the median region~$\mathbb{C}(0)$ 
%%:=\{{\bf x}\vert  \nabla\psi({\bf x})={\bf 0}\}$
% is compact and has Lebesgue measure zero;% $\mathbb{C}(0)=\mathcal{C}(0)$;
%(\nabla\Psi)^{-1}(\{0\})$ is a compact set with Lebesgue measure zero, and
\item[{\it (ii)}] the quantile contours ${\mathcal C}(q)$ and regions ${\mathbb C}(q)$, $0<q<1$ defined by~${\bf Q}_\pm$ are such that $\bigcap_{0<q<1}{\mathbb C}(q)=\partial\psi(\{{\bf 0}\})=K\vspace{0.5mm}$; $K$ thus qualifies as the median set~$\mathbb{C}(0)$ of $\rm P$ as defined in \eqref{mediandef}. \vspace{-2mm}
\end{enumerate}
If, moreover,  $f\in{\cal C}^{k,\alpha}_{\text{\rm loc}}({\text{\rm spt}}({\rm P}_{\!f}))$ for some $k\geq~\! 0$,~then\vspace{-1mm} 
\begin{enumerate}  
%\noindent (ii) 
%the restriction of $\nabla\phi$ to $\mathbb{R}^d\! \setminus\!  \mathbb{C}(0)$%(\nabla\Psi)^{-1}(\{0\})$
%  is a homeomorphism from~$\mathbb{R}^d\!\setminus\! \, \mathbb{C}(0)$ to~$\mathbb{S}_d\!\setminus\!  \{{\bf 0}\}$, with inverse (defined on $\mathbb{S}_d\!\setminus\! \{{\bf 0}\}$)  $\nabla\psi$, where $\psi=\phi^*$ is the Legendre transform of $\phi$; for $d=1, 2$, however,~$\mathbb{C}(0)$ consists of a single point and~$\nabla\phi$ is a homeomorphism from~$\mathbb{R}^d$ to $\mathbb{S}_d$;\vspace{1mm}
\item[{\it (iii)}] 
\begin{enumerate}  
%\noindent (iii)
% if, moreover,  $\rm P$ has Lebesgue density $f\in{\cal C}^{k,\alpha}_{\text{\rm loc}}\big(\mathbb{R}^d\big)$ for some $k\geq 0$, then %\vspace{-1mm}  %for ${\bf x}\in \mathbb{R}^d\setminus K$, 
%\begin{enumerate}
\item[{\it (a)}]   ${\bf Q}_\pm$ and ${\bf F}_\pm$   are  diffeomorphisms of class ${\cal C}^{k+1,\alpha}_{\text{\rm loc}}$  between $\mathbb{S}_d\!\setminus\!  \{{\bf 0}\}$ and~${\text{\rm spt}}({\rm P})\!\setminus\!  \mathbb{C}(0)$;
\item[{\it (b)}] $\displaystyle{f({\bf z})={c_d^{-1}}\, \text{\rm det}\big[{\bf H}_{\psi}\! \big( \nabla \phi ({\bf z})\big)\big]\Vert \nabla\phi ({\bf z}) \Vert ^{1-d} I\big[{\bf z}\in{\text{\rm spt}}({\rm P}_{\!f})\!\setminus\!  \mathbb{C}(0)\big]}$   
where~$c_d$ is the area $2\pi^{d/2}/\Gamma (d/2)$   of the unit sphere~${\cal S}_{d-1}$ and ${\bf H}_{\phi^*} ({\bf u})$  the Hessian\footnote{That Hessian exists since $k\geq 0$ and $\nabla \phi ({\bf z})\neq {\bf 0}$ for ${\bf z}\in{\text{\rm spt}}({\rm P}_{\!f})\!\setminus\!  \mathbb{C}(0)$.} of $\psi$ computed at $\bf u$. \vspace{-2mm}
\end{enumerate}
\end{enumerate}
\end{Prop}

Denote by ${\mathcal P}_d^{\pms}\subset{\mathcal P}_d$ the class of all distributions of the form ${\rm P}=\nabla{\footnotesize\text{$\Upsilon$}}$ where~${\footnotesize\text{$\Upsilon$}}$ is convex and 
% is resulting from pushing~${\rm U}_d$ forward by a  gradient of convex function
  $\nabla{\footnotesize\text{$\Upsilon$}}$ a homeomorphism from $\mathbb{S}_d\setminus\{\mathbf{0}\}$ to~$\nabla{\footnotesize\text{$\Upsilon$}} \left(\mathbb{S}_d\setminus\{\mathbf{0}\}\right)$ such that~$\nabla{\footnotesize\text{$\Upsilon$}} \left(\{\mathbf{0}\}\right)$ is a compact convex set of Lebesgue measure zero.  By construction, such $\rm P\in{\mathcal P}_d^{\pms}$ has center-outward quantile function~${\bf Q}_\pms=\nabla{\footnotesize\text{$\Upsilon$}}$, center-outward distribution function ${\bf F}_\pms({\bf x})=(\nabla{\footnotesize\text{$\Upsilon$}})^{-1}$ for $\bf x$ in the range of~$\nabla{\footnotesize\text{$\Upsilon$}}$ and~$\|\mathbf{F}_\pms(\mathbf{x})\|=1$  outside that range, and satisfies Proposition~\ref{Figalliprop}; the latter actually can be rephrased as ${\cal P}_d^{\text{\rm conv}}\!\subset{\mathcal P}_d^{\pms}$, with the following immediate corollary in terms of quantile regions and contours.\vspace{-1mm}

\begin{cor}\label{Figallicor}  For any ${\rm P}\in{\mathcal P}_d^{\pms}$ (hence, any ${\rm P}\in{\cal P}_d^{\text{\rm conv}}\!$) 
 and $q\in[0,1)$, the quantile regions $\mathbb{C}(q)$ 
%of ${\rm P}\in{\cal P}_d^+$
 are closed, connected, and nested, with continuous boundaries~$\mathcal{C}(q)$ satisfying~$\mu_d(\mathcal{C}(q))=0$. 
\end{cor}

 For any  distribution ${\rm P}\in{\mathcal P}_d^{\pms}$, %{\cal P}_d^+$,
  ${\bf F} _{\!{\pms}}$ thus induces a (partial) ordering of~$\mathbb{R}^d$ similar to the ordering induced on the unit ball by  the system of polar coordinates, and actually coincides with the ``vector rank transformation" considered in Chernozhukov et al.~(2017) when the reference distribution is ${\rm U}_d$. The quantile contours~$\mathcal{C}(q)$ 
  also have the interpretation of depth contours associated with   their  Monge-Kantorovich depth.  Their  assumption of a compact support satisfying  Cafarelli regularity conditions   are  sufficient (not necessary)   for ${\rm P}\in{\mathcal P}_d^{\pms}$.\vspace{-1mm} 

\subsection{Center-outward ranks and signs in $\mathbb{R}^d$}\label{gridsec}\vspace{-1mm} 

Turning to the sample situation, let    ${\bf Z}\n :=\big({\bf Z}\n_1,\ldots , {\bf Z}\n_n\big)$ denote an $n$-tuple of random vectors--- observations or residuals associated with some parameter $\pmb\theta$ of interest. We throughout consider the case that the~${\bf Z}\n_i$'s  are (possibly, under parameter value $\thetab$) i.i.d.\   with   density~$f\in{\cal F}^d$, distribution ${\rm P}$ and center-outward distribution function ${\bf F} _{\!{\pms}}$.  

For  the empirical counterpart ${\bf F} _{\!{\pms}}\n$ of ${\bf F} _{\!{\pms}}$, we propose the following extension of the univariate concept  described in Appendix~\ref{1dimsec}.  Assuming $d\geq 2$,  let~$n$ factorize into\vspace{-1mm} 
\begin{equation}\label{factorization}n= n_Rn_S + n_0,\qquad  n_R,\ n_S, \  n_0\in\mathbb{N}, \quad 0\leq n_0<\min ( n_R, n_S) % ,\quad n_r\to\infty , \ n_s\to\infty
\vspace{-1mm} \end{equation}
where $n_R\to\infty$ and  $n_S\to\infty$ as $n\to\infty$ (implying $n_0/n\to 0\vspace{0.5mm}$); \eqref{factorization} is extending to $d\geq 2$ the  factorization of $n$ into~$n=\lfloor\frac{n}{2}\rfloor 2 + n_0\vspace{0.5mm}$ with $n_0= 0$ ($n$ even) or~$n_0=1$ ($n$ odd) that leads, for $d=1$, to the grids~\eqref{evenoddgrid}.% and \eqref{evenoddgrid}.  

Next, consider a sequence of  ``regular grids" of $n_Rn_S$ points in the unit ball~$\mathbb{S}_d$ obtained as the intersection between% \smallskip
\begin{enumerate}
\item[--] 
%\noindent -- 
a ``regular" $n_S$-tuple ${\mathfrak S}^{(n_S)}:=({\bf u}_1,\ldots {\bf u}_{n_S})$ of unit vectors, and %\vspace{-2mm}
\item[--] 
%
%\noindent --
 the $n_R$ hyperspheres centered at $\bf 0$, with radii $\dfrac{j}{n_R  +1}$, 
$j=1,\ldots ,n_R$,\vspace{-1mm} %, \dfrac{2}{n_R  +1}, \ \ldots ,\  \dfrac{n_R }{n_R +1}$, 
\end{enumerate}

\noindent along with $n_0$ copies of the origin whenever $n_0>0$. In theory, by  a ``regular" $n_S$-tuple ${\mathfrak S}^{(n_S)}=({\bf u}_1,\ldots {\bf u}_{n_S})$, we only mean that   the sequence of uniform discrete distributions over~$\{{\bf u}_1,\ldots {\bf u}_{n_S}\}$   converges weakly, as~$n_S\to\infty$, to the uniform distribution over~${\cal S}_{d-1}$. In practice, each $n_S$-tuple should be ``as uniform as possible". For $d=2$, perfect regularity can  be achieved by  dividing the unit circle into~$n_S$ arcs of equal length~$2\pi/n_S$. Starting with~$d=3$,  however, this typically is no longer possible. A random array of $n_S$ independent and uniformly distributed unit vectors does satisfy (almost surely) the weak convergence requirement. More regular deterministic arrays (with faster convergence) can be considered, though, such as  the {\it low-discrepancy sequences}  of the type considered in numerical integration and Monte-Carlo methods (see, e.g., Niederreiter~(1992),   Judd~(1998), Dick and Pillichshammer (2014), or Santner et al.~(2003)), which are current practice in numerical integration and the design  of computer experiments. 

The resulting  grid of $n_Rn_S$ points  then is such that the discrete distribution with probability masses $1/n$ at each gridpoint and probability mass~$n_0/n$ at the origin 
%---call it {\it uniform over the augmented grid}---
converges weakly to the uniform ${\rm U}_d$ over the ball~${\mathbb{S}}_d$.
%---recall that, by uniform,  we mean {\it sphetically uniform}.
 That grid, along with the $n_0$ copies of the origin, is called the {\it augmented grid} ($n$ points).

We then define ${\bf F}_{\pms}\n({\bf Z}\n_i)$, $i=1,\ldots , n$ as the solution of an optimal coupling problem between the observations and the augmented grid. Let~$\cal T$ denote the set of all possible bijective mappings between ${\bf Z}\n_1,\ldots , {\bf Z}\n_n$ and  the~$n$ points of the augmented  grid just described.  Under the assumptions made, the ${\bf Z}\n_i$'s are all distinct with probability one, so that $\cal T$   contains~$n!/n_0!$ classes of $n_0!$ indistinguishable couplings each (two couplings $T_1$ and~$T_2$ are indistinguishable if~$T_1({\bf Z}\n_i)=T_2({\bf Z}\n_i)$ for all $i$). \vspace{-1mm}

\begin{defin}\label{empFpm} {\rm Call {\it empirical center-outward distribution function} any of the
%(a.e.\ equal)
 mappings %\vspace{-1mm}
% $
 ${\bf F} _{\!{\pms}}\n\!\! : %\ {\bf Z}\n :=
 \big({\bf Z}\n_1,\ldots , {\bf Z}\n_n\big)\!\mapsto\! \big({\bf F} _{\!{\pms}}\n({\bf Z}\n_1),\ldots ,{\bf F} _{\!{\pms}}\n({\bf Z}\n_n)
\big)=:{\bf F} _{\!{\pms}}\n({\bf Z}\n)$ % \vspace{-1mm}$$
satisfying\vspace{-2mm}\vspace{-1mm}  
\begin{equation}\label{optass1} 
\sum_{i=1}^n\big\Vert {\bf Z}\n_i - {\bf F} _{\!{\pms}}\n({\bf Z}\n_i)
\big\Vert ^2%)\pr\big({\bf Z}\n_i - {\bf F}\n({\bf Z}\n_i)
%\big) 
=
 \min_{T\in{\cal T}} \sum_{i=1}^n\big\Vert {\bf Z}\n_{i} - T({\bf Z}\n_i)
\big\Vert ^2\vspace{-1mm}
\end{equation}\setcounter{equation}{9}
or, equivalently, \vspace{-2.5mm}
\begin{equation}\label{optass1} 
\sum_{i=1}^n\big\Vert {\bf Z}\n_i - {\bf F} _{\!{\pms}}\n({\bf Z}\n_i)
\big\Vert^2 
=
 \min_\pi \sum_{i=1}^n\big\Vert {\bf Z}\n_{\pi (i)} - {\bf F} _{\!{\pms}}\n({\bf Z}\n_i)
\big\Vert ^2\vspace{-2.5mm}
\end{equation}
where the set $\{{\bf F} _{\!{\pms}}\n({\bf Z}\n_i)\vert\ i=1,\ldots ,n\}$ consists of the $n$ points of the augmented  grid and 
$\pi$ ranges over the $n!$ possible permutations of $\{1,2,\ldots ,n\}$.} \end{defin}\vspace{-0.5mm}%\medskip

Determining such a coupling is a standard optimal assignment problem, which   takes the form of a linear program for which efficient   algorithms are available (see  Peyr\' e and Cuturi~(2019) for a recent survey).   

  Call {\it order statistic} ${\bf Z}\n_{{\scriptscriptstyle{(\, .\, )}}}$ of ${\bf Z}\n$  the {\it un}-ordered $n$-tuple of~${\bf Z}\n_i$ values---equivalently, an arbitrarily   ordered  version of the same. To fix the notation, let~${\bf Z}\n_{{\scriptscriptstyle{(\, .\, )}}}\!\!:=  \big({\bf Z}\n_{(1)},\ldots ,{\bf Z}\n_{(n)}\big)$, where~${\bf Z}\n_{(i)}\vspace{-1mm}$ is such that its first component is the~$i$th order statistic of the $n$-tuple of~${\bf Z}\n_i$'s first components. 
%  The set of points ${\bf z}\in\mathbb{R}^{nd}$ sharing the same order statistic value ${\bf z}_{{\scriptscriptstyle{(\, .\, )}}}\!$ is the set of all permutations of  
%
Under this definition, the   points ${\bf z}\in\mathbb{R}^{nd}$
 at which %two solutions   may differ are those where
  \eqref{optass1} possibly admits two minimizers or more %. Those points
   lie in the union~$N$ of a finite number of linear subspaces of  $\mathbb{R}^{nd}$ where some equidistance  properties hold between ${\bf Z}\n_i$'s and gridpoints; therefore, $N$ is ${\bf Z}\n_{{\scriptscriptstyle{(\, .\, )}}}$-measurable and has  Lebesgue measure zero. Such multiplicities have no practical impact, thus, since (for a given grid) they take place on a unique null set $N$.
 % such that~${\rm P}^n(N)=0$ for any~${\rm P}\in{\mathcal P}_d$. 
   
Another type of multiplicity occurs, even over $\mathbb{R}^{nd}\!\setminus\! N$: each of the  minimizers~${\bf F} _{\!{\pms}}\n({\bf Z}\n)$ of \eqref{optass1} indeed is such that  the $n$-tuple \vspace{-2mm} 
\begin{equation}\label{optcouples} \big\{\big({\bf Z}\n_1, {\bf F} _{\!{\pms}}\n({\bf Z}\n_1)\big),\ldots , \big({\bf Z}\n_n, {\bf F} _{\!{\pms}}\n({\bf Z}\n_n)\big)\big\}\vspace{-1mm}
\end{equation}
 is  one of the $n_0!$ indistinguishable couplings between the $n$ observations and the~$n$ points of the augmented   grid  that minimize, over the $n!$ possible couplings,  the sum %(the mean)
  of within-pairs squared distances. %---a trivial and purely formal (since indiscernible)
  That multiplicity, which involves $n_0$ tied observations,  does not occur for~$n_0=0$ or~$1$: the mapping ${\bf z}\mapsto\big({\bf z}_{{\scriptscriptstyle{(\, .\, )}}},{\bf F} _{\!{\pms}}\n({\bf z})\big)$ then is injective over $\mathbb{R}^{nd}\!\setminus\! N$. For $n_0>1$,  it is easily taken care of by replacing, in the grid,  the  $n_0>1$ copies of~$\bf 0$
%         the origin
          with~$n_0$ i.i.d.~points uniformly distributed over~$(n_R+1)^{-1}{\mathbb S}_d$% or over $(2n_R+1)^{-1}{\mathcal S}_{d-1}$
---a convenient tie-breaking device (see footnote 9 in Appendix~\ref{DFPropsec}) restoring the injectivity over $\mathbb{R}^{nd}\!\setminus\!N$  of~${\bf z}\mapsto\big({\bf z}_{{\scriptscriptstyle{(\, .\, )}}},{\bf F} _{\!{\pms}}\n({\bf z})\big)$. 

Reinterpreting \eqref{optass1}  as an  expected (conditional on the order statistic---see Section~\ref{DFsec} for a precise definition) 
 transportation cost, the same optimal coupling(s)  also constitute(s) the optimal~L$^2$ transport mapping  the   empirical distribution to the uniform discrete distribution over the augmented  grid (and, conversely, the two  problems being entirely symmetric, the optimal~L$^2$ transport mapping   the uniform discrete distribution over the augmented  grid to the  empirical distribution). Classical results (McCann~(1995) again) then show that  optimality is achieved (i.e., \eqref{optass1} is satisfied) iff   the so-called {\it cyclical monotonicity} property holds for the $n$-tuple \eqref{optcouples}. \vspace{-2mm}
% Except for a set $N_0^{nd}$ with Lebesgue measure zero in~$\mathbb{R}^{nd}$ (those points for which the minimal distance, in {optass1}, is  the same for at least two permutations of the grid---a finite collection of linear subspaces with Hausdorff dimension less than $nd$), and apart from the trivial multiplicity just mentioned, the solution is unique.\vspace{-1mm}   

\begin{defin}\label{cycmon} {\rm A subset  $S$ of $\mathbb{R}^d\times \mathbb{R}^d$ is said to be {\em cyclically monotone} if, for any finite collection of  points $\{({\bf x}_1, {\bf y}_1),\ldots ,({\bf x}_k, {\bf y}_k)\}\subseteq S$,\vspace{-1mm} 
\begin{equation}\label{cyclmonot}
 \langle {\bf y}_1,\ {\bf x}_{2}-{\bf x}_{1}\rangle + \langle {\bf y}_2,\ {\bf x}_{3}-{\bf x}_{2}\rangle +\ldots + \langle {\bf y}_k,\ {\bf x}_{1}-{\bf x}_{k}\rangle  \leq 0.\vspace{-1mm}
 \end{equation}
}\end{defin} 
The subdifferential of  a convex function does enjoy cyclical monotonicity, which heuristically  can be interpreted as a discrete version of the fact that a smooth convex
function has a positive semi-definite second-order differential. 

Note that a finite subset $S=\{({\bf x}_1,{\bf y}_1),\ldots ,  ({\bf x}_n,{\bf y}_n)\}$ of $\mathbb{R}^d\times \mathbb{R}^d$ is cyclically monotone iff~\eqref{cyclmonot} holds for $k=n$---equivalently,   iff, among all pairings of~$({\bf x}_1,\ldots ,{\bf x}_n)$ and $({\bf y}_1,\ldots ,{\bf y}_n)$,~$S$ maximizes~$\sum_{i=1}^n \langle {\bf x}_i, {\bf y}_i\rangle $ (an empirical covariance), or minimizes  $\sum_{i=1}^n \Vert {\bf y}_i -  {\bf x}_i\Vert ^2$ (an empirical distance).  In other words, a finite subset $S$ is cyclically monotone iff  the couples~$({\bf x}_i,{\bf y}_i)$ are a solution of the optimal assignment problem with assignment cost $\Vert {\bf y}_i -  {\bf x}_i\Vert ^2$. The L$^2$ transportation cost considered here is thus closely related to the concept of convexity and the geometric property of cyclical monotonicity; it does not play the statistical role of  an estimation loss function, though---the L$^2$ distance between the empirical transport and its population counterpart  (the expectation of which might be infinite), indeed, is never considered.

Associated with our definition of an empirical center-outward distribution function ${\bf F}\n _{\pms}$ are the following concepts  of\vspace{-1.5mm}
\begin{enumerate}
\item[--] 
%\noindent -- 
{\it center-outward ranks} 
 $R\n _{{\pms} ,i}:= (n_R +1)\Vert {\bf F}\n _{\pms}({\bf Z}\n_i)\Vert$,
 \item[--] 
%\noindent  -- 
{\it empirical center-outward quantile contours} and {\it regions} 
\end{enumerate}%\begin{eqnarray*}
 $${\cal C}\n_{\pms;{\bf Z}\n}\!\Big(\frac{j}{n_R +1}\Big)\!\!:=\!\big\{{\bf Z}\n_i\vert R\n _{{\pms} ,i} = j \big\} \text{ 
%&& \text{ and} \\ 
and } 
\mathbb{C}\n_{\pms;{\bf Z}\n}\!\Big(\frac{j}{n_R +1}\Big)\!\!:=\! \big\{{\bf Z}\n_i\vert R\n _{{\pms} ,i} \leq j\big\},$$
%&&\text{ respectively;}
\begin{enumerate}
\item[] respectively, where  $j/(n_R +1)$, $j=0,1,\ldots ,n_R$,  is an empirical probability contents, to be interpreted as a quantile order,  
 % \noindent --  
 \item[--]  {\it center-outward signs} \vspace{-1mm}
 ${\bf S}\n _{{\pms} ,i}\!:={\!\bf F}\n _{\pms}({\bf Z}\n_i) I\Big[{\bf F}\n _{\pms}({\bf Z}\n_i)\!\neq\!{\bf 0}\Big]/\Vert {\bf F}\n _{\pms}({\bf Z}\n_i)\Vert$, and {\it center-outward sign curves} $\{{\bf Z}\n_i\vert  {\bf S}\n _{{\pms} ,i} ={\bf u}\}$, ${\bf u}\in
 {\mathfrak S}^{(n_S)}$.  
 % \item[--] 
\vspace{-1.0mm}\end{enumerate}

\noindent The 
  %center-outward quantile
   contours, curves, and regions defined here are finite collections of observed points; the problem of turning them into continuous contours enclosing compact regions and continuous lines is treated in Section~\ref{sec3}.% and~\ref{barFpm}.  

%\medskip

Up to this point, we have defined  multivariate generalizations of the univariate concepts of center-outward distribution and quantile functions, center-outward ranks and signs, all reducing to their univariate analogues in case $d=1$. However, it remains to show that those multivariate extensions are adequate in the sense that they enjoy in $\mathbb{R}^d$ the characteristic   properties that make the inferential success of their univariate counterparts---namely, \vspace{-2mm}
\begin{enumerate}
\item[(GC)] a Glivenko-Cantelli-type asymptotic relation between   ${\bf F} _{\!{\pms}}\n$ and ${\bf F} _{\!{\pms}}$, and 
\item[(DF$^+$)] the (essential) maximal ancillarity property described for $d=1$ in Section~\ref{orderRd}. 
%maximal  distribution-freeness (with respect to~$f\in{\cal F}^d$) property associated with the Basu factorization  of the sample $\sigma$-field.
\vspace{-1.5mm}%, and
%\item[(HW$^*$)] the maximal invariance property leading to semiparametric efficiency preservation (HW). \vspace{-2mm}
\end{enumerate} 
%Establishing those  properties %---for (HW), actually, the maximal invariance property (HW$^*$)  underlying (HW)---
This is the objective of Sections~\ref{GCsec} and \ref{DFsec}. % and Appendix \ref{INVsec}.%; see Appendix~I for the proofs.
 \vspace{-2.5mm}

  \subsection{Glivenko-Cantelli}\label{GCsec}\vspace{-1.5mm}
    
  With the definitions adopted in Sections~\ref{defsubsec} and \ref{gridsec}, the traditional Glivenko-Cantelli theorem, under  center-outward  form \eqref{empiricalpm}, holds, essentially {\it ne varietur}, in~$\mathbb{R}^d$ under ${\rm P}\in{\mathcal P}_d^{\pms}$.\vspace{-2.5mm} 
  
  \begin{Prop}\label{PropGC0} Let ${\bf Z}\n_1,\ldots ,{\bf Z}\n_n$ be i.i.d.\ with distribution~${\rm P}\in{\mathcal P}_d^{\pms}$. Then, \vspace{-1mm}
\begin{equation}\label{GCmaxeq}
\max_{1\leq i\leq n}\Big\Vert   {\bf F}\n _{\pms}({\bf Z}\n_i) - {\bf F} _{\!{\pms}}({\bf Z}\n_i) \Big\Vert  \longrightarrow 0\quad \text{a.s.\  as } n\to\infty\vspace{-2mm} .
\end{equation}%\vspace{-5mm}
  \end{Prop}
  The particular  case of elliptical distributions is considered in~Appendix~\ref{ellsec}.

Proposition~\ref{PropGC0} considerably reinforces, under more general assumptions (no second-order moments), an early strong consistency result by Cuesta-Albertos et al.~(1997). It readily follows from the more general  Proposition~\ref{PropGC},  which   extends \eqref{GCmaxeq}  under $\sup$ form to cyclically monotone interpolations of~${\bf F}\n _{\pms}\!$.
%; for the proof of Proposition~\ref{PropGC},  see Appendix~\ref{GCproofsec}. 
 \vspace{-2.5mm} % (see~\eqref{supGC}).%The proof of~\eqref{GCmaxeq} is postponed to Section~\ref{GCproofsec}. 
  
   \subsection{Distribution-freeness and maximal ancillarity}\label{DFsec}\vspace{-1mm}
  
%Denote by ${\cal B}_{{\scriptscriptstyle{(\, .\, )}}}\n$ the sub-$\sigma$-field generated by ${\bf Z}\n_{{\scriptscriptstyle{(\, .\, )}}}$---namely, the sub-$\sigma$-field of permutationally invariant Borel sets in ${\mathcal B}_d^n$. 
 Proposition~\ref{DFProp} provides the multivariate extension of 
% extends to the center-outward case
  the usual %finite-sample
   distributional properties of  univariate order statistics and    ranks. Note that, contrary to Proposition~\ref{PropGC0}, it holds for ${\rm P}\in
% $ in the %general
%   class~$
   {\mathcal P}_d$. 
%    of absolutely continuous distributions; s
   See Appendices~\ref{DFPropsec}  and~\ref{Basuapp} for   a proof and details on  sufficiency, ancillarity, %(strong) essential equivalence,
    and (strong) essential maximal ancillarity.    
%    Denote by~$X$ the observation in a statistical model $\big({\mathcal X}, {\mathcal A},{\mathcal P}
%    \big)$: a  
%    statistic $T(X)$ is called {\it strongly ${\mathcal P}$-essentially equivalent} to $X$ if there exists $N\in{\mathcal A}$ such that ${\rm P}(N)=0$ for any ${\rm P}\in{\mathcal P}$ and  $x\mapsto T(x)$ is injective over %the restriction to
%     ${\mathcal X}\!\setminus\! N$.
\vspace{-2mm}
  
    \begin{Prop}\label{DFProp} Let ${\bf Z}\n_1,\ldots , {\bf Z}\n_n$ be i.i.d.~with distribution ${\rm P}\in{\cal P}_d$, center-outward distribution function ${\bf F}_{\pms}$, order statistic ${\bf Z}\n_{{\scriptscriptstyle{(\, .\, )}}}$,  and empirical center-outward distribution function~${\bf F}\n_{\!{\pms}}\!$.    Then,\vspace{-3.5mm}\\ 
%    
%    , and empirical center-outward distribution function ${\bf F}\n_{\!{\pms}}$. Then,
% \noindent  
 
    {\it (i)}~${\bf Z}\n_{{\scriptscriptstyle{(\, .\, )}}}$     is sufficient and complete, hence minimal sufficient, for ${\cal P}_d\n$;

% \noindent 
 {\it (ii)}{\rm (DF)}  ${\bf F}\n_{\!{\pms}}({\bf Z}\n)
 :=\big({\bf F}\n_{\!{\pms}}({\bf Z}\n_{1}),\ldots , {\bf F}\n_{\!{\pms}}({\bf Z}\n_{n}) \big)\vspace{-0.4mm}$ is    %${\cal P}_d\n$-a.s.\ 
 uniformly distributed over the $n!/n_0!$ permutations with repetitions (the origin  counted as $n_0$ indistinguishable points) of  the %augmented
  grid described in Section~\ref{gridsec}\vspace{-0.3mm};
%           \item[(iii)]  f

%\noindent
 {\it (iii)}~for $n_0=0$, %or after tie-breaking, 
  the vectors of center-outward ranks $\big(R\n_{{\pms} ,1}, \ldots , R\n_{{\pms} ,n}\big)\vspace{-0.5mm}$ and signs $\big({\bf S}\n_{{\pms} ,1}, \ldots , {\bf S}\n_{{\pms} ,n}\big)$ are mutually independent; for $n_0>0\vspace{0.3mm}$, the same independence  holds for the $(n_Rn_S)$-tuple of ranks and signs associated with the (random)  set $\{i_1<\ldots <i_{n_Rn_S}\}$ such that ${\bf F}\n_\pms ({\bf Z}\n_{i_j})\neq{\bf 0}$;\vspace{-0.5mm}
%        \item[(iv)]
        
%\noindent
 {\it (iv)} for all~${\rm P}\in~\!{\mathcal P}_d$,  ${\bf Z}\n_{{\scriptscriptstyle{(\, .\, )}}}$ and  ${\bf F}\n_{\!{\pms}}({\bf Z}\n)$ % $\big({\bf F}\n_{\!{\pms}}({\bf Z}\n_{1}),\ldots , {\bf F}\n_{\!{\pms}}({\bf Z}\n_{n}) \big)$
 are mutually $\rm P$-inde\-pendent, and 

%\noindent 
{\it (v)} for $n_0\leq 1$ or after adequate  tie-breaking (cf.  comment below),  ${\bf F}\n_{\!{\pms}}({\bf Z}\n)$ %{\it (va)}
 is strongly ${\cal P}_d\n\!$-essentially maximal ancillary.
% ,  and
%  {\it (vb)}~enjoys the Basu factorization property  {\rm (DF$^+$)}, that is,  ${\bf Z}\n\!$ and~$\big({\bf Z}\n_{{\scriptscriptstyle{(\, .\, )}}}, {\bf F}\n_{\!{\pms}}({\bf Z}\n)\big)$ are (strongly) ${\mathcal P}_d\n\!$-essentially   equivalent, with~${\bf Z}\n_{{\scriptscriptstyle{(\, .\, )}}}$ satisfying~{\it (i)},  ${\bf F}\n_{\!{\pms}}({\bf Z}\n)$ satisfying~{\it (ii)}, and   {\it (iv)} their joint distribution factorizing into the product of its ${\bf Z}\n_{{\scriptscriptstyle{(\, .\, )}}}$- and  ${\bf F}\n_{\!{\pms}}({\bf Z}\n)$-marginals.
      \end{Prop}\vspace{-1mm}

In {\it (iii)} and {\it (v)}, $n_0$ plays a special role. In  {\it (iii)},   the fact that the sign, for the~$n_0$ observations mapped to the origin, is not a unit vector  induces, for~$n_0\geq1$,  a (very mild) dependence between signs and ranks which, however,  does not affect joint distribution-freeness.   In {\it (v)},   $n_0\leq 1$ implies that~${\bf z}\mapsto\big({\bf z}_{{\scriptscriptstyle{(\, .\, )}}},{\bf F}\n_{\pms}({\bf z})\big)$  is   injective over ${\mathbb R}^{nd}\!\setminus\! N$. % and the~$n!/n_0!$ permutations with repetitions  of the gridpoint in {\it (ii)} reduce to the $n!$ ``ordinary"  ones.   %plays an $n_0$-tuple role in the matching between the observations and the grid. 
%      ; that dependence rapidly fades away as $n$ increases.
%        Also, note that, still for $n_0\leq 1$, % (hence,~$n_0!=1$),
         As previously explained, injectivity is easily restored   via a simple tie-breaking device:    {\it (v)} then is satisfied  irrespective of $n_0$. %: see Appendix~\ref{gridapp} for a discussion. 
          Note  that the proportion~$n_0/n$ of points involved anyway tends to zero as~$n\to\infty$. 
         
         More important is the interpretation of essential maximal ancillarity in terms of finite-sample semiparametric efficiency   in case ${\bf Z}_i$ is the ${\boldsymbol\theta}$-residual ${\bf Z}_i({\boldsymbol\theta})$ in some semiparametric model with parameter of interest ${\boldsymbol\theta}$ and nuisance %the residual density
          $f$ (see Section~\ref{orderRd}). %, the same  ``finite-sample semiparametric efficiency" property  (see Section~\ref{orderRd}) as in dimension $d=1$.     
  Another crucial consequence of~\!{\it (v)} is the following corollary. \vspace{-1.5mm}
          \begin{cor}\label{gridcor} Denote by $\tilde{\cal B}\n_{\!\pms}$  the sub-$\sigma$-field  generated by the mapping $\tilde{\bf F}\n_{\!\pms}$    associated with some other deterministic\footnote{Deterministic here means nonrandom or randomly generated from a probability space that has no relation to the observations.} $n$-points grid---whether over the unit ball, the unit cube, or any  other fixed domain. Then, there exists $M\in\mathcal B ^n_d$ such that~ ${\rm P}\n(M) = 0$ for all ${\rm P}\in {\mathcal P}_d$ and ${\cal B}\n_{\!\pms}\cap\big({{\mathbb{R}}^{nd}\!\setminus\! M}\big)=\tilde{\cal B}\n_{\!\pms}\cap\big({{\mathbb{R}}^{nd}\!\setminus\! M}\big)$.\vspace{-1.5mm}% are essentially equivalent $\sigma$-fields.
          \end{cor}
          
 \color{black}         
          
%          that, denoting by $\tilde{\cal B}\n_{\!\pms}$ \vspace{-0.5mm} the sub-$\sigma$-field  generated by the mapping $\tilde{\bf F}\n_{\!\pms}$    associated with some other deterministic\footnote{Deterministic here means nonrandom or randomly generated from a probability space that has no relation to the observations.} $n$-points grid---whether in the unit ball, the unit cube, or any  other fixed domain---it holds that ${\cal B}\n_{\!\pms}=\tilde{\cal B}\n_{\!\pms}$. Indeed, if both ${\bf F}\n_{\!\pms}$ and $\tilde{\bf F}\n_{\!\pms}$, jointly with ${\bf Z}\n_{{\scriptscriptstyle{(\, .\, )}}}$, are observationally equivalent to ${\bf Z}\n\!$, they are in a one-to-one correspondence, hence generate the same sub-$\sigma$-fields.  
 It follows (see Appendix~\ref{Basuapp}) that        ${\cal B}\n_{\!\pms}$ and $\tilde{\cal B}\n_{\!\pms}$ are strongly essentially equivalent $\sigma$-fields.
  Ranks and signs associated with  distinct grids, thus, essentially generate the same sub-$\sigma$-fields, which 
%           (except for the interpretation in terms of ranks and signs)
            considerably attenuates the impact of  grid choices; see Appendix~\ref{DFPropsec} for details and a proof.   \vspace{-1.5mm}
       
\setcounter{equation}{0}

  \section{Smooth interpolation under  cyclical monotonicity constraints}\label{sec3} \vspace{-1mm} 
  
  So far, Definition~\ref{empFpm} only provides a value of ${\bf F}\n _{\pms}$  at the sample values~${\bf Z}\n_i\!$.   If~${\bf F}\n _{\pms}$ is to be extended to ${\bf z}\in\mathbb{R}^d$, an interpolation $\overline{\bf F}\n _{\pms}$, similar for instance to  the one shown, for $d=1$, in Figure~\ref{Ffig} of Appendix~\ref{1dimsec}, has to be constructed. Such  interpolation   should   belong to the class of gradients of convex functions from~$\mathbb{R}^d\!$  to~${\mathbb{S}}_d$, so that   the resulting contours ${\cal C}\n_{\pms;{\bf Z}\n}$   have the nature of continuous quantile contours. 
%, namely, be  the images of the corresponding hyperspheres (with radius $\Vert {\bf F}\n _{\pms}({\bf Z}\n_i) \Vert$) by some gradient of convex function.
 Moreover, it still should enjoy (now under a $\sup_{{\bf z}\in\mathbb{R}^d}$ form similar to~\eqref{GC1}) the  Glivenko-Cantelli %strong  consistency
  property.\footnote{It should be insisted, though, that the $\max_{1\leq i\leq n}$ form~\eqref{GCmaxeq} of Glivenko-Cantelli is not really restrictive, as  interpolations do not bring any additional information, and are mainly intended for    (graphical  or virtual) depiction of quantile contours.}   Constructing such  interpolations    is considerably  more delicate for~$d\geq 2$ than in the univariate case. 
  %, and  is the subject of Part~II of this paper, where we also refer to for numerical implementation and pictures. 

Empirical center-outward distribution functions ${\bf F}_\pms\n\!\!$, as defined in Definition~\ref{empFpm}, are cyclically monotone (discrete) mappings from the random sample (or $n$-tuple of residuals) ${\bf Z}\n_1,\ldots ,{\bf Z}\n_n$ to a (nonrandom) regular   grid of~${\mathbb{S}_d}$; hence,~${\bf F}_\pms\n$  is defined at the observed points only. Although such discrete~${\bf F}_\pms\n$ perfectly fulfills its statistical role as a sufficient sample summary    carrying the same information as the sample itself, one may like to define an empirical center-outward distribution function as an object of the same nature---a smooth cyclically monotone mapping from~$\mathbb{R}^d$ to~${\mathbb{S}_d}$---as  its population counterpart~${\bf F}_\pms$. This brings into the picture the problem of the existence and construction, within the class of gradients of convex functions,   of a continuous  extension~${\bf x}\mapsto~\!\overline{\bf F}\n_\pms(x)$ of the discrete~${\bf F}\n_\pms$, yielding a Glivenko-Cantelli theorem of the $\sup_{{\bf x}\in\mathbb{R}^d}$ form---namely,~
%\begin{equation}\label{supGC}
$\sup_{{\bf x}\in\mathbb{R}^d}\Vert  \overline{\bf F}\n_\pms({\bf x}) - {\bf F}_\pms({\bf x})
  \Vert \to 0$ a.s.\  as $n\to\infty
$%\end{equation}
---rather than the $\max_{1\leq i\leq n}$ form established in Proposition~\ref{PropGC0}.  That problem reduces to the more general problem of %the existence and construction of
 smooth interpolation under  cyclical monotonicity (see Definition~\ref{cycmon}) constraints, which we now describe.

Let $\pmb{\mathcal{X}}_n=
 \{{\bf x}_1,\ldots,{\bf x}_n\}$ and  $
\pmb{\mathcal{Y}}_n=
\{{\bf y}_1,\ldots,{\bf y}_n\}$ denote two $n$-tuples of points in~$ \mathbb{R}^d$. Assuming that  there exists a {\it unique} bijection  $T:\pmb{\mathcal{X}}_n\rightarrow \pmb{\mathcal{Y}}_n$ such that the set~$\big\{\!\big({\bf x}, T({\bf x})\big)\vert\ {\bf x}\in  \pmb{\mathcal{X}}_n \big\}$ is cyclically monotone, there is no loss of generality in relabeling the elements of~$\pmb{\mathcal Y}_n$ so that~${\bf y}_i=T({\bf x}_i)$. Accordingly, we throughout   are making the following assumption.\smallskip

\noindent {\bf Assumption (A).}  The $n$-tuples $\pmb{\mathcal X}_n$ and $\pmb{\mathcal Y}_n$ are such that $T\!: {\bf x}_i\mapsto T({\bf x}_i)={\bf y}_i$ for  $i=1,\ldots ,n$ is the unique cyclically monotone bijective map from $\pmb{\mathcal X}_n$ to~$\pmb{\mathcal Y}_n$.\smallskip

\noindent Our goal, under  Assumption (A), is to construct a smooth (at least continuous) cyclically monotone map~$\overline{T}\!:\mathbb{R}^d\!\to \mathbb{R}^d$ such that
 $\overline{T}({\bf x}_i)={T}({\bf x}_i)~\!=~\!{\bf y}_i$,~$i=~\!1,\ldots,n$. 

It is well known that the subdifferential of a convex function~$\psi$ from~$\mathbb{R}^d$ to~$\mathbb{R}$ enjoys cyclical monotonicity. A classical result by Rockafellar (1966) establishes the converse: any finite cyclically monotone subset $S$ 
%=\{({\bf x}_i, {\bf y}_i)\vert i=1,\ldots ,n\}$
 of~$\mathbb{R}^d\times~\!\mathbb{R}^d$ lies in the subdifferential of some convex function. 
Our result reinforces this characterization by restricting to differentiable convex functions. Note that a differentiable convex function $\psi$ is automatically continuously differentiable, with unique   (at all ${\bf x}$) subgradient~$\nabla\psi({\bf x})$ and subdifferential~$\{({\bf x},\nabla\psi({\bf x}))\vert {\bf x}\in\mathbb{R}^d\}$. When $\psi$ is convex and differentiable,  the mapping~$x\mapsto\nabla\psi(x)$ thus enjoys cyclical monotonicity. We show in Corollary~\ref{Rock+} that, conversely,  any   subset $S=\{({\bf x}_i, {\bf y}_i)\vert i=1,\ldots ,n\}$ of~$\mathbb{R}^d\times\mathbb{R}^d$ enjoying cyclical monotonicity  is  the subdifferential (at ${\bf x}_1,\ldots ,{\bf x}_n$) of some (continuously) differentiable convex function $\psi$.

Note that Assumption (A) holds if and only if identity is the unique minimizer of %\vspace{-3mm}
%\begin{equation}\label{assumption2}
%\frac 1 n 
$\sum_{i=1}^n \|{\bf x}_i-{\bf y}_{\sigma(i)}\|^2$  among the set of all permutations $\sigma$ of $\{1,\ldots,n\}$.   
%\vspace{-2mm}\end{equation}
 Letting~$c_{i,j}:=\|{\bf x}_i-{\bf y}_j\|^2$, the same condition can   be recast in terms of uniqueness of the solution of  the linear program\vspace{-1mm}
\begin{equation}\label{assumption3}
%\begin{aligned}
\min_{\pi}  % &
 \sum_{i=1}^n \sum_{j=1}^n c_{i,j}\pi_{i,j} \quad%  \\
\mbox{s.t. } %&
 \sum_{i=1}^n \pi_{i,j}=\sum_{j=1}^n \pi_{i,j}=\frac 1 n, \ %\\
%&
 \pi_{i,j}\geq 0,\  i,j=1,\ldots,n.
% \end{aligned}
\vspace{-1mm}\end{equation}
Clearly,  $\sigma(i)=i$ minimizes $\sum_{i=1}^n \|{\bf x}_i-{\bf y}_{\sigma(i)}\|^2$   iff 
$\pi_{i,i}=  1/ n$, $\pi_{i,j}=0$ for~$j\ne i$ is the unique solution of (\ref{assumption3}).

Our solution to the cyclically monotone interpolation problem  is constructed in two steps. First (Step 1), we extend $T$ to a piecewise constant cyclically monotone map defined on a set in $\mathbb{R}^d$ whose complementary has Lebesgue measure  zero. Being piecewise constant, that  
map cannot be smooth. To fix this, we apply (Step 2) a regularization procedure  yielding the required smoothness while keeping the 
interpolation feature. For Step 1,  we rely on the following result  (see  Appendix~\ref{Proofsec31} for the proof).\vspace{-1.5mm}

\begin{Prop} \label{step1} Assume that ${\bf x}_1,\ldots,{\bf x}_n \in\mathbb{R}^d$ and $ {\bf y}_1,\ldots,{\bf y}_n \in\mathbb{R}^d$ are such that~$i\ne j$ implies~${\bf x}_i\neq {\bf x}_j$ and ${\bf y}_i\neq{\bf y}_j$. Then,
\begin{enumerate}%\noindent
\item[{\it (i)}] the map $T({\bf x}_i)={\bf y}_i$, $i=1,\ldots,n$ is cyclically monotone if and only if there exist real numbers $\psi_1,\ldots,\psi_n$ such that\vspace{-1mm}
$$\langle {\bf x}_i, {\bf y}_i\rangle-\psi_i =\max_{j=1,\ldots,n} (\langle {\bf x}_i, {\bf y}_j\rangle-\psi_j),\quad i=1,\ldots,n;\vspace{-2mm}$$
\item[{\it (ii)}]  furthermore, $T$ is the unique cyclically monotone map from $\{{\bf x}_1,\ldots,{\bf x}_n\}$ to~$\{{\bf y}_1,\ldots,{\bf y}_n\}$ if and only if 
there exist real numbers $\psi_1,\ldots,\psi_n$ such that \vspace{-1mm}
\begin{equation}\label{keycondition}
\langle {\bf x}_i, {\bf y}_i\rangle-\psi_i>\max_{j=1,\ldots,n, j\ne i} (\langle {\bf x}_i, {\bf y}_j\rangle-\psi_j),\quad i=1,\ldots,n.
\vspace{-2mm}\end{equation}
\end{enumerate}
\end{Prop}

\begin{rem}\label{Comment1}{\em 
The condition,  in Proposition \ref{step1}, that ${\bf y}_1,\ldots,{\bf y}_n$ are distinct in general is not satisfied in the case of empirical center-outward distribution functions, where, typically, ${\bf y}_1=\cdots={\bf y}_{n_0}$  with  ${\bf y}_1\ne {\bf y}_i$ for~$i>n_0 $ and $n_0$ ranging between $0$ and $\min(n_R, n_S)-1$.  This can be taken care of by means of  the tie-breaking device described in Section~\ref{gridsec}. The proof (see Appendix~\ref{Proofsec31}), however, is easily adapted to show that the map $T({\bf x}_i)={\bf y}_i$, $i=1,\ldots,n$ is cyclically monotone if and only if there exist 
real numbers $\psi_1,\psi_{n_0+1},\ldots,\psi_n$ such that, setting $\psi_i=\psi_1$, $i=2,\ldots,n_0$,\vspace{-1mm}
$$\langle {\bf x}_i, {\bf y}_i \rangle-\psi_i=\max_{j=1,\ldots,n} (\langle {\bf x}_i, {\bf y}_j \rangle-\psi_j),\quad i=1,\ldots,n.\vspace{-2mm}$$
Similarly, the map $T({\bf x}_i)={\bf y}_i$, $i=1,\ldots,n$ is the unique cyclically monotone map from~$\pmb{\mathcal X}_n$ 
%$\{{\bf x}_1,\ldots,{\bf x}_n\}$
 to $\{{\bf y}_1,{\bf y}_{n_0+1}\ldots,{\bf y}_n\}$ mapping
$n_0$~points in $\pmb{\mathcal X}_n$ 
%$\{{\bf x}_1,\ldots,{\bf x}_n\}$
 to ${\bf y}_1$ if and only if there exist real numbers $\psi_1,\psi_{n_0+1},\ldots,\psi_n$ such that\vspace{-1mm}
$$\langle {\bf x}_i, {\bf y}_1\rangle-\psi_1>\langle {\bf x}_i, {\bf y}_j \rangle-\psi_j,\quad i=1,\ldots,n_0,\, j=n_{0}+1,\ldots,n,\vspace{-2mm}$$
$$\langle {\bf x}_i, {\bf y}_i \rangle-\psi_i>\langle {\bf x}_i, {\bf y}_j \rangle-\psi_j,\quad i=n_0+1,\ldots,n,\, j=1,n_{0}+1,\ldots,n,\, j\ne i.$$
Details are omitted.\vspace{-0mm} %This can be proved with straightforward changes to the proof of Proposition~\ref{step1}. We omit details.
}
\end{rem}

%\smallskip
As a consequence of Proposition \ref{step1}, we can extend $T$ to a cyclically monotone map from~$\mathbb{R}^d$ to~$\mathbb{R}^d$ as follows. Under Assumption (A),
we can choose $\psi_1,\ldots,\psi_n$ such that (\ref{keycondition}) holds. Consider the convex map
\begin{equation}\label{DefPhi}
{\bf x}\mapsto \varphi({\bf x}):=\max_{1\leq j\leq n} (\langle {\bf x}, {\bf y}_j\rangle -\psi_j).\vspace{-1mm}
\end{equation}
Now the sets $C_i=\{x\in\mathbb{R}^d\vert\, (\langle {\bf x}, {\bf y}_i\rangle-\psi_i)>\max_{j\ne i}(\langle {\bf x}, {\bf y}_j\rangle-\psi_j)\}$ are open convex sets such that
$\varphi$ is differentiable in $C_i$, with $\nabla \varphi ({\bf x})={\bf y}_i$, ${\bf x}\in C_i$. The complement of~$\bigcup_{i=1}^n C_i$ has  Lebesgue
measure zero. Thus, we can extend $T$ to~${\bf x}\in \bigcup_{i=1}^n C_i$, hence to almost all~${\bf x}\in\mathbb{R}^d$, by setting
 $\overline{T}({\bf x}):=\nabla \varphi ({\bf x}).$

By construction, ${\bf x}_i\in C_i$, hence $\overline{T}$ is an extension of $T$.  Theorem~12.15 in Rockafellar and Wets~(1998) implies that $\overline{T}$ 
is cyclically monotone. We could  (in case~$\bigcup_{i=1}^n C_i\varsubsetneq \mathbb{R}^d$)  extend $\overline{T}$ from $\bigcup_{i=1}^n C_i$ to  $\mathbb{R}^d$ while preserving cyclical monotonicity, but such extension of~$\overline{T}$ cannot be continuous. Hence, we 
do not pursue that idea and, rather, try to find a smooth extension of~$T$. For this,  consider the Moreau envelopes\vspace{-2mm} 
\begin{equation}\label{DefPhiEpsilon}
\varphi_\varepsilon({\bf x}):=\inf_{{\bf y}\in\mathbb{R}^d}\Big[\varphi({\bf y})+\frac{1}{2\varepsilon} \|{\bf y}-{\bf x}\|^2 \Big],\quad {\bf x}\in\mathbb{R}^d,\ \varepsilon >0\vspace{-2mm}
\end{equation}
of $\varphi$ (as defined in (\ref{DefPhi})):  see, e.g.,
 Rockafellar and Wets~(1998). The following theorem shows that,   for sufficiently small $\varepsilon>0$, $\nabla \varphi_\varepsilon$---the so-called Yosida regularization  of $\nabla \varphi$ ({Yosida}~1964)---provides a continuous, cyclically monotone interpolation of~$({\bf x}_1,{\bf y}_1),\ldots,({\bf x}_n,{\bf y}_n)$, as desired. \vspace{-2mm}%Summing up, we have the following theorem.

\begin{Prop}\label{InterpolationTheorem}
Let Assumption (A) hold, and consider $\varphi$ as in (\ref{DefPhi}), 
with constants~$\psi_1,\ldots,\psi_n$ satisfying (\ref{keycondition}). Let $\varphi_\varepsilon$ as in (\ref{DefPhiEpsilon}). Then, there exists~$e~\!>~\!0$ such that, 
for every $0<\varepsilon\leq e$, the map $\varphi_\varepsilon$ is continuously differentiable \linebreak and~$T_\varepsilon :=\nabla \varphi_\varepsilon$
is a continuous, cyclically monotone map such that
 $T_\varepsilon({\bf x}_i)={\bf y}_i$ for all  $i=1,\ldots,n$ 
and $\|T_\varepsilon({\bf x})\|\leq \max_{i=1,\ldots,n}\| {\bf y}_i\|$ for all ${\bf x}\in\mathbb{R}^d$. \vspace{-2mm}
\end{Prop}   
The main conclusion of Proposition~\ref{InterpolationTheorem}  (see Appendix~\ref{InterpolationTheoremProofSec} for the proof) remains true in the setup of Remark~\ref{Comment1}, and we still can guarantee the existence of a convex, continuously differentiable~$\varphi$
such that $\nabla \varphi({\bf x}_i)={\bf y}_1$ for  $i=1,\ldots,n_0$ and~$\nabla \varphi({\bf x}_i)={\bf y}_i$ for $i=n_0+1,\ldots,n$.  %in that case.
More generally,   the following corollary, which heuristically  can be interpreted as a discrete version of the fact that a smooth convex function has a positive semi-definite second-order differential, is an immediate consequence.\vspace{-1.5mm}
\begin{cor}\label{Rock+}
Any cyclically monotone   subset~$%S=
\{({\bf x}_i, {\bf y}_i)\vert i=1,\ldots ,n\}$ of~$\,\mathbb{R}^d\times~\!\mathbb{R}^d$ such that ${\bf x}_i\ne {\bf x}_j$ for $i\ne j$ lies in the subdifferential (at ${\bf x}_i$, $i=1,\ldots ,n$) of some (continuously) differentiable convex function $\psi$. \vspace{-2mm}
\end{cor}
\begin{rem}\label{rem1}
{\rm
It is important to note that, in spite of what intuition may suggest, and except for  the univariate case ($d=1$), 
linear interpolation does not work in this problem; see Remark~\ref{notaA1} in the appendix for a counterexample.\vspace{-1.5mm}}
\end{rem}
\begin{rem}\label{rem2}
{\rm
The interpolating function~$T_\varepsilon$ given by the proof of Proposition~\ref{InterpolationTheorem} is not only
continuous but, in fact, Lipschitz with constant $1/ \varepsilon$ (see, e.g., Exercise 12.23 in \cite{RockafellarWets}). 
%Rockafellar and Wets~(1998)). 
Looking for the smoothest possible interpolation
we should, therefore, take the largest possible~$\varepsilon$ for which the interpolation result remains valid. Let us assume that
$\|{\bf y}_i\|\leq 1$, $i=1,\ldots,n$ (this does not imply any loss of generality; we could adequately normalize the data to get this satisfied,  then backtransform the interpolating
function). Set\vspace{-1mm} 
\begin{equation}\label{maxepsilon}
\varepsilon_0:=\frac 1 2 \min_{1\leq i\leq n} \Big( (\langle {\bf x}_i, {\bf y}_i \rangle-\psi_i)-\max_{j\ne i} (\langle {\bf x}_i, {\bf y}_j \rangle-\psi_j)\Big).\vspace{-1mm}
\end{equation}
Then, arguing as in the proof of Proposition~\ref{InterpolationTheorem}, we see that $B({\bf x}_i,\varepsilon_0)\subset C_i$. Let~$\varepsilon>0$ and~$\delta>0$ be such that $\varepsilon+\delta<\varepsilon_0$. Then, for ${\bf x}\in B({\bf x}_i,\delta)$, we have~${\bf x}-\varepsilon {\bf y}_i\in B({\bf x}_i,\varepsilon_0)$, and we can mimic the argument in the proof 
 to conclude that, 
for ${\bf x}\in B({\bf x}_i,\delta)$, we have 
 $\varphi_\varepsilon({\bf x})=\langle {\bf x}, {\bf y}_i \rangle-\psi_i-\frac \varepsilon 2 \|{\bf y}_i\|^2$, and, consequently, 
$T_\varepsilon({\bf x}_i)={\bf y}_i$ for every~$\varepsilon<\varepsilon_0$ with~$\varepsilon_0$ given by~(\ref{maxepsilon}).  By continuity of the Yosida regularization
(see Theorem 2.26 in Rockafellar and Wets~(1998)), we conclude that~$T_{\varepsilon_0}({\bf x}_i)={\bf y}_i$, $i=1,\ldots,n$. We summarize our findings in the following result.\vspace{-2mm}

\begin{cor}\label{CorollaryEpsilon0}
%{\hspace{-1mm}
%}
Let Assumption (A) hold. 
%Assume that $x_1,\ldots,x_n,y_1,\ldots,y_n\in \mathbb{R}^d$ are such that the map $T(x_i)=y_i$ is the unique cyclically monotone
%map from $\{x_1,\ldots,x_n\}$ to $\{y_1,\ldots,y_n\}$.
 Assume further that $\|{\bf y}_i\|\leq~\!1$ for all~$i=~\!1,\ldots,n$.
Let $\varphi({\bf x}):=\max_{1\leq j\leq n} (\langle {\bf x}, {\bf y}_j \rangle-\psi_j)$ with $\psi_1,\ldots,\psi_n$ defined as in~(\ref{keycondition}),~$\varphi_\varepsilon$  as in (\ref{DefPhi}), 
%\color{red} shouldn't this be ``is a minimizer for the right-hand side linear program (\ref{dual1})"? \color{black} 
and
$\varepsilon_0$ as in (\ref{maxepsilon}). Then $T_{\varepsilon_0}:=\nabla \varphi_{\varepsilon_0}$ is a Lipschitz continuous, cyclically monotone map, with Lipschitz constant
$1 /{\varepsilon_0}$, such that~$T_{\varepsilon_0}({\bf x}_i)={\bf y}_i$, $ i=1,\ldots,n$ {%\color{red}
and $\|T_{\varepsilon_0}({\bf x})\|\leq 1$ for every~${\bf x}\in\mathbb{R}^d$.}
\end{cor}

To conclude, let us turn  to the choice of the weights $\psi_i$ that satisfy condition~(\ref{keycondition}), as required by our construction. In view of
Corollary \ref{CorollaryEpsilon0} and the discussion in Remark~\ref{rem2}, choosing the weights  that maximize $\varepsilon_0$ in (\ref{maxepsilon})  
results  in smoother interpolations. 
%This maximization problem can be recast as  the linear program 
 The optimal smoothing value  then is half of the maximum in the linear program\vspace{-1mm} 
\begin{equation}\label{otherLP}
%\begin{aligned}
%&
 \max_{\psi,\varepsilon}  \,  \varepsilon\qquad% \\
\mbox{s.t. }\  %&% \quad
  \langle {\bf x}_i, {\bf y}_i-{\bf y}_j\rangle \geq \psi_i-\psi_j+\varepsilon,\quad i,j\in\{1,\ldots,n\},\  i\ne j ;
%\end{aligned}
\vspace{-2mm}\end{equation}
the optimal $\psi_j$'s are the corresponding weights. The dual of (\ref{otherLP}) is\vspace{-2mm} 
%ity in linear programming shows that the maximal value in (\ref{otherLP}) equals the minimal value in
%\begin{equation}
\begin{align}
\min_{z_{i,j}, i\ne j}   \; & \sum_{i,j=1,\ldots,n;\, i\ne j} \!\!\! z_{i,j} \langle {\bf x}_i, {\bf y}_i-{\bf y}_j\rangle\label{otherLPdual}  \\
\mbox{s.t. } &  \sum_{j=1,\ldots,n;\, j\ne i} \!\!\!  (z_{i,j}-z_{j,i})=0,  %\\
%&
  \sum_{i,j=1,\ldots,n;\, i\ne j} \!\!\!  z_{i,j}=1,\ \  z_{i,j}\geq 0,  \  i, j=1,\ldots, n.\nonumber
\end{align}\vspace{-4mm}
%\end{equation}

Now, (\ref{otherLPdual}) is a circulation problem over a complete graph with $n$ vertices. By the Flow Decomposition 
Theorem (see, e.g., Theorem 3.5 and Property 3.6 in  Ahuja et al.~\!(1993)), any circulation is of the form 
%can be expressed as
 $%\displaystyle
 {z_{i,j}\!=\!\sum_{W\in\mathcal{W}}\delta_{ij}(W) f(W)}$ 
where $\mathcal{W}$ denotes the set of all cycles in the graph, $\delta_{ij}(W)=1$ if the arc connecting $i$ and $j$ belongs to cycle $W$ 
($\delta_{ij}(W)=0$ otherwise), and $f(W)\geq 0$ is the flow along cycle $W$. Writing $c_{i,j}=\langle {\bf x}_i, {\bf y}_i-{\bf y}_j\rangle$ and 
$c(W)=\sum_{i,j} \delta_{ij}(W) c_{i,j}$ (where~$c(W)$ is the cost of moving one mass unit  along the cycle $W$), the objective function in (\ref{otherLPdual})
takes the form\vspace{-2mm} 
$$\sum_{i,j=1,\ldots,n;\, i\ne j} \!\!\! c_{i,j}z_{i,j}=\sum_{W\in\mathcal{W}} c(W)f(W), \vspace{-1.5mm}$$
with the constraint $\sum_{W\in\mathcal{W}}|W|f(W)=1$ where $|W|$ denotes the length (number of arcs) in the cycle $W$.  Putting $\tilde{f}(W):=|W|f(W)$, (\ref{otherLPdual}) can be rewritten as\vspace{-1.5mm}
$$%\begin%{aligned}
%&
\min_{\tilde{f}(W)}   \; \sum_{W\in\mathcal{W}} \!\!\! \tilde{f}(W)\frac {c(W)}{|W|} \qquad % \\
\mbox{s.t. } %&
  \sum_{W\in\mathcal{W}}\tilde{f}(W)=1,\quad \tilde{f}(W)\geq 0.
%\end{aligned}
\vspace{-2.5mm}$$
It follows that the optimal solution to (\ref{otherLPdual}) is %given by
 $z_{i,j}= {\delta_{ij}(\widehat{W})}/{|\widehat{W}|}$, 
where $\widehat{W}$ is a \textit{minimum mean cost cycle}, that is, a minimizer  among all cycles of ${c(W)}/{|W|}$. The computation of the 
minimum mean cost cycle can be carried out in polynomial time using, for instance, Karp's algorithm (Karp~(1978)). For this, we fix a vertex in the graph
(vertex 1, say; this choice does not affect the final ouput) and  write $d_{k,i}$ for the length of the shortest path from $1$ to $i$ in $k$ steps (where the lentgh of
the path $(i_1, i_2,\cdots ,i_k)$ is $c_{i_1,i_2}+\cdots+c_{i_{k-1},i_k}$ and $d_{k,i}=+\infty$ if there is no path with $k$ steps from $1$ to $i$). 
The lengths 
$d_{k,i}$ for $0\leq k\leq n$ and~$1\leq i\leq n$ can be computed  recursively starting from $d_{0,1}=0$, $d_{0,i}=\infty$ for~$i\ne 1$, and~$d_{k+1,i}=\min_{j} (d_{k,j}+c_{j,i})$ 
with $c_{i,i}=\infty$. Then, the minimum cycle mean is 
$\varepsilon^*=\min_{1\leq i\leq n}\max_{0\leq k\leq n-1}({d_{n,i}-d_{k,i}})/({n-k})$, 
which can be computed in $O(n^3)$ steps (see Theorem 1 and subsequent comments in  Karp~(1978)).  We observe that Assumption (A) is equivalent to $\varepsilon^*>0$.

We still need to compute the optimal weights $\psi_i$. For this, we can consider the graph with modified costs $\tilde{c}_{i,j}:=c_{i,j}-\varepsilon^*$ and compute the length $\tilde{d}_i$  of the shortest path (of any length) from vertex $1$ to $i$. It is easy to see that a shortest path of length at most $(n-1)$ exists. Hence we can compute
the shortest $k$-step distances $\tilde{d}_{k,i}$ as above, and  $\tilde{d}_i=\min_{0\leq k\leq n-1}\tilde{d}_{k,i}$. Finally, we set $\psi=-\tilde{d}_i$.
Now, by optimality,  $\tilde{d}_j\leq \tilde{d}_i+\tilde{c}_{i,j}$, that is,  
 $c_{i,j}\geq \psi_i-\psi_j+\varepsilon^*$. 
This shows that~$(\psi_1,\ldots,\psi_n,\varepsilon^*)$ is an optimal solution to (\ref{otherLP}) which, moreover,  can be computed in $O(n^3)$ computer time.

For 
%In the particular case~
$n=2$, 
 it is easily  seen that the optimum in~(\ref{otherLPdual}) 
(hence  in (\ref{otherLP}))\linebreak  is~$\varepsilon_0= {\langle {\bf x}_1-{\bf x}_2,{\bf y}_1-{\bf y}_2\rangle}/4>0$. The optimal weights can be chosen\linebreak 
as~$\psi_i= \langle({\bf x}_1+{\bf x}_2) , {\bf y}_i\rangle/2$, $i=1,2$. In the one-dimensional case,  if $n=2$,   uniqueness of $T$ holds iff  ${x}_1<x_2$ and~$y_1<~\!y_2$. A simple computation  
 yields
 \[T_\varepsilon(x)=\left\{
 \begin{array}{clrcl}
 y_1&\text{ for }&&  \big(x- {(x_1+x_2)}/2\big)/\varepsilon\!\!\!&\leq y_1, \\ 
   \big(x- {(x_1+x_2)}/2\big)/\varepsilon&\text{ for }& y_1\leq\!\!\! & \big(x- {(x_1+x_2)}/2\big)/\varepsilon\!\!\!&\leq y_2\\ 
 y_2&\text{ for }&y_2\leq\!\!\! &  \big(x- {(x_1+x_2)}/2\big)/\varepsilon.
 \end{array}
 \right.
 \]
% 
% 
% 
%%$T_\varepsilon(x)=y_1$ for 
%%%\quad\text{ if }\quad
%%$\frac 1 \varepsilon \big(x- {(x_1+x_2)}/2\big)\leq y_1$,  \linebreak 
%$T_\varepsilon(x)=y_2$ for 
%%\quad\text{ if }\quad
% $\frac 1 \varepsilon \big(x- {(x_1+x_2)}/2\big)\geq y_2$,  
%and  
% $T_\varepsilon(x)=\frac 1 \varepsilon \big(x- {(x_1+x_2)}/2\big)$ for~$y_1\leq \frac 1 \varepsilon \big(x- {(x_1+x_2)}/2\big)\leq y_2$. 
 We
see that $T_\varepsilon$ is an extension of %the map 
$x_i\mapsto y_i\vspace{1mm}$, $i=1,2\vspace{-0.5mm}$ if and only if~$x_2-x_1\geq~\!-2\varepsilon y_1$\linebreak  and~$x_2-x_1\geq 2\varepsilon y_2$,
which implies that $\varepsilon \leq  {(x_2-x_1)}/{(y_2-y_1)}$---equivalently,~$1 /\varepsilon$ larger than or equal to 
% \geq  {(y_2-y_1)}/{(x_2-x_1)}$ (note that~${(y_2-y_1)}/{(x_2-x_1)}$ 
%is
${(y_2-y_1)}/{(x_2-x_1)}$, the minimal Lipschitz constant of any Lipschitz extension of~$x_i\mapsto y_i$. This yields,  for~$y_1=-1$, $y_2=1$, \vspace{-1mm}
%we get~
$$\varepsilon_0={(x_2-x_1)}/2= {(y_2-y_1)}/{(x_2-x_1)}\vspace{-1mm}$$
 and   
%and we see that
% the interpolating function
  $T_{\varepsilon_0}$ is  the Lipschitz extension of~$x_i\mapsto y_i\vspace{0mm}$ with minimal Lipschitz constant.
\vspace{-1mm}}
\end{rem}

%\setcounter{equation}{0}

%\section{Interpolation of  empirical center-outward distribution functions}\label{barFpm0}%\vspace{-1mm}

%\subsection{Smooth interpolation of ${\bf F}_\pms\n$}\label{barFpm}\vspace{-1mm}

We now turn back to the smooth extension of the %(discrete)
 empirical center-outward distribution function ${\bf F}_\pms\n$  of Section~\ref{gridsec}. 
 Proposition~\ref{InterpolationTheorem} (and subsequent comments in case $n_0>1$) allows us to 
extend~${\bf F}_\pms\n$ to a Lipschitz-continuous gradient of  convex function over $\mathbb{R}^d\vspace{-0.5mm}$,   denoted as~$\overline{\bf F}_\pms\n$. The following result (proof in Appendix~\ref{GCproofsec}) extends to~$\overline{\bf F}_\pms\n$ the Glivenko-Cantelli result of Proposition~\ref{PropGC0}. We state (and prove) it for the  value $\varepsilon_0$ \eqref{maxepsilon} of the smoothing constant; with obvious modifications, it also holds for any admissible $\varepsilon$.  \vspace{-1.5mm} 
\begin{Prop}{\!\!(Glivenko-Cantelli)\!% for center-outward distribution functions)
}\label{PropGC} %{\color{magenta}
Let $\overline{\bf F}_\pms\n$ denote the smooth interpolation, with smoothing constant $\varepsilon_0$, of ${\bf F}_\pms\n$ computed from a sample of observations with  distribution ${\rm P}\in{\mathcal P}_d^{\pms}%$, \supsetneq {\mathcal P}_d^{\text{\em conv}}
$ and  center-outward distribution function ${\bf F}_\pms$. Then, \vspace{-1.5mm}
$$% \[
 \sup_{{\bf x}\in\mathbb{R}^d}\|\overline{\bf F}_\pms\n({\bf x})-{\bf F}_{\pms}({\bf x})\|\to~\!\!0\quad  \text{{\rm a.s.} as $n\to\infty$.}
% \]
\vspace{-2.5mm}$$% with probability one.
\end{Prop}
\begin{rem}\label{rem3}
{\rm Throughout, we focused on a smooth interpolation of ${\bf F}_\pms\n$, applying Proposition~\ref{InterpolationTheorem} to the cyclically monotone $n$-tuple $\big({\bf Z}\n_i, {\bf F}_\pms\n({\bf Z}\n_i)\big)$, %\qquad 
$i=1,\ldots ,n.$ %, for which Theorem \ref{PropGC} holds.
 For $n_0\leq 1$ (or after implementing the tie-breaking device described in Section~\ref{gridsec}), the resulting $\overline{\bf F}_\pms\n$ is invertible, yielding a smooth interpolation---denote it as $\overline{\bf Q}_\pms\n\!\!:=\big(\overline{\bf F}_\pms\n\big)^{-1}\!\!$---of the empirical quantile function~${\bf Q}_\pms\n\!\!$. For~$n_0>1$,  the restriction of $\overline{\bf F}_\pms\n$ to $\mathbb{R}^d\!\setminus\!\big({\bf F}_\pms\n\big)^{-1}({\bf 0})$ (which has Lebesgue measure one) can be considered instead.  In all cases, strong consistency still holds for $\overline{\bf Q}_\pms\n$;  uniformity is lost, however, unless spt$({\rm P})$ itself is compact.
 % Glivenko-Cantelli uniformity  is lost to uniformity over arbitrary closed balls of the form $q\,\bar{\mathbb{S}}_d$, with~$0<q<1$. %Alternatively, one also could consider applying Proposition~\ref{InterpolationTheorem} to the cyclically monotone $n$-tuple 
% $\big({\bf F}_\pms\n({\bf Z}\n_i), {\bf Z}\n_i\big)$,  $i=1,\ldots ,n$. 
\vspace{-2mm}}
\end{rem}
\begin{rem}\label{rem4}
{\rm Another interpolation  of ${\bf Q}_\pms\n\!\!$ is considered in Chernozhukov\linebreak et~al.~(2017), based on the so-called {\it $\alpha$-hull} method (see, e.g.,  Pateiro-L\'opez and Rodr\'{\i}guez-Casal~(2010)). Although producing visually nice results (Figure~2, same reference), that method does not take into account any cyclical monotonicity constraints. The resulting contours therefore do not have the nature of   quantile contours. Moreover,  contrary to $\overline{\bf Q}_\pms\n$,  the $\alpha$-hull interpolation does not yield a homeomorphism; %over  the full domain~$\mathbb{S}_d$ of a quantile function;
 $\alpha$-hull  contours need not be closed, and the resulting quantile regions need not be connected: see  Appendix~\ref{Secalphahull} for an example.\vspace{-2mm}
}
\end{rem}

%\begin{rem}\label{rem5}
%{\rm 
An alternative  ``multivariate step function" extension  of~${\bf F}_\pms\n\!\!$ is proposed in Appendix~\ref{stepF}. \vspace{-2mm}
%}
%\end{rem}

\section{Some numerical results}\label{numsec}\vspace{-1.5mm}

This section provides some two-dimensional numerical illustrations of the results of this paper. The codes we used were written in R, and can handle sample sizes as high as $n=20000$ (with $n_R=100$ and $n_S=200$, for instance) on a computer with 32Gb RAM.   The algorithm consists of three main steps.

(Step 1) Determine the optimal assignment between the sample points and the regular grid. This could be done with a cubic implementation of the Hungarian algorithm like the one included in the  \texttt{clue} R package (for a detailed account of the Hungarian algorithm and the complexity of different implementations, see, e.g., Chapter~4 in Burkhard et al.~(2009)). Faster algorithms are available, though, as  Bertsekas' {\it auction algorithm} or its variant, the {\it forward/reverse auction algorithm}, (Chapter~4 in Bertsekas~(199),   implemented in the R package \texttt{transport}. These auction algorithms depend on some parameter  $\epsilon>0$ and give  in $O(n^2)$ time a solution to the assignment problem which is within~$n\epsilon$ of being optimal. If the costs are integers and $n\epsilon<1$, the solution given by the auction algorithm is optimal. Else, Step 2 below provides a check for the optimality of the solution given by the auction algorithm. If the check is negative, the algorithm is iterated with a smaller value of $\epsilon$.

(Step 2) Compute the optimal value $\varepsilon_0$ of the regularization parameter and the optimal weights $\psi_i$. This is achieved via Karp's algorithm and the subsequent computation of shortest path distances as described in the discussion after Corollary~\ref{CorollaryEpsilon0}. If $\varepsilon^*<0$, then the solution of the assignment problem
was not optimal and we return to Step 1 with a smaller value of $\epsilon$. If not, we go to Step~3. %\vspace{-4mm}

(Step 3) Compute the Yosida regularization based on a projected gradient descent method.% \vspace{-2mm} 

%The next sections  investigate (Section~\ref{secConv}) the convergence of our method  and (Section~\ref{secShape1})  its ability to recover the ``shape" of a distribution for~$d=2$  and various Gaussian mixtures.   Appendix~\ref{secShape2} provides similar results for compact supports. %\vspace{-6mm}%For obvious graphical reasons, we only consider bivariate observations. %It should be noted, however, that the complexity of the algorithms 

% \subsection{Convergence}\label{secConv}\vspace{-1mm} 
 
\begin{figure}[h!]%[htbp]
\begin{center} 
\arraycolsep = 5pt \renewcommand{\arraystretch}{1}
$\noindent\begin{array}{ccc}
\framebox{\includegraphics*[scale=0.3,trim=30 30 20 30,clip]{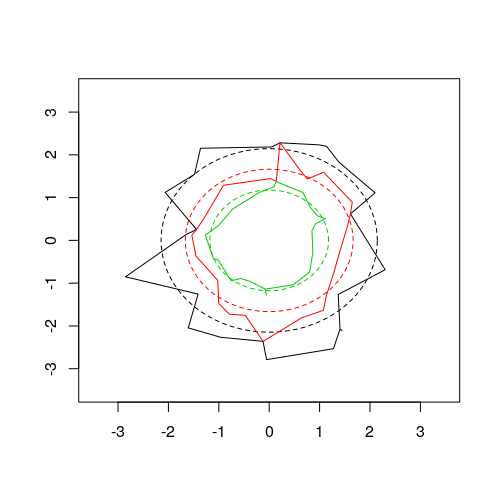}}
&
\framebox{\includegraphics*[scale=0.3,trim=30 30 20 30,clip]{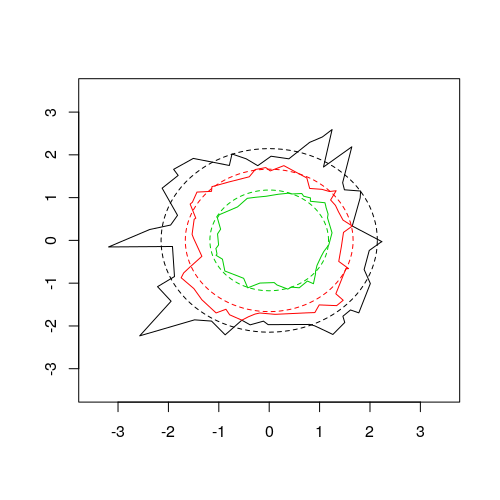}}
&
\framebox{\includegraphics*[scale=0.3,trim=30 30 20 30,clip]{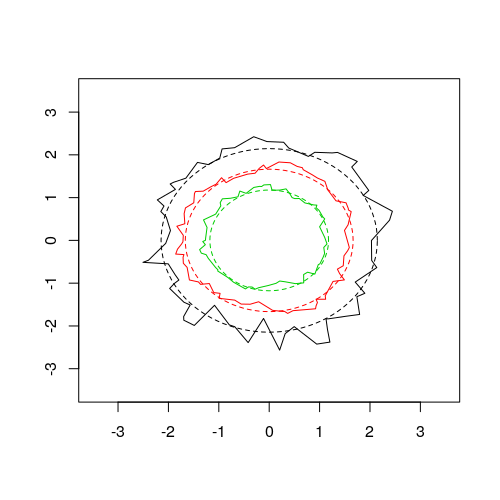}}\vspace{1mm}\\
$n=200$ & $n=500$ & $n=1000$ 
\\
\framebox{\includegraphics*[scale=0.3,trim=30 30 20 30,clip]{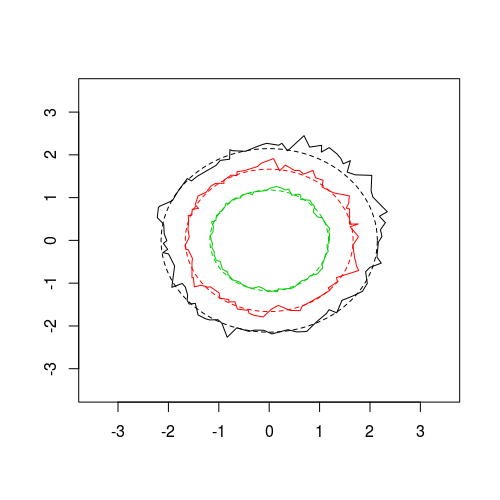}}
&
\framebox{\includegraphics*[scale=0.3,trim=30 30 20 30,clip]{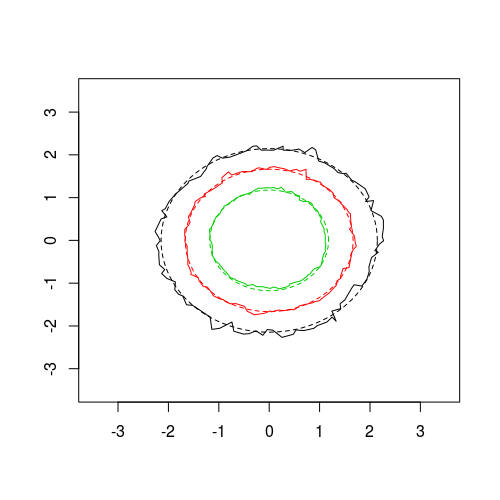}}
&
\framebox{\includegraphics*[scale=0.3,trim=30 30 20 30,clip]{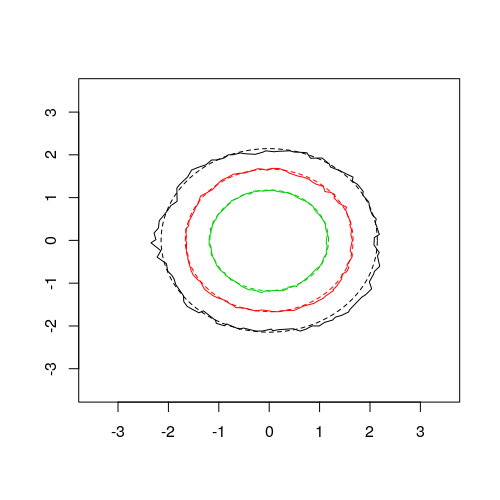}}\vspace{1mm}\\
$n=2000$ & $n=5000$ & $n=10000$ \vspace{-1mm}

\end{array}$
\caption{Smoothed empirical center-outward quantile contours (probability contents   .50   (green),   .75  (red), .90 (black)) computed from $n=$ 200, 500, 1000, 2000, 
5000, 10000 i.i.d.\ observations from a bivariate~${\cal N}( {\bf 0},{\bf I})$ distribution, along with their %(spherical)
 theoretical counterparts.}\label{FigConv}    
\end{center}\vspace{-2mm}
\end{figure}

 In Figure~\ref{FigConv}, % this section, 
 we illustrate the convergence (as formulated by the Glivenko-Cantelli result of Proposition~\ref{PropGC}),  of empirical contours to their population counterparts as the sample size increases. The problem is that  analytical expressions for the population contours are not easily derived, except for spherical distributions. We therefore investigate the case of i.i.d. observations with bivariate~${\cal N}(0,\!\text{ Id})$ distributions, and increasing  samples  sizes~$n=$ 200, \ldots 
 %500, 1000, 2000, 5000,
  10000.

  Inspection of Figure~\ref{FigConv}  clearly shows the expected consistency. Empirical contours are nicely nested, as they are supposed to be. For sample sizes as big as~$n=500$, and despite the fact that the underlying distribution is light-tailed, the $.90$ empirical contour still exhibits significant ``spikes" out and in the theoretical circular contour. Those spikes reflect the intrinsic variability of an empirical quantile of order $.90$ based on about $n_R$ observations;   they rapidly %and uniformly disappear from $n=1000$ on.  Appendix~\ref{secShape2} provides similar results for compact supports.%\vspace{-3mm}
  
%\subsection{Gaussian mixtures}\label{secShape1}\vspace{-1mm}

\begin{figure}[h!]%[htbp]\label{FigGM} 
\begin{center} 
\arraycolsep = 5pt \renewcommand{\arraystretch}{1}
$\noindent\begin{array}{ccc}
\framebox{\includegraphics*[width=3.5cm, height =3.5cm, trim= 40 180 20 180, clip]{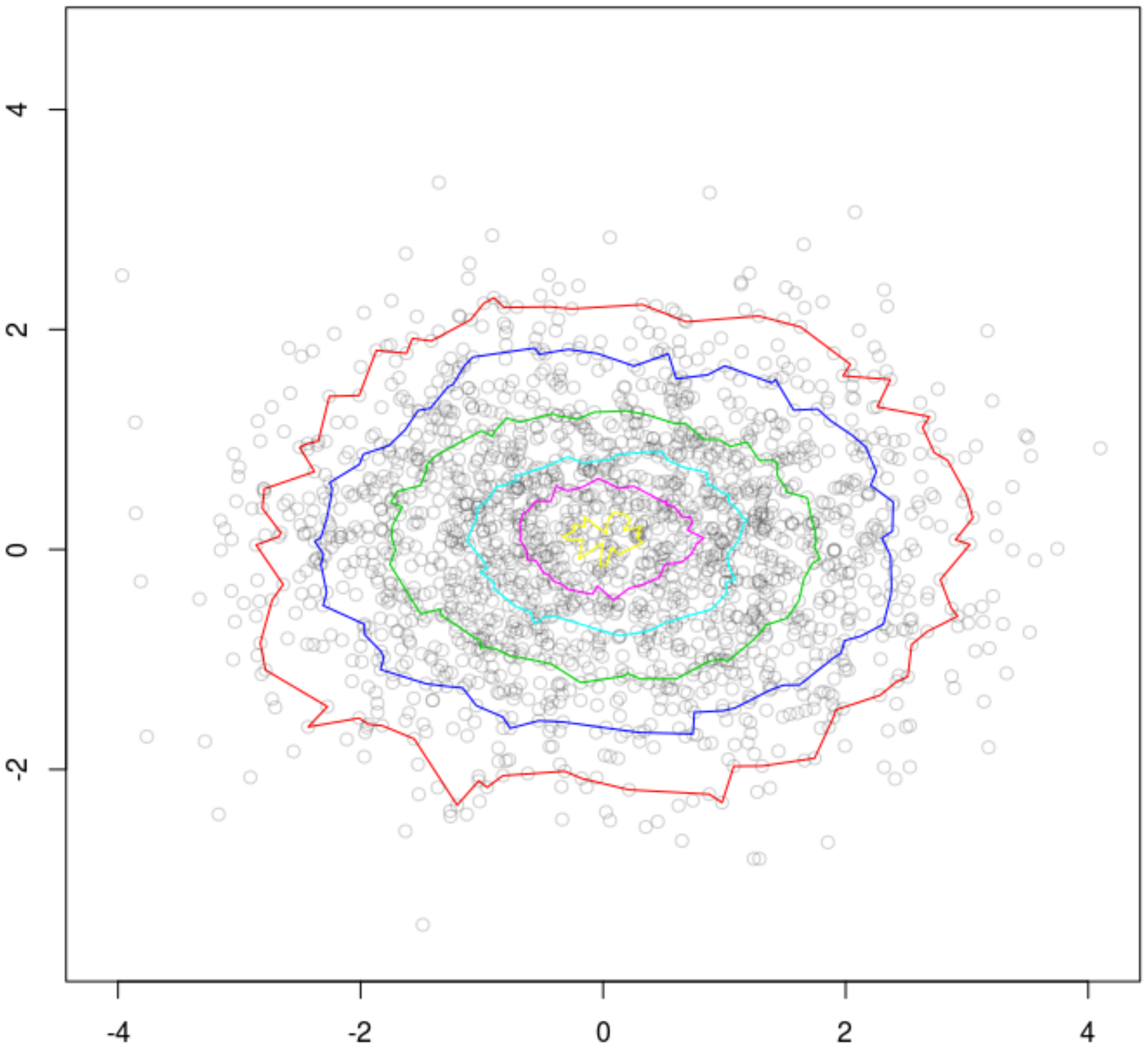}}
&
\framebox{\includegraphics*[width=3.5cm, height =3.5cm, trim= 40 180 20 180, clip]{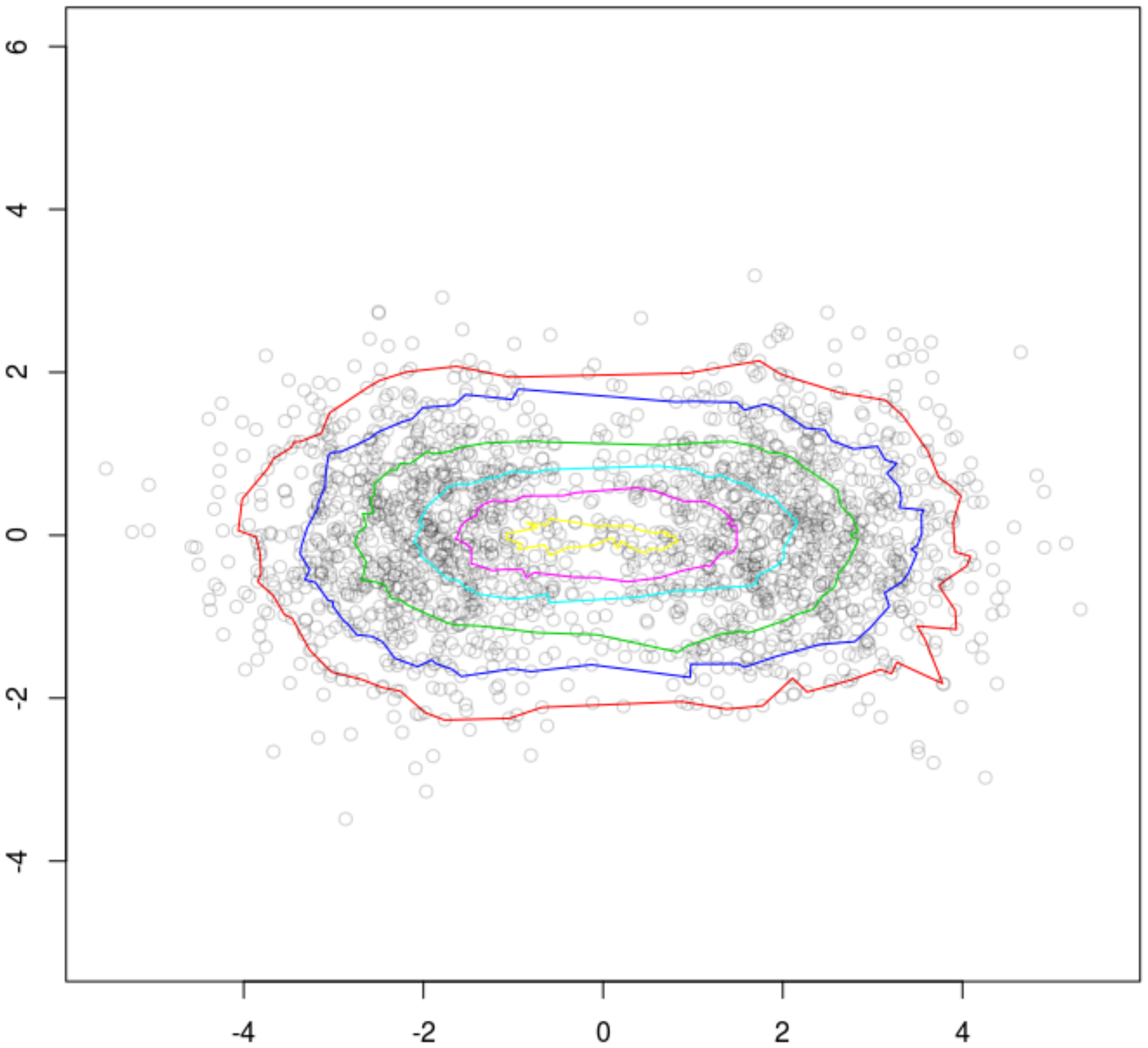}}
&
\framebox{\includegraphics*[width=3.5cm, height =3.5cm, trim= 40 180 20 180, clip]{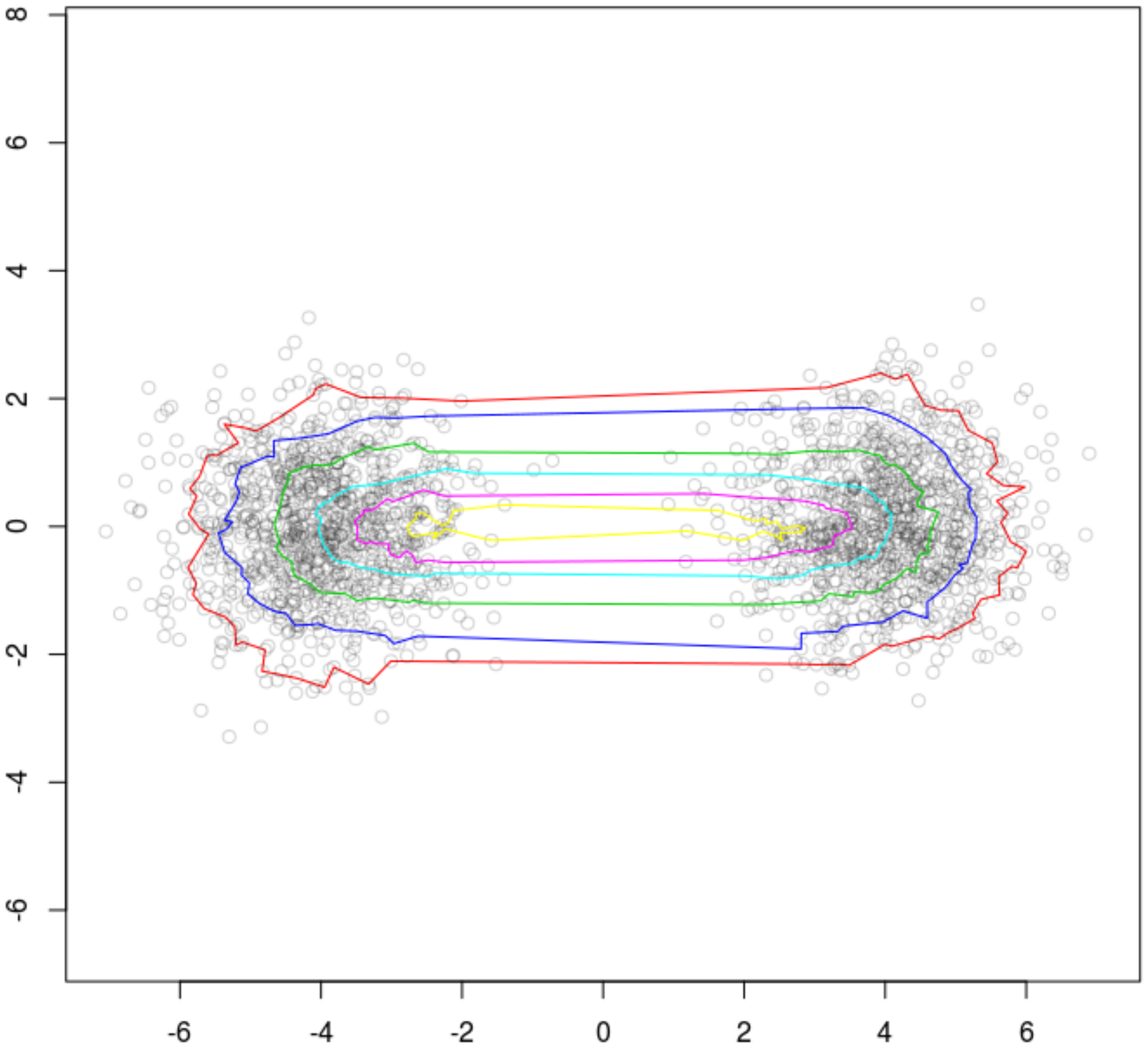}}\vspace{2mm}\\
\scriptstyle
 \frac{1}{2}
{\cal N}
\Big(\!
\Big(\!\!\!\begin{array}{c}\scriptstyle -1\vspace{-1mm} \\ \scriptstyle 0\end{array}
\!\!\!\Big), \text{Id}\Big)
+ 
\frac{1}{2}
{\cal N}
\Big(\!
\Big(\!\!\!\begin{array}{c} \scriptstyle1\vspace{-1mm}  \\ \scriptstyle 0\end{array}
\!\!\!\Big), \text{Id}\Big) 
&\scriptstyle
\frac{1}{2}
{\cal N}
\Big(\!
\Big(\!\!\!\begin{array}{c}\scriptstyle -2 \vspace{-1mm} \\ \scriptstyle 0\end{array}
\!\!\!\Big), \text{Id}\Big)
+ 
\frac{1}{2}
{\cal N}
\Big(\!
\Big(\!\!\!\begin{array}{c}\scriptstyle 2\vspace{-1mm}  \\ \scriptstyle 0\end{array}
\!\!\!\Big), \text{Id}\Big) 
 & \scriptstyle
 \frac{1}{2}
{\cal N}
\Big(\!
\Big(\!\!\!\begin{array}{c}\scriptstyle -4\vspace{-1mm}  \\ \scriptstyle 0\end{array}
\!\!\!\Big), \text{Id}\Big)
+ 
\frac{1}{2}
{\cal N}
\Big(\!
\Big(\!\!\!\begin{array}{c}\scriptstyle 4\vspace{-1mm}  \\ \scriptstyle 0\end{array}
\!\!\!\Big), \text{Id}\Big)\vspace{-1mm} 
\end{array}$
\caption
{
Smoothed empirical center-outward quantile contours (probability contents .02 (yellow), .20 (cyan), .25 (light blue)    .50   (green),   .75  (dark blue), .90 (red)) computed  from~$n=2000$ i.i.d.\ observations from  mixtures of  two bivariate Gaussian distributions.%\vspace{-4mm}
}\label{FigGMsym}   
\end{center}
\end{figure}

\begin{figure}[htbp]
\begin{center}
\arraycolsep = 3.1pt \renewcommand{\arraystretch}{1}
$\noindent\begin{array}{ccc}
\framebox{\includegraphics*[width=3.5cm, height =3.5cm, trim= 40 180 20 180,
clip]{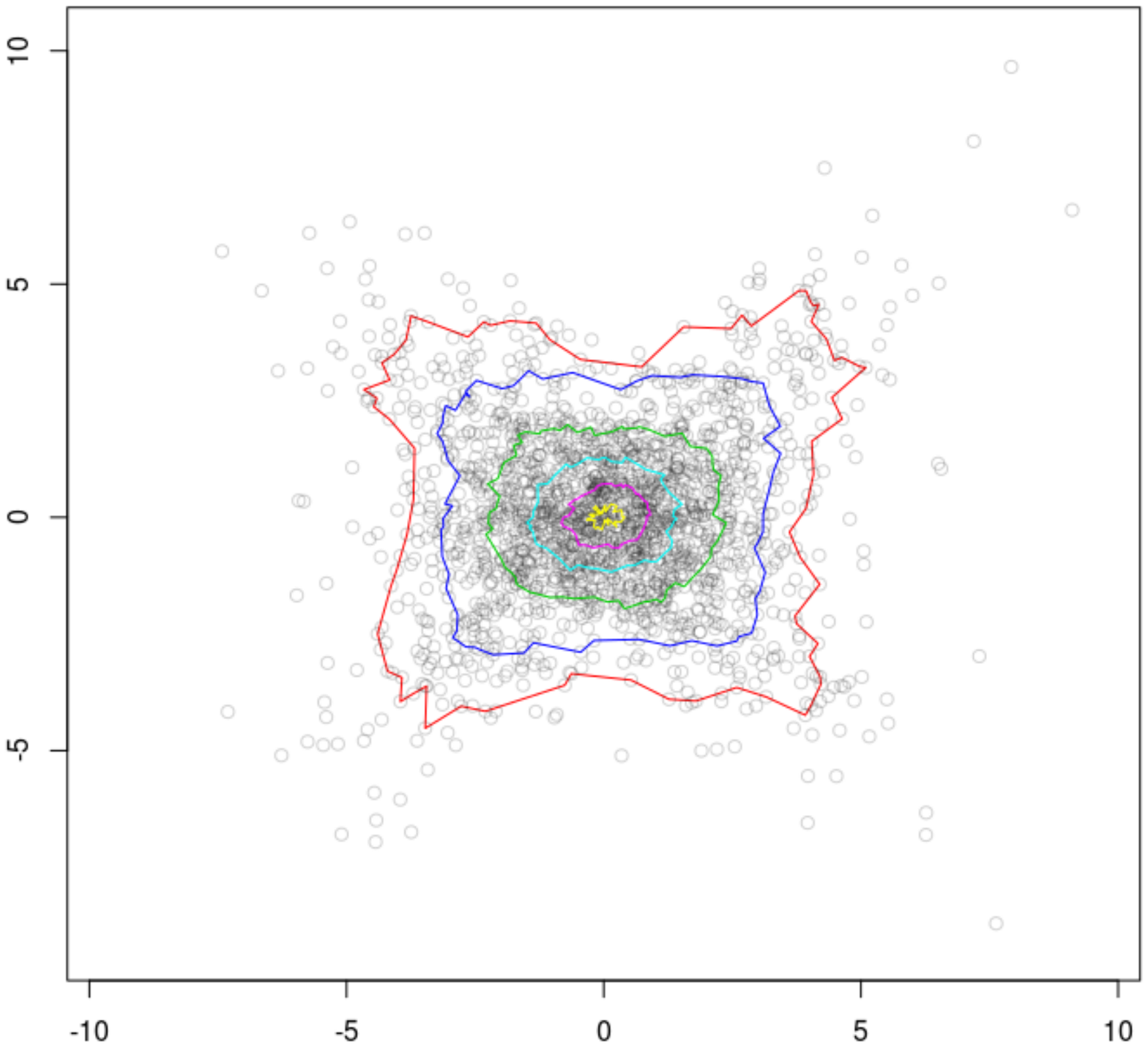}}
&
\framebox{\includegraphics*[width=3.5cm, height =3.5cm, trim= 40 180 20 180,
clip]{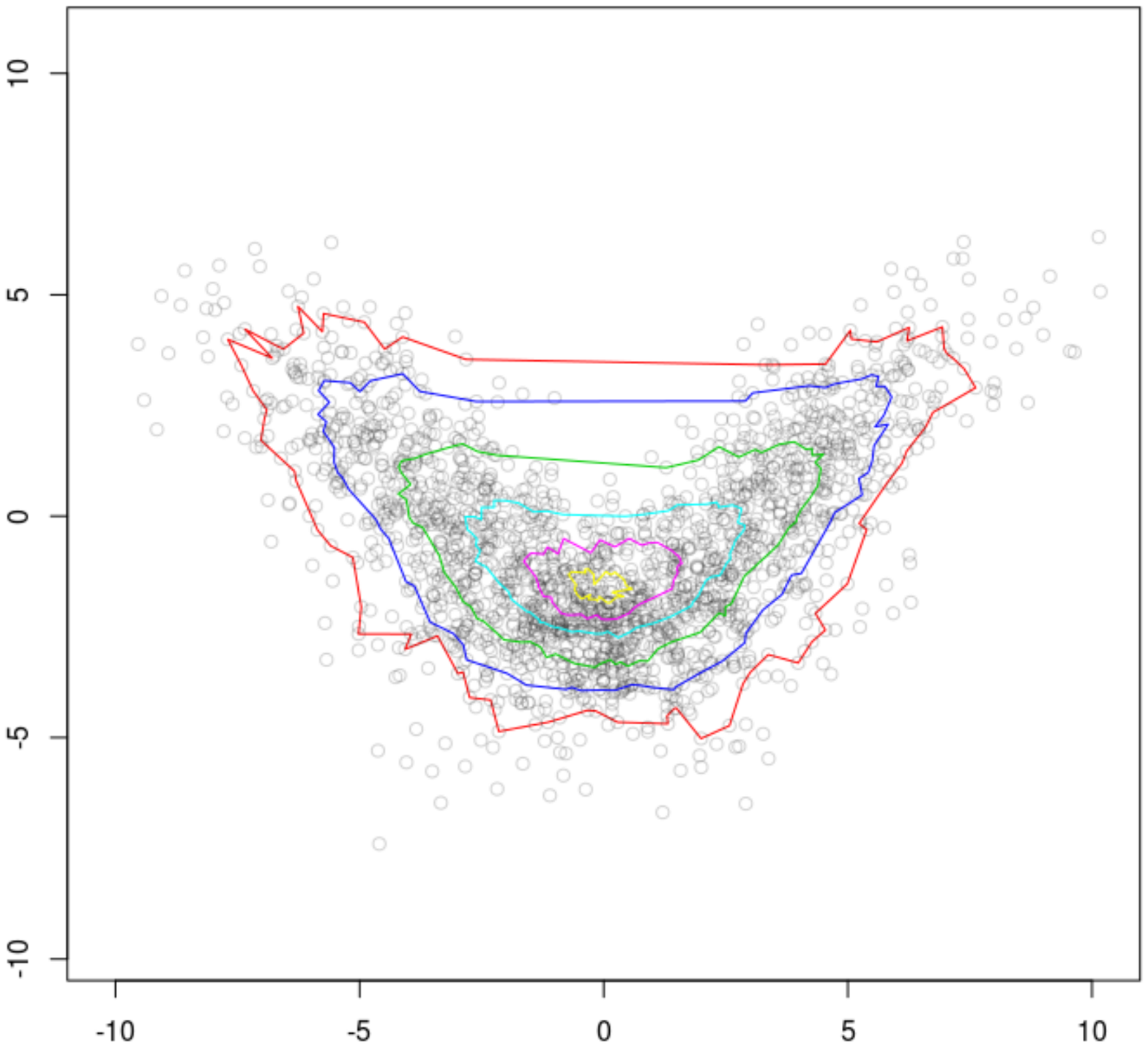}}
&
\framebox{\includegraphics*[width=3.5cm, height =3.5cm, trim= 40 180 20 180,
clip]{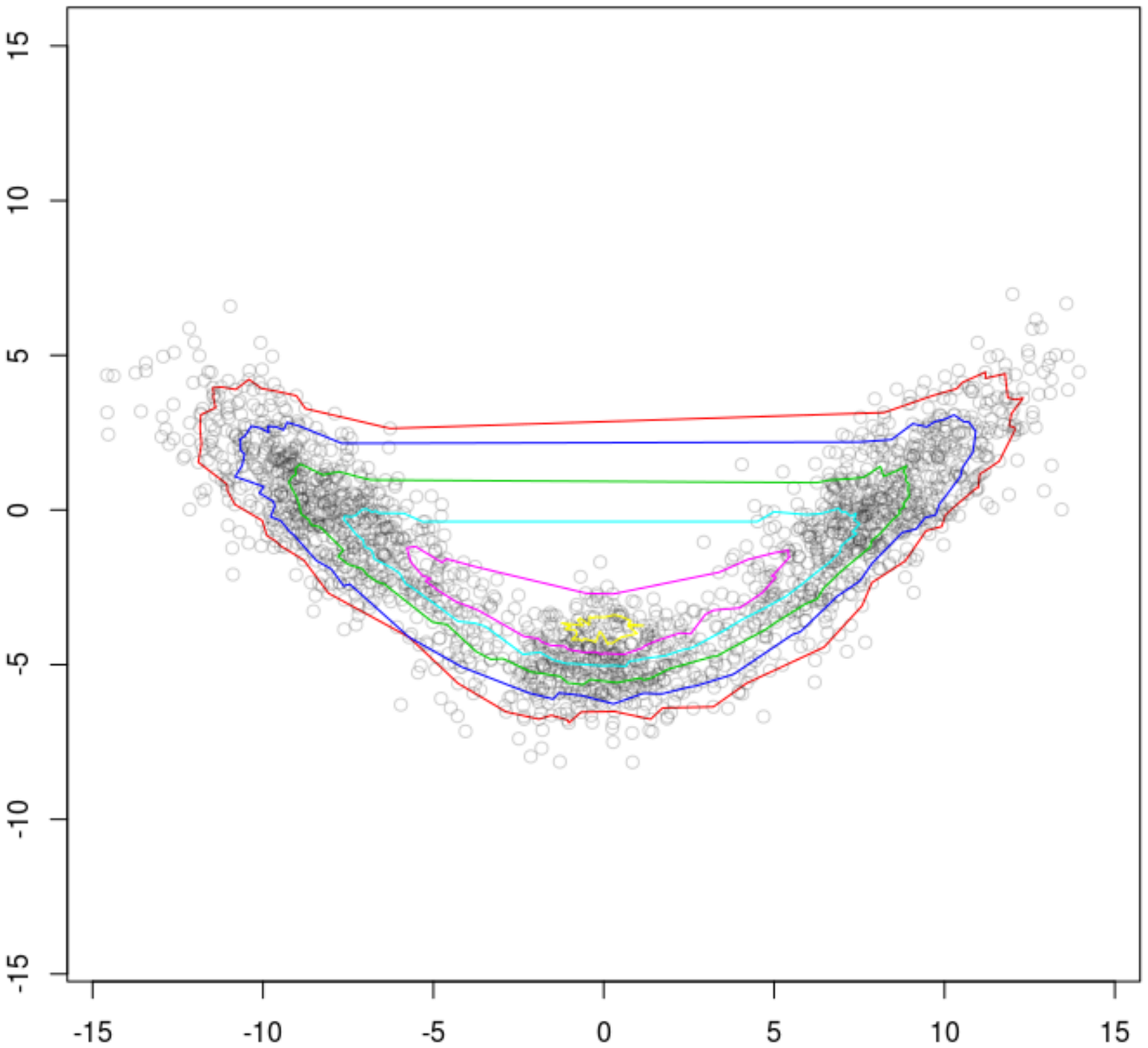}}
\vspace{2mm}\\
\scriptstyle
\frac{3}{8}{\cal N}({\boldsymbol\mu}_0,\mathbf{\Sigma}_1)+\frac{3}{8}{\cal
N}({\boldsymbol\mu}_0,\mathbf{\Sigma}_2)
&
\hspace{-2mm}
\scriptstyle
\frac{3}{8}{\cal N}(-3{\boldsymbol\mu}_h,\mathbf{\Sigma}_1)+\frac{3}{8}{\cal
N}(3{\boldsymbol\mu}_h,\mathbf{\Sigma}_2)
&
\hspace*{-2mm}
\scriptstyle
\frac{3}{8}{\cal N}(-8{\boldsymbol\mu}_h,\mathbf{\Sigma}_1)+\frac{3}{8}{\cal
N}(8{\boldsymbol\mu}_h,\mathbf{\Sigma}_2)
\\
\scriptstyle
\hspace{14mm}
+
\frac{1}{4}{\cal N}({\boldsymbol\mu}_0,\mathbf{\Sigma}_3)
&\scriptstyle
\hspace{14mm}
+
\frac{1}{4}{\cal N}(-\frac 5 2{\boldsymbol\mu}_v,\mathbf{\Sigma}_3)
&\scriptstyle
\hspace{14mm}
 +
\frac{1}{4}{\cal N}(-5{\boldsymbol\mu}_v,\mathbf{\Sigma}_3)
\end{array}$
\caption{
Smoothed empirical center-outward quantile contours  (probability
contents .02 %\protect\linebreak
 (yellow), .20 (cyan), .25 (light blue)    .50   (green),   .75
(dark blue), .90 (red))  computed from~$n=2000$   i.i.d.~observations from
mixtures of  three bivariate Gaussian distributions, with 
$
{\boldsymbol\mu}_0=\left(
\protect\begin{smallmatrix} \scriptstyle 0 \\ 
 \scriptstyle
0\protect\end{smallmatrix}
\right)
$, 
$
{\boldsymbol\mu}_h=\left(
\protect\begin{smallmatrix}
\scriptstyle 1 \\ 
 \scriptstyle 0\protect\end{smallmatrix}
 \right)
 $, 
$
{\boldsymbol\mu}_v=\left(
\protect\begin{smallmatrix} \scriptstyle 0 \\ 
 \scriptstyle
1\protect\end{smallmatrix}
\right)
$,  
$
\mathbf{\Sigma}_1=\left(
\protect\begin{smallmatrix}
\scriptstyle 5 & \scriptstyle -4 \\ 
 \scriptstyle -4 & \scriptstyle 5
\protect\end{smallmatrix}
\right)
$, 
$\mathbf{\Sigma}_2=\left(
\protect\begin{smallmatrix}
\scriptstyle 5 & \scriptstyle 4 \\  \scriptstyle 4 & \scriptstyle 5
\protect\end{smallmatrix}
\right)
$, 
$
\mathbf{\Sigma}_3=\left(
\protect\begin{smallmatrix} \scriptstyle 4 & \scriptstyle 0 \\
 \scriptstyle 0 & \scriptstyle 1 
 \protect\end{smallmatrix}
 \right)
 $.
 }\label{FigGM}
\end{center}
\end{figure}

\begin{figure}[h!]%[htbp]
\begin{center}
\arraycolsep = 5pt \renewcommand{\arraystretch}{1}
$\noindent\begin{array}{c}
\framebox{\includegraphics*[width=9.4cm,height=6.4cm,trim=30 30 20 30,clip]
{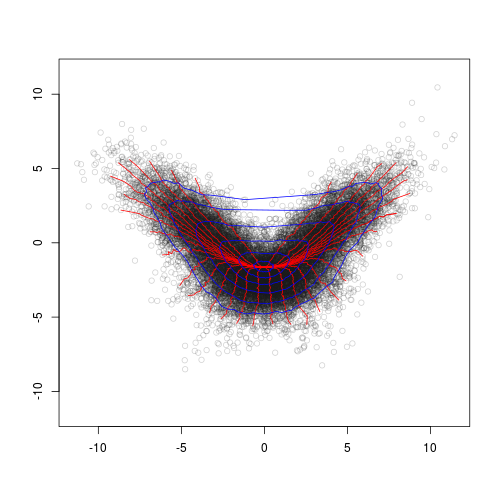}}
%&
%\framebox{\includegraphics*[width=7.5cm,height=7.4cm,trim=30 30 20 30,clip]
%{Fig6b.png}}
\end{array}$
\caption{Center-outward quantile contours and  sign curves for the same Gaussian mixture as in the middle panel of Figure~\ref{FigGM},  with
% $n=10000$ (left) and
  $n=20000$ (right).%\vspace{-4mm}
  }
  \label{BananaRays}   
\end{center}
\end{figure}

Figures~\ref{FigGMsym}--\ref{BananaRays} consider various Gaussian mixtures. Gaussian mixtures generate a variety of   possibly multimodal and non-convex empirical  dataclouds. In Figure~\ref{FigGMsym}, we simulated $n=2000$ observations from a symmetric mixture of two spherical Gaussians. Figure 3 clearly demonstrates the quantile contour  nature of our interpolations, as opposed to level contours. Level contours in the right-hand panel clearly would produce disconnected regions separating the two modes of the mixture. Here, the contours remain nested---a fundamental monotonicity property of quantiles.     The low-probability region between the two component populations is characterized by a ``flat profile" of %empirical quantile contours: the whole region in between the two modes is ``quite central", and one can move, for instance, from one mode to the other without crossing the .5  contour. 

Figure~\ref{FigGM} similarly considers a mixture of three
 Gaussian distributions producing, in the central and right panels,  distinctively nonconvex datasets. Picking that nonconvexity is typically difficult, and none of the traditional depth contours (most of them are intrinsically convex) are able to do it. Our interpolations do pick it,  the inner contours much faster than the outer ones, as $n$   increases. The very idea of a smooth interpolation indeed leads to bridging empty regions with nearly piecewise linear solutions. This is particularly clear with the .90 contour in the right-hand panel: the banana shape of the distribution is briefly sketched at the inception of the concave part, but rapidly turns into an essentially linear interpolation in the ``central part of the banana".  That phenomenon %, though,
 disappears as $n\to\infty$ %tends to infinity as
  and  the ``empty" regions eventually fill~in.

Attention so far has been given to quantile contours, neglecting an important feature of center-outward quantile functions: being  vector-valued, they also carry essential directional information. That information is contained in the empirical sign curves---the images,  by the interpolated  empirical quantile function, of the radii of the underlying regular grid. 
%Depth-based quantiles, which are scalar-valued, have nothing equivalent to offer: see Appendix~\ref{secTukey}.
 In the spherical case, those sign curves are
%, more or less, straight halflines running through the center-outward median and  uniformly distributed among   all directions: they are
 quite uninformative %, thus,
  and we did not plot them in Figures1 and~2. In the highly non-spherical   Gaussian mixture of Figure~3, those sign curves are conveying an essential information.

 Figure~\ref{BananaRays} is providing the full picture for $n=20000$ (see also Figure~\ref{BananaRays'} in Appendix~\ref{secTukey}). % and $n=20000$.
  The sign curves to the left and to the right of the vertical direction are vigorously combed to the left and  the right. Since each curvilinear sector comprised between two consecutive sign curves roughly has the same probability contents, Figure~\ref{BananaRays}   provides graphical evidence of a very low density in the central concavity bridged by the contours, thus producing a clear  visualization of the banana shape of the dataset. Such figures, rather than contours alone, are the descriptive plots associated  with  empirical center-outward quantile functions. See Appendix~\ref{secTukey} for a comparison with Tukey depth.\vspace{-1.5mm}

  \section{Conclusions and perspectives}\label{Consec}\vspace{-1mm}

 Unlike the earlier proposals, our concepts of distribution and quantile functions, ranks, and signs are satisfying the properties that make their univariate counterparts efficient and meaningful tools for statistical inference. 
 In principle, they   are paving the way   to a solution of  the long-standing open problem of  distribution-free  inference in multivariate analysis,  
%  in the absence of ``any" distributional assumptions,
   offering a unique combination of  strict distribution-freeness and semiparametric efficiency. A preliminary version (Hallin~2017) of this paper already triggered several important applications:   De Valk and  Segers~(2018), Shi et al.~(2019), Deb and Sen~(2019), Ghosal and Sen~(2019), Hallin, La Vecchia, and Liu~(2019),  Hallin, Hlubinka, and Hudecov\' a~(2020), ...  A number of  questions remain open, though. In particular,\linebreak {\it (i)} Several issues remain to be studied about the concepts themselves: how in finite samples should we choose the factorization of $n$ into $n_Rn_S + n_0$? %should we combine several of them?
   should we consider cross-validation? how do those grids compare to random grids?  

\vspace{-5mm}
  %\vspace{-12mm}

%\color{blue}

%\subsection{Sign curves}\label{signcurvesec}

%\vspace{-8mm}
%
%
% 
%%\color{blue}
% 
% %   \color{red}
%
%
% 
% 
%% ome of those problems  
%% 
%% 
%% open the door to a  new theory of empirical processes, calling for further results such as   Donsker  and iterated logarithm theorems, or Bahadur representations.   They also pave   the way     Many questions remain open, though,  until those objectives can be attained.
%% %\vspace{-2mm}
%%% \begin{enumerate}
%%% \item[--] 
%%
%% {\it (i)} Several issues remain to be studied about the concepts themselves: how in finite samples should we choose the factorization into $n_Rn_S + n_0$? should we combine several of them? should we consider cross-validation? how?  
%%    what happens if we drop the assumption of   nonvanishing densities?% \vspace{-2mm}
%%%  \item[--]  
%
{\it (ii)} How should we construct efficient rank tests in specific problems? Proposition~\ref{GCell} suggests  replacing,   in the optimal test statistics derived under elliptic symmetry,  %by Hallin, Paindaveine, and Verdebout (see the references below),
 the Mahalanobis   ranks and signs with the center-outward ones.  Can we similarly construct one-step R-estimators? 
%   (a problem which, for $d\geq 2$, so far is solved under elliptical symmetry only)? 
   This, which requires H\' ajek-type asymptotic representation results, would result in a fairly   complete toolkit of distribution-free (hence ``universally valid") semiparametrically efficient-at-elliptical-densities rank-based  %inference
 procedures for multivariate analysis and multivariate time series. % problems.% \vspace{-2mm}
%    \item[--]

  {\it (iii)}  Can 
  goodness-of-fit tests be based, e.g.~on Kolmogorov-Smirnov or Cram\' er-von Mises  distances between center-outward distribution functions?%\vspace{-2mm}
  
%   \item[--] 
{\it (iv)} Turning to quantiles, what are the properties of ${\bf Q}\n_{{\pms}} ({\bf 0})$ (for $n_0\neq 0$) as a multivariate median? can we construct multivariate median or sign tests? can we, on the model of Carlier et al.~(2016) or Hallin et al.~(2010, 2015),  perform multiple-output quantile regression? 
% (reconstruction of conditional center-outward quantile  contours as a function of covariates)?
 construct multivariate growthcharts (as in McKeague~et al.~(2011))? How?%\vspace{-2mm}
   
%    \item[--] 
{\it (v)} Center-outward quantile contours are obvious candidates as multivariate value-at-risk concepts, playing a central role in risk management; in that context, still in dimension $d=1$, the primitives of ordinary distribution or quantile functions enter the definitions of a number of relevant notions such as Lorenz curves, average values at risk, or expected
shortfall,  see Gushchin and Borzykh~(2017). The potential functions $\psi$ and $\phi$ 
%characterizing the underlying ${\bf F}_\pms$ and its Legendre transform
 are
 natural multivariate extensions of those primitives, and likely to provide useful multivariate extensions. 
     
{\it (vi)}    What happens in high dimension ($d\to\infty$)? in functional spaces? on spheres (directional data) and other Riemannian manifolds?%\vspace{-2mm}

Finally, these new empirical distribution and quantile functions   are calling for a study  of the corresponding empirical processes with further results such as   Donsker  and iterated logarithm theorems, or Bahadur representations. \vspace{-2.5mm}

 %    \end{enumerate}

\color{black}
%\medskip

%\vspace{-3mm}
%More references are provided in the  online appendix. 

\newpage

\appendix

\setcounter{footnote}{0}
\section%n{Ranks, signs, and measure transportation}
{\!Measure transportation in a nutshell}\label{transpsec}%{ranksec}

 \subsection{\!Measure transportation, from Monge to McCann}\label{transpsec1}

Starting from a very practical problem---{\it How should one best  move given piles of sand to fill up given holes of the same total volume?}---Gaspard Monge~(1746-1818), with his 1781 {\it M\' emoire sur la Th\' eorie des D\' eblais et des Remblais}, initiated a  profound mathematical theory anticipating different areas of differential geometry, linear programming, nonlinear partial differential equations, fluid mechanics, and probability.

In modern notation, the simplest  and most intuitive---if not  most general---formulation of Monge's problem is (in probabilistic form) as follows.   Let  ${\rm P}_1$ and~${\rm P}_2$ denote two %long to the family $\cal P$ of
 probability measures over (for simplicity)~$(\mathbb{R}^d, \mathcal{B}^d)$ and\linebreak  let~$L : \mathbb{R}^{2d} \to[0, \infty ]$  be a Borel-measurable loss function such that  $L({\bf x}_1,{\bf x}_2)$ represents the cost of transporting ${\bf x}_1$ to ${\bf x}_2$. The objective is to find a measurable  (transport) map $T_{{\rm P}_1 ; {\rm P}_2} : \mathbb{R}^d\to \mathbb{R}^d$ solving the  minimization problem 
\begin{equation}\label{inf}
\inf _T\int_{\mathbb{R}^d} L\big({\bf x}, T(\mathbf{x})\big)\mathrm{dP}_1
\qquad \text{subject to} \quad T\# {\rm P}_1 = {\rm P}_2
\end{equation} 
where~$T$ ranges over the set of measurable map from $\mathbb{R}^d$ to $\mathbb{R}^d$ and  $T\# {\rm P}_1$  is the so-called {\it push forward of ${\rm P}_1$ by  $T$}.\footnote{In statistics, a more classical   but heavier notation for $T\# {\rm P}_1$ would  be ${\rm P}_1^{T{\bf X}}  $ or $\overline{T} {{\rm P}_1}$, where~$\overline{T} $ is the transformation of $\cal P$ \textit{induced by}~$T$; see Chapter~6 of Lehmann and Romano~(2005).} For simplicity, and with a slight abuse of language, we will say that $T$ {\it  is mapping ${\rm P}_1$ to}  ${\rm P}_2$.  A map~$T_{{\rm P}_1 ; {\rm P}_2}$ achieving the infimum in~(\ref{inf})   is called an {\it optimal transport map}, in short,  an {\it optimal transport}, of ${\rm P}_1$ to~${\rm P}_2$.  In the sequel, we shall restrict to  the quadratic (or L$^2$) loss function~$L({\bf x}_1,{\bf x}_2)=\Vert {\bf x}_1- {\bf x}_2 \Vert^2_2$.

The problem looks simple but it is not. Monge himself (who moreover was considering the more delicate loss $L({\bf x}_1,{\bf x}_2)=\Vert {\bf x}_1- {\bf x}_2 \Vert_2$) did not solve it, and relatively little progress was made until the 1940s
%, when  renewed interest in the topic was triggered   by
and  the pathbreaking duality approach of %Leonid Vitalievitch
 Kantorovich. Relaxing the problem into %as %under with a more flexible formulation . % and, in particular, his  duality approach.  
 \begin{equation}\label{Kantorovich}
 \inf _\gamma\int_{\mathbb{R}^d\times\mathbb{R}^d } L\big({\bf x}, {\bf y}\big)\mathrm{d}\gamma ({\bf x}, {\bf y})
\qquad \text{subject to} \quad \gamma\in\Gamma({\rm P}_1, {\rm P}_2)
 \end{equation}
 where $\Gamma({\rm P}_1, {\rm P}_2)$ denotes the collection of all distributions over $\mathbb{R}^d\times\mathbb{R}^d$ with marginals ${\rm P}_1$ and~${\rm P}_2$, Kantorovich established that the solutions of \eqref{Kantorovich} are of the form~$\big(\text{Identity} \times T\big)\#{\rm P}_1$ where $T$ solves Monge's problem \eqref{inf}. 

The topic attracted a renewed surge of interest some thirty years ago. %, and has become one of the most active subjects of contemporary pure and applied mathematics,
  For the  L$^2$ transportation cost,  Cuesta-Albertos and Matr\'an (1989) showed (under the assumption of  finite second-order moments) the existence of   solutions of  the Monge problem   and  
%(this result working in Hilbert spaces and relaxing the absolute continuity of the involved distributions), 
Rachev and R\"uschendorf~(1990) characterized them in terms of gradients of convex (potential) functions.  Brenier (1991) with his celebrated {\it polar factorization theorem}  independently obtained the same results and, moreover, established  the (a.s.) uniqueness of the solution. 
% Among  the most powerful ensuing results is the {\it Polar Factorization Theorem} by Brenier~(1987, 1991;  see Chapter~3 in Villani~(2003))  which implies, among other things,  that for~L$^2$ loss, if~$ {\rm P}_1$ and~${\rm P}_2$ are absolutely continuous with  finite second-order moments, the solution of Monge's problem exists, is (a.e.) unique, and the gradient of a convex (potential) function---a form of multivariate monotonicity.

Measure transportation  ever since  has been among the most active domains of mathematical analysis, with applications in various fields, from fluid mechanics to economics (see   Galichon~(2016)), learning, and statistics (Carlier et al.~(2016); Panaretos and Zemel~(2016, 2018); \'Alvarez et al.~(2018) and del Barrio et al.~(2018)). It was popularized recently by the  French Fields medalist C\' edric Villani, with two authoritative monographs (Villani~2003, 2009), where we refer to for background reading, along with  the two volumes by Rachev and R\" uschendorf~(1998), where the scope is somewhat closer to probabilistic and  statistical concerns.  
 
 Whether described as in (\ref{inf}), or relaxed into the more general coupling form~\eqref{Kantorovich}, the so-called Monge-Kantorovich problem remains an optimization problem, though, which  only makes sense under densities for which expected costs are finite---under finite variances, thus, for quadratic loss. When defining concepts of distribution and quantile  functions, ranks and signs, one clearly would like to avoid such assumptions. This is made possible thanks to   a  remarkable  result by  McCann~(1995), hereafter     {\it McCann's theorem}, the   nature of  which is geometric rather than analytical. Contrary to Monge, Kantorovitch, and Brenier,   McCann (1995) does not require any moment restrictions and avoids using Kantorovich duality.  McCann's main theorem implies  that, for any given absolutely continuous distributions~${\rm P}_1$ and~${\rm P}_2$ over~$\mathbb{R}^d$, there exists  convex functions~$\psi : \mathbb{R}^d\to \mathbb{R}$ with a.e. gradients\footnote{Recall that a convex function is a.e. differentiable.}~$\nabla\psi$ pushing~$ {\rm P}_1$ forward to~${\rm P}_2$; although $\psi$ may not be unique,~$\nabla\psi$ is~${\rm P}_1$-a.s. uniquely determined\footnote{This means that, if $\psi_1$ and $\psi_2$ are convex and such that $\nabla\psi_1\#{\rm P}_1= {\rm P}_2= \nabla\psi_2\#{\rm P}_1$, then~${\rm P}_1\left[\{
{\bf x}:\, \nabla\psi_1({\bf x})\neq \nabla\psi_2({\bf x}) \right]=0$.}.    Under the existence  of   finite moments of order two,  $\nabla\psi$ moreover is with a~L$^2$-optimal (in the Monge-Kantorovich sense) transport  pushing~${\rm P}_1$ forward to~${\rm P}_2$.  
 
 \subsection{Measure transportation, quantiles, and ranks: a review of the literature}\label{reviewsec}
Measure transportation  ideas  only recently  made their way to statistical applications. Most of  them are related to Wasserstein distance (see Panaretos and Zemel~(2019)) and somewhat disconnected from the problems considered here.  They are the basis, however,  of Carlier et al.~(2016)'s method of {\it vector quantile regression} and  Chernozhukov et al.~(2017)'s concept of {\it Monge-Kantorovich depth} and related quantiles, ranks and signs, two papers of which   Ekeland et al.~(2012) can be considered a precursor.  While Carlier et al.~(2016) consider mappings to the unit cube, Chernozhukov et al.~(2017)   deal with mappings to general reference distributions, including the uniform over the unit ball. On the other hand, they %However,    Chernozhukov et al.~(2017)
  emphasize the consistent estimation of Monge-Kantorovich depth/quantile contours, with techniques that  strongly exploit  Kantorovich's duality approach and  require   compactly supported distributions, hence finite moments of all orders. 
  
  In the present paper, we  privilege mappings to the (spherical) uniform   over the unit ball, which  enjoys better invariance/equivariance properties than the unit cube---the latter indeed  is not coordinate-free, and possesses edges and vertices, which are ``very special points". Mappings to the unit ball naturally extend the structure of elliptical models, which is central to classical Multivariate Analysis, and is induced by  linear sphericizing transformation---transports to spherical distributions. The same spherical structure  also is the basis of the Mahalanobis ranks and signs approach developed in Hallin and Paindaveine~(2002a, b, c, etc.). Adopting   McCann's geometric point of view, we manage to waive moment assumptions   which, as we already stressed,    are  not natural  in the context.   Moreover, we are  focusing on the  inferential virtues  of   ranks and signs, which are rooted in their independence with respect to the order statistic. The focus, applicability and decision-theoretic nature of our approach, in that respect,  are quite different from    Chernozhukov et~al.~(2017). 
  
  This paper results from  merging two working papers, Hallin~(2017) (essentially, Sections~\ref{introintro} and~\ref{MKsec}, with the Glivenko-Cantelli and Basu factorization results of Sections~\ref{GCsec} and~\ref{DFsec}) and del Barrio et al.~(2018) (essentially, Sections~\ref{sec3} and~\ref{numsec}, with the cyclically monotone interpolation of Section~\ref{sec3}, the extended Glivenko-Cantelli result  of  Proposition~\ref{PropGC}, and the numerical illustrations of Section~\ref{numsec}). 
  
  Inspired by Chernozhukov et al.~(2017), Boeckel et al.~(2018) propose, under the name of {\it $\nu$-Brenier distribution function} ($\nu$ a distribution over $\mathbb{R}^d$ with convex compact support\footnote{The authors suggest the Lebesgue-uniform  rather than the spherical uniform distribution over the unit ball. }), a very general concept the empirical version of which satisfies a Glivenko-Cantelli property under  compactly supported absolutely continuous distributions.   Their empirical  $\nu$-Brenier distribution functions, however, are  obtained by mapping the sample to an independent  random sample of $\nu$ and therefore do not provide (even for $d=1$) a neat interpretation in terms of ranks and signs.   Yet another approach is taken in a recent paper by Faugeras and R\" uschen\-dorf~(2018), who propose combining  a mapping in the Chernozhukov et al.~(2017) style with a preliminary  copula transform. This copula transform takes care of the compact support/second-order moment restriction, but results in a concept that crucially depends on the original coordinate system. 
  
The ideas developed in Chernozhukov et al.~(2017) and Hallin~(2017), on the other hand, have  been successfully adopted by Shi, Drton, and Han~(2019), who exploit the distribution-freeness properties of center-outward ranks in the construction of distribution-free tests of independence between random vectors (a long-standing open problem).  Deb and Sen~(2019)  obtain similar results
%\footnote{The results in Deb and Sen~(2019) are not exactly the same as ours, though, due to slight differences in the definitions. For instance, their definition of the quantile function as a $\rm P$-a.s. inverse of ${\bf F}_\pms$ explains the fact that $\{\bf 0\}$ and  ${\bf Q}_\pms(\{\bf 0\})$ do not play the same special role as in Figalli~(2018) or del Barrio et al.(2019).} 
 using  different reference uniform distributions, different empirical transports, and different asymptotic techniques. In both cases, the key properties  are distribution-freeness (and  the Basu factorization property (DF$^+$) which, however, is not explicitly mentioned) of center-outward ranks.   Ghosal and Sen~(2019) also propose population concepts of distribution and quantile functions that are  similar  to those of Hallin~(2017). Their empirical versions, however, are quite different, as their objective, contrary to this paper, is  quantile reconstruction rather than a multivariate theory of rank-based inference. In particular, their empirical distribution and quantile functions are based on semi-discrete transportation (pushing the empirical distribution of the sample forward  to a continuous reference such as~${\rm U}_d$ or the Lebesgue-uniform over the unit cube). The resulting ranks and signs then are losing the distribution-freeness properties that are central to our approach. The computational benefit is that their empirical quantile functions, contours, and regions   are obtained directly via the semi-discrete optimal transport\footnote{As a consequence,    their empirical distribution functions  only are continuous, while ours are at least Lipschitz-continuous   (see Corollary~\ref{CorollaryEpsilon0}) and the related discussion.} instead of  cyclically monotone interpolation as in Section~\ref{sec3}.  

Recently, optimal {\it center-outward R-estimators}  have been derived (Hallin, La Vecchia, and Lu~2019) for
VARMA models, optimal {\it center-outward rank tests} are proposed by Hallin, Hlubinka, and Hudecova~(2019) for multiple-output regression and MANOVA, while center-outward quantile-based methods for the measurement of multivariate
risk are proposed in del Barrio, Beirlant, Buitendag, and Hallin~(2019). Applications to the study of tail behavior and extremes can be found  in De Valk and Segers~(2018).

  \setcounter{equation}{0}

\section{Distribution and quantile functions, ranks, and signs in~dimension one}\label{1dimsec}
\subsection{Traditional univariate concepts%distribution and quantile function in $\mathbb{R}$
}\label{Classconcsec}
The population and empirical concepts of   distribution function, hence those   of ranks, signs,  order statistics, and quantiles, are well understood  in  dimension one. 
% the family ${\cal F}^1$ of all distributions with nonvanishing densities on the real line $\mathbb R$. 
%, and the family~${\cal F}^d_{\text{\tiny{ell}}}$ of all elliptical distributions over $\mathbb{R}^d$ with nonvanishing radial densities. 
Before introducing   multivariate extensions, let us  briefly revisit some of their essential properties. % the traditional versions  of those   fundamental concepts and some of  their main properties. %  in order to facilitate further multivariate extensions.  

%\subsection{Traditional univariate concepts}\label{rpropsec}
Let $F$ denote the distribution function of a random variable $Z$ with  distribution ${\rm P}\in{\cal P}_1$. %  and density~$f$.  %and %connected  
%support $\overline{\text{spt}}({\rm P}\!_f)$ with interior spt$({\rm P}\!_f)$.
 It is well known that~$F$ is a {\it probability-integral transformation} ($Z\sim{\rm P}\!_f$ iff $F(Z)\sim{\rm U}_{[0,1]}$, where~${\rm U}_{[0,1]}$ is the uniform over~$[0,1]$), that is, in the terminology of measure transportation,~$F$ {\it pushes~${\rm P}$ forward to}~${\rm U}_{[0,1]}$:~$F\#{\rm P}={\rm U}_{[0,1]}$. %Moreover, 
%If, moreover, for any $D>0$, there exists~$0<\lambda_D\leq \Lambda_D$ such that $\lambda_D\leq f(z)\leq\Lambda_D$ for any $z$ in~spt$({\rm P}\!_f)\cap [-D,D]$, then
 %$F$  is a homeo\-morphism between spt$({\rm P}\!_f)$ and $(0,1)$ and so is its inverse~$Q:=F^{-1}$.
 % between~[0,1] and spt$({\rm P}\!_f)$.  

Denote by  ${\bf Z}\n :=\big(Z_1\n,\ldots , Z\n_n\big)$ an $n$-tuple of  random variables---observa\-tions or residuals associated with some parameter $\pmb\theta$ of interest (see Section~\ref{orderRd}). Denoting by $R\n_i$  the rank of $Z\n_i$ among $Z_1\n,\ldots , Z\n_n$, the value  at $Z\n_i$ of the empirical distribution $F\n$ of ${\bf Z}\n$ is~$F\n(Z\n_i):=R\n_i /(n+1)$, where 
  the denominator $(n+1)$ is adopted rather than $n$ in order for $F\n(Z\n_i)$ to take values in $(0,1)$ rather than $[0,1]$. Note that the mapping $Z\n_i\mapsto~\!R\n_i /(n+~\!1)$ is  monotone nondecreasing   from the sample to the regular grid\begin{equation}\label{classicalgrid}\{1/(n+1),\ 2/(n+1),\ldots , n/(n+~\!1)\}. 
\end{equation}
 The empirical distribution function $F\n$ then can be defined over $\mathbb{R}$ as an arbitrary non-decreasing interpolation of this discrete mapping.  
 %e $n$ couples~$\big(Z\n_i, F\n(Z\n_i)\big)$: 
   Usual practice is adopting a right-continuous step function interpolation, but  that choice carries no information and has no %particular 
   particulat statistical justification: any other choice is as legitimate. 
  
    Intimately related with the concept of ranks is the {\it dual} concept of {\it order statistic} ${\bf Z}\n_{{\scriptscriptstyle{(\,\normalsize{\bf .}\, )}}}:=\big(Z\n_{(1)},\dots,Z\n_{(n)}\big)$, with the $r$th order statistic~$Z\n_{(r)}$, $r=1\ldots ,n$ implicitly
 defined by~$Z\n_{(R\n_i)} = Z\n_i$,  
$ i=1,\ldots ,n$. Assume that ${\bf Z}\n$ is an i.i.d.~$n$-tuple with unspecified  %density~$f$ and %,  distribution function~$F$, and 
distribution ${\rm P}\in{\cal P}_1$. %, (joint distribution ${\rm P}\n_{\!f}$).
 Then, ${\bf Z}\n_{{\scriptscriptstyle{(\,\normalsize{\bf .}\, )}}}$ is {\it minimal sufficient}  and {\it complete} for $f$, while the vector ${\bf R}\n:= \big(R\n_1,\ldots ,R\n_n\big)$ of ranks is uniform over the $n!$ permutations of~$\{1,\ldots ,n\}$, hence distribution-free. Clearly, there is a one-to-one correspondence between $\left({\bf Z}\n_{{\scriptscriptstyle{(\,\normalsize{\bf .}\, )}}},{\bf R}\n\right)$ and  ${\bf Z}\n$, so that~(DF$^+$) follows from  Basu's Theorem (see Section~\ref{orderRd}). 
  The Glivenko-Cantelli theorem moreover   tells us that, irrespective of the nondecreasing interpolation $F\n$ adopted,  
\begin{equation}\label{GC1}\sup_{z\in\mathbb{R}} \Big\vert  F\n(z) - F(z)\Big\vert \longrightarrow 0\quad \text{a.s.\  as } n\to\infty 
\end{equation}
which, for ${\rm P}\in{\cal P}_1$,    is equivalent to the apparently weaker forms %property (GC)  that 
\begin{equation}\label{GC1'}\sup_{z\in\text{spt}({\rm P})} \Big\vert  F\n(z) - F(z)\Big\vert \longrightarrow 0\quad \text{a.s.\  as } n\to\infty 
\end{equation}
where spt$({\rm P})$ denotes the support of $\rm P$ and
\begin{equation}\label{empirical}
\max_{1\leq i\leq n}\Big\vert   F\n(Z\n_i) - F(Z\n_i) \Big\vert  \longrightarrow 0\quad \text{a.s.\  as } n\to\infty .
\end{equation}
%Actually, $F\n$ is entirely determined by its restriction to the~$Z\n_i$'s---the $n$ couples~$(Z\n_i,F\n(Z\n_i))$, $i=1,\ldots ,n$. All other values of $F\n$ constitute~an arbitrary  interpolation carrying no further information: any choice of a~nonde- creasing interpolation would be equally legitimate and %, in particular, 
%does satisfy the~same Glivenko-Cantelli property \eqref{GC1}. From now on, we use the notation  $F\n$ for that restriction;  
%%(a data-driven mapping of the observations to the grid~\eqref{classicalgrid});
% any monotone nondecreasing interpolation will be denoted by~$\bar{F}\n$. 

 Finally, note that ${\rm P}\in{\cal P}_1$ (as well as $F$) is entirely characterized by the restriction of $F$ to spt$({\rm P})$ and the fact that it is monotone nondecreasing (i.e., the gradient of a convex function).

\subsection{Univariate center-outward  concepts%distribution and quantile function in $\mathbb{R}$%, quantiles, ranks, and signs
}\label{pmranksec}
The strong  left-to-right  orientation of the real line  underlying  the definition of~$F$, the ranks, and $F\n$, however,  cannot be expected to extend to dimension~two and higher. 
For the purpose of multidimensional generalization,  we therefore  consider  slightly modified concepts, based on a center-outward orientation. Define the {\it center-outward distribution function}    of    $Z\sim {\rm P}\!_f\in{\cal P}_1$ as~${\bf F} _{\!{\pms}} := 2F-1$.

  Clearly,  being  linear transformations of  each other, $F$ and ${\bf F} _{\!{\pms}}$ carry the same information about~${\rm P}\!_f$. Just as~$F$,%  the center-outward distribution function
  ~${\bf F} _{\!{\pms}}$ is a probability-integral transformation, now to  the uniform distribution ${\rm U}_1$ over the 
 % one-dimensional
   unit ball~$\mathbb{S}_1\! =\! (-1,1)$: 
%\begin{equation}\label{FpmU}
%$Z\!\sim\!{\rm P}\!_f$ iff~$U\! :={\bf F} _{\!{\pms}}(Z)\!\sim\! {\rm U}_1$ or, equivalently,
${\bf F} _{\!{\pms}}\# {\rm P}\!_f = {\rm U}_1$. 
%\end{equation}

 Boldface is used in order to emphasize the interpretation of ${\bf F} _{\!{\pms}}$ as a vector-valued quantity:  while $\Vert {\bf F} _{\!{\pms}} (z)\Vert =\vert2F(z)-1\vert$ is the ${\rm U}_1$-probability contents of the interval $(\pm\Vert {\bf F} _{\!{\pms}} (z)\Vert$) (the one-dimensional ball with radius $\Vert {\bf F} _{\!{\pms}} (z)\Vert$),  the unit vector ${\bf S} _{\pms}(z):={\bf F} _{\!{\pms}} (z)/\Vert {\bf F} _{\!{\pms}} (z)\Vert$ (${\bf S} _{\pms}(0)$ can be defined arbitrarily) is a direction (a point on the unit sphere~${\cal S}_0=\{-1, 1\}$) or a sign---the sign of the devia\-tion %~$z~{\!-\!}~\text{Med}({\rm P}\!_f)$
  of~$z$ from the median $\text{Med}({\rm P}):=F^{-1}(1/2)={\bf F} _{\!{\pms}} ^{-1}(0)$ of ${\rm P}$ (possibly, an interval~$[\text{Med}^-({\rm P}), \text{Med}^+({\rm P})]$ that does not intersect spt$({\rm P})$). 
  %Those interpretations, as we shall see, will carry over to dimension~$d\geq 2$. 
 
%A quantile function usually is defined as the  inverse of a distribution function.
Inverting %the restriction to spt$({\rm P})$ of 
${\bf F} _{\!{\pms}}\!$ yields
%\footnote{That restriction is continuous and strictly increasing}
 a  (possibly set-valued)  {\it center-outward quantile function}~${\bf Q} _{{\pms}}$. % mapping $u\in\mathbb{S}_1$ to ${\bf Q} _{{\pms}}(u):={\bf F} _{\!{\pms}}^{-1}(u)\in\text{spt}({\rm P})$. 
%;     ${\bf F} _{\!{\pms}}\!$ and ${\bf Q} _{{\pms}}\!$ are homeomorphisms between spt$({\rm P}\!_f)$   and ${\mathbb S}_1$. 
% 
% $${\bf Q} _{\pms}: {\bf u}\in \mathbb{S}_1=(-1,1)\mapsto 
%{\bf Q} _{\pms} ({\bf u}):={\bf F} _{\!{\pms}}^{-1}({\bf u}).$$ 
%Quantile  are indexed by the points %$\bf u$
% of the unit ball $\mathbb{S}_1$; %=(-1,1)$; 
%  $u=\Vert{\bf u}\Vert\in(0,1]$ is to be interpreted as a {\it quantile level}. 
%%=\{z^-_{\Vert{\bf u}\Vert}, z^+_{\Vert{\bf u}\Vert}\}\qquad 
%%$$
%%\vspace{1.5mm}. 
  The sets~$\big\{{\bf Q} _{\pms} (u)  \big\vert \vert u\vert\! =\! p \big\}~\!=~\!\big\{ z^-_{p}\! , z^+_{p} \big\}%\quad u\in[0,1]
  $ 
 and   inter\-vals~$\big\{{\bf Q} _{\pms} (u)  \big\vert \vert{u}\vert \leq p \big\}~\!=~\!\big[ z^-_{p}\! , z^+_{p} \big]%,\quad u\in[0,1]
    $, 
 with $z^-_{p}$ and $z^+_{p}$  such that~${\rm P}\big([z^-_{p},\text{Med}^-({\rm P})]\big) = {\rm P}\big([\text{Med}^+({\rm P}),z^+_{p}]\big) = p/{2}$,     accordingly have the interpretation of {\it quantile contours} and {\it quantile regions}, respectively, with quantile level (probability contents) $0\leq p <1$. %at quantile level~$u$.% $u\in[0,1]$.   
 Those quantile regions are closed, connected, and nested.  
 
% While traditional distribution and quantile functions are associated with nested  half-lines of the form $(-\infty, z_u]$ carrying probability $u\in(0,1)$, the center-outward ones are about nested intervals~$[z^-_{u},z^+_{u}]$ (containing $\text{Med}({\rm P})$)    with ${\rm P}$-probability contents $u \in[0,1)$, the geometry of which, unlike the traditional collection of half-lines (which is fixed), is adapted to the underlying~${\rm P}$. 
% The translation of the center-outward  concept in terms of the traditional one is straightforward, though, as $z^-_{u} =z_{u}$ and $z^+_{u} = z_{1- {u}}$, where $z_\alpha := F^{-1}(\alpha)$. 
%\vspace{2mm}

%\subsection{Center-outward ranks and signs in $\mathbb{R}.$\!\!\!\!}\label{c-orandssec}
Turning to a sample $Z\n_1\!\!,\ldots ,Z\n_n\!\!$ (with probability one, $n$ distinct values), consider the $\lfloor n/2\rfloor $ observations sitting to the right of the median. Ordering them from  left to  right yields  ranks $R\n_{{\pms} ;i}$, say, with values $1,\ldots,\lfloor n/2\rfloor $; give them  sign~${\bf S}\n_{{\pms} ;i}={\bf 1}$ (the positive unit vector). Similarly rank the $\lfloor n/2\rfloor $~observa\-tions sitting to the left of the median  from right to left, obtaining  ranks~$R\n_{{\pms} ;i}$;~give them  sign~${\bf S}\n_{{\pms} ;i}=-{\bf 1}$.\footnote{In case $n$ is odd and the median is~$Z\n_{i_0}$, put ${\bf S}\n_{{\pms} ;i_0}={\bf 0}$ and~$R\n_{{\pms} ;i_0}=0$.} Call   $R\n_{{\pms} ;i}$ and   ${\bf S}\n_{{\pms} ;i}$  {\it center-outward ranks} and {\it signs}, respectively, and define the {\it empirical center-outward distribution function}~as 
\begin{equation}\label{Fnpm}
{\bf F}\n _{\pms}(Z\n_i):= {\bf S}\n_{{\pms} ;i}\dfrac{R\n _{{\pms};i}}{\left\lfloor n/2\right\rfloor +1}=
\left\{
\begin{array}{cl}
2F\n(Z\n_i)-1
&\text{$n$  odd}\vspace{1mm}\\ 
 \dfrac{n+1}{n+2}\left(2F\n(Z\n_i)-1\right) +\dfrac{1}{n+2}&\text{$n$  even,}
\end{array}
\right.
\end{equation}
with values on the regular grids 
\begin{eqnarray}\nonumber
\frac{-\left\lfloor n/2\right\rfloor}{\left\lfloor n/2\right\rfloor +1} , \ldots , \frac{-2}{\left\lfloor n/2\right\rfloor +1}, \frac{-1}{\left\lfloor n/2\right\rfloor +1},\!\!&\!\!\!\!\!\!\! 0\!\!\!\!\!\!\!&\!\!,\! \frac{1}{\left\lfloor n/2\right\rfloor +1}, \frac{2}{\left\lfloor n/2\right\rfloor +1},  \ldots , \frac{\left\lfloor n/2\right\rfloor}{\left\lfloor n/2\right\rfloor +1}\vspace{2mm} \\ 
\label{evenoddgrid}\text{($n$ odd), and  \phantom{($n$ even)}\hspace{24mm}} \!\!&\!\!\!\!\!\!\! \!\!\!\!\!\!\!&\!\!\!   \\ %\end{equation} 
%($n$ odd), and 
%\begin{eqnarray}\label{evenoddgrid}
\nonumber
\frac{-\left\lfloor n/2\right\rfloor}{\left\lfloor n/2\right\rfloor +1} , \ldots , \frac{-2}{\left\lfloor n/2\right\rfloor +1}, \frac{-1}{\left\lfloor n/2\right\rfloor +1}, \!\!\!\!&&\!\!\!\!\frac{1}{\left\lfloor n/2\right\rfloor +1}, \frac{2}{\left\lfloor n/2\right\rfloor +1}, \ldots ,  \frac{\left\lfloor n/2\right\rfloor}{\left\lfloor n/2\right\rfloor +1} 
\end{eqnarray} 
 ($n$ even), respectively.  Those grids are the intersection of the two unit vectors~${\bf u}=\pm{\bf 1}$ and the collection of  $\left\lfloor n/2\right\rfloor$ ``circles" with  center at the origin and radii ${R}/{(\left\lfloor n/2\right\rfloor +1)}$, $R=1,\ldots, \left\lfloor n/2\right\rfloor$---along  ($n$  odd) with the origin itself.

Under the assumptions made, each sign ${\bf S}\n_{{\pms} ;i}\vspace{-1mm}$ is uniform over the unit sphere~${\cal S}_0$, and independent of the ranks $R\n_{{\pms} ;i}$; each rank is uniformly distributed  over the integers $(0, 1,2,\ldots , \lfloor n/2\rfloor )$ or $(1,2,\ldots ,\lfloor n/2\rfloor =n/2)$ according as $n$  is odd or even; 
% ($n$  odd), the integers $(1,2,\ldots ,\lfloor n/2\rfloor =n/2)$ ($n$  even), while
  the $n$-tuple~$\big({\bf F}\n _{\pms}(Z\n_1),\ldots ,{\bf F}\n _{\pms}(Z\n_n)\!\big)$ is uniform over the $n!$ permutations of the grids \eqref{evenoddgrid}.  
  
\begin{figure}[htbp]
%\begin{center}
$\hspace{-72mm}$\includegraphics
[width=60mm, height=77mm]
{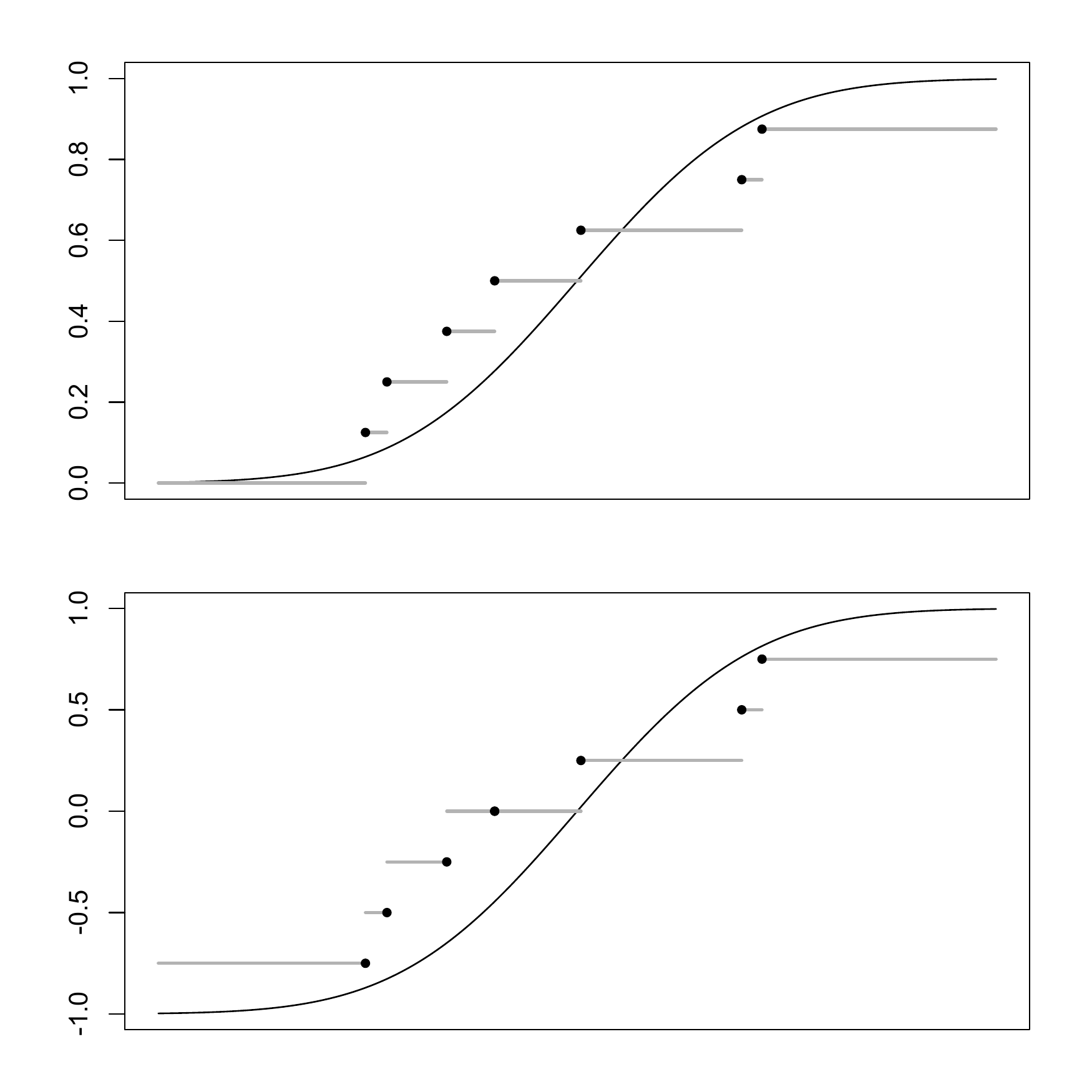}\vspace{-74.7mm}

$\hspace{56mm}$\framebox{{\includegraphics*[angle= -89.5, scale=.315, trim= 0 113 10 110, clip]{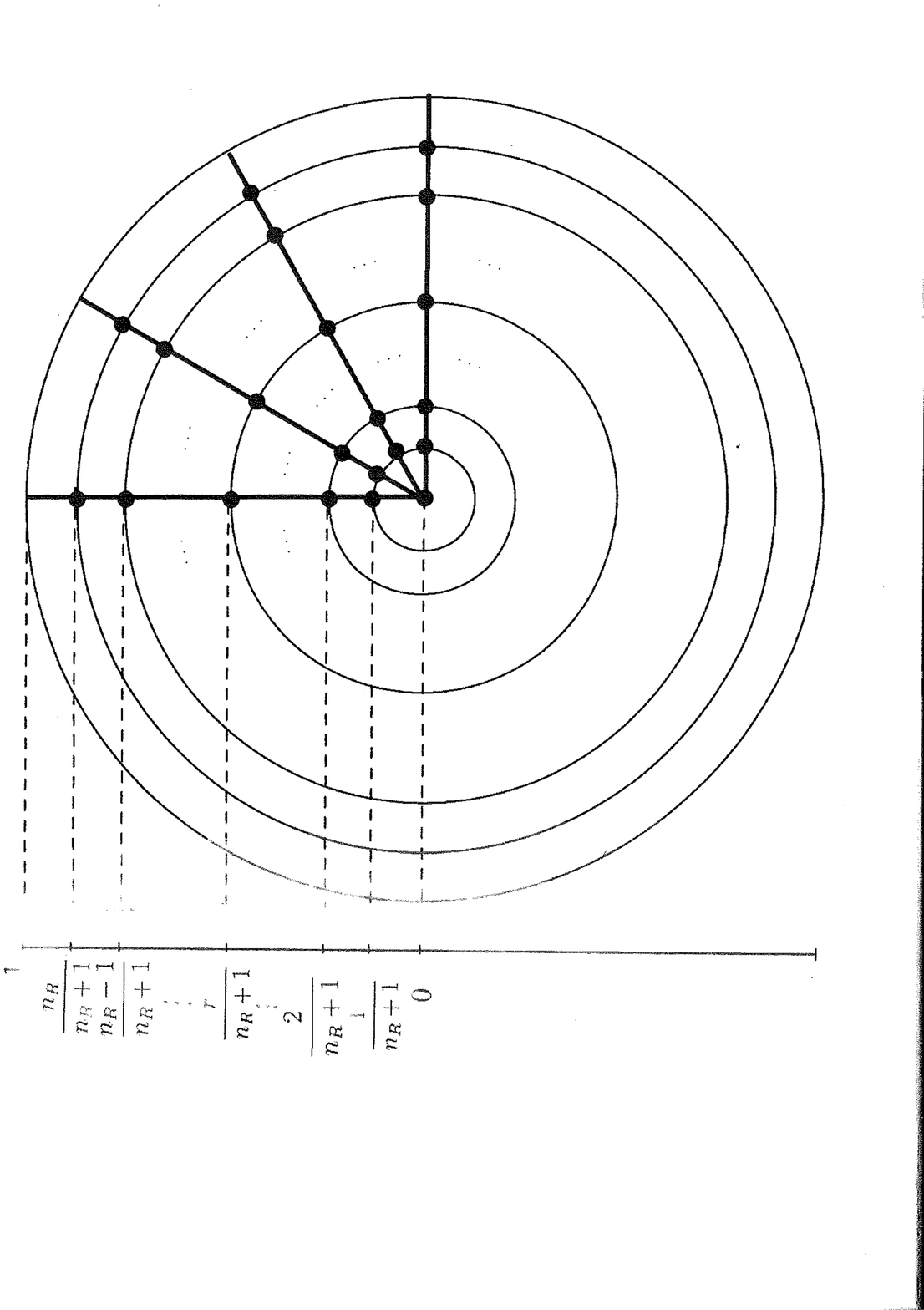}}}
%\end{center}
\begin{caption}{A~classical distribution function $F$  and its empirical counterpart $F\n$ for~$n=7$ (top left panel), along with (bottom left panel) their center-outward versions ${\bf F}_{{\!\pms}}$ and  ${\bf F}\n_{{\!\pms}}$,  the latter with  left-continuous piecewise  constant  interpolation on the left-hand side of the (empirical) median,  right-continuous piecewise constant  interpolation on  the right-hand side of the median; a regular grid of~$n=n_Rn_S$ points over $\mathbb{S}_2$ (right panel). }\label{Ffig}
\end{caption}
\end{figure}

  In view of \eqref{Fnpm} and \eqref{evenoddgrid}, the   Glivenko-Cantelli result~\eqref{empirical} for $F\n$ straightforwardly extends to ${\bf F}\n _{\pms}$:
\begin{equation}\label{empiricalpm}
\max_{1\leq i\leq n}\Big\Vert   {\bf F}\n _{\pms}(Z\n_i) - {\bf F} _{\!{\pms}}(Z\n_i) \Big\Vert  \longrightarrow 0\quad \text{a.s.\  as } n\to\infty % \quad .
\end{equation}
If ${\bf F}\n _{\pms}$ is to be defined over the whole real line, any nondecreasing interpolation~$\overline{\bf F}\n _{\pms}$ of the~$n$ couples~$(Z\n_i, {\bf F}\n _{\pms}(Z\n_i))$ provides a solution, %. Clearly, infinitely many choices are possible and
 all of them yielding  Glivenko-Cantelli statements under sup$_{z\in\mathbb{R}}$ or  sup$_{z\in\text{spt}({\rm P})}$ form (similar to \eqref{GC1} and  \eqref{GC1'}). 
%Some are continuously differentiable, some are simply continuous (e.g., a linear interpolation), some are discontinuous (e.g., step functions).
 Among them is the continuous-from-the-left on the left-hand side of the (empirical) median, and continuous-from-the-right  on the right-hand side of the median piecewise constant interpolation shown in Figure~\ref{Ffig}.% (bottom left).
 
    Clearly, the traditional ranks $R\n_i$ and the empirical 
    center-outward values~${\bf F}\n_{\!{\pms}}(Z\n_i)$,~$i=1,\ldots ,n$, generate the same $\sigma$-field and both enjoy (DF$^+$):    all classical rank statistics thus  can be rewritten in terms of    ${\bf F}\n_{\!{\pms}}$. %(Z\n_i)$'s. 
%     center-outward ranks and signs.
      Traditional ranks  and center-outward ranks and signs, %(the ${\bf F}\n_{\!{\pms}}(Z\n_i)$'s), % complemented with the signs,
         therefore, are strictly equivalent statistics. 
         
          \subsection{Relation to measure transportation}\label{meastranspsec}
 
 The probability-integral transformation $z\mapsto {\bf F}_{\!{\pms}}(z)$  from $\mathbb{R}$ to the unit ball~$\mathbb{S}_1$ (the interval $(-1, 1)$) is pushing %the distribution
  ${\rm P}_{\!f}\in{\cal P}_1$ forward to the uniform distribution~${\rm U}_1$ over $\mathbb{S}_1$.  As a continuous monotone non-decreasing function, it is the  gradient (here, the derivative) of a convex function $\psi_{\!f}$, say, which,  therefore, is everywhere continuously differentiable. Actually, it is the unique monotone function pushing~$\rm P$ forward to ${\rm U}_d$. It follows from McCann's theorem (see Section~\ref{transpsec1}) that~$\nabla\psi_{\!f}$ coincides, ${\rm P}_{\!f}$-almost surely---hence, over spt$({\rm P}_{\!f})$---with any monotone nondecreasing function (any gradient of a convex function)  %s of convex function
   $\nabla\psi$ pushing~${\rm P}_{\!f}$ forward to ${\rm U}_1$.   It follows that $\psi_{\!f} -\psi$ is a constant, hence that $\nabla\psi_{\!f} =\nabla \psi$ everywhere. It follows that such a gradient is uniquely determined on spt$({\rm P}_{\!f})$, and that    ${\bf F}_{\!{\pms}}$ on spt$({\rm P}_{\!f})$    can be characterized as the  unique gradient of a convex function 
     pushing~${\rm P}_{\!f}$ forward to~${\rm U}_1$. The (noninformative) values of ${\bf F}_{\!{\pms}}$ outside~spt$({\rm P}_{\!f})$ then are easily obtained by imposing monotonicity and  range $[0,1]$.
     
     The huge advantage of this characterization is that it does not involve the canonical ordering of~$\mathbb{R}$, hence   readily extends  to dimension $d\geq 2$. The extension, actually, would be entirely straightforward for distributions ${\rm P}_{\!f}$ with nonvanishing densities $f$ (hence support $\mathbb{R}^d$). More general cases require some additional care with ${\rm P}_{\!f}$-a.s.\ uniqueness, though---while the support of ${\rm P}_{\!f}\in{\cal P}_1$ consists at most of a countable collection of intervals, the support of of ${\rm P}_{\!f}\in{\cal P}_d$ is potentially  much weirder.  Everywhere continuous differentiability of the potential $\psi$, in particular, will not survive in higher dimension.

   \section{Center-outward and Mahalanobis ranks and signs }\label{ellsec}
Recall that  a $d$-dimensional random vector ${\bf X}$ has elliptical distribution~${\rm P}_{{\boldsymbol\mu},\boldsymbol{\Sigma},f}$ with location~${\boldsymbol\mu}\in\R^d$, positive definite symmetric $d\times d$ scatter matrix ${\boldsymbol\Sigma}$ and radial density $f$ iff ${\bf Z}:={\boldsymbol\Sigma}^{-1/2}({\bf X}-{\boldsymbol\mu})$ has spherical distribution ${\rm P}_{{\boldsymbol 0},\boldsymbol{I},f}$, which holds iff 
%\begin{equation}
%\label
${\bf F}_{\scriptscriptstyle\text{ell}}({\bf Z}):= {\bf Z}F\big(\Vert {\bf Z}\Vert\big)/{\Vert {\bf Z}\Vert }
\sim {\rm U}_d,
$ %\end{equation}
where $F$, with density $f$, is the distribution function of $\Vert {\bf Z}\Vert$ (the {\it radial distribution} and {\it radial density}). Elliptical distributions with nonvanishing radial densities clearly belong to the class ${\cal P}_d^{\text{conv}}$, with support  $\mathbb{R}^d$. 

The mapping ${\bf Z}\mapsto {\bf F}_{\scriptscriptstyle\text{ell}}({\bf Z})$ is   a probability-integral transformation.    Chernozhukov et al.~(2017) show (Section~2.4)  that it actually coincides with~${\bf Z}$'s center-outward distribution function~${\bf F}_{\pms}$. Letting ${\bf X}\n_i\!$, $i=1,\ldots ,n$ be   i.i.d.~with elliptical distribution~${\rm P}_{{\boldsymbol\mu},\boldsymbol{\Sigma},f}$, denote by $\hat{\boldsymbol\mu}\n\!$ and $\hat{\boldsymbol\Sigma}^{(n)}\!$  
 consistent estimators of~${\boldsymbol\mu}$ and~${\boldsymbol\Sigma}$, respectively: the empirical version of ${\bf F}_{\scriptscriptstyle\text{ell}}$, based on Mahalanobis ranks and signs   (the ranks $R\n_i$  of the   residual moduli $\Vert {\bf Z}\n_i\Vert\!\!:=\Vert\hat{\boldsymbol\Sigma}^{(n)-1/2}({\bf X}\n_i\!~\!-~\!\hat{\boldsymbol\mu}\n)\Vert$ and the corresponding
 unit vectors ${\bf Z}\n_i/\Vert {\bf Z}\n_i\Vert$)
% ${\bf U}\n_i\!\!:=\hat{\boldsymbol\Sigma}^{(n)-1/2}({\bf X}\n_i-\hat{\boldsymbol\mu}\n)/\Vert\hat{\boldsymbol\Sigma}^{(n)-1/2}({\bf X}_i-\hat{\boldsymbol\mu}\n)\Vert$
  is, for the $i$th  observa-\linebreak tion,~$ {\bf F}_{\scriptscriptstyle\text{ell}}\n({\bf Z}\n_i):= \big({R\n_i}/{(n+1)}\big){\bf U}\n_i
$.\vspace{-1mm}
  \begin{Prop}\label{GCell} Let   ${\bf X}\n_i$, $i=1,\ldots ,n$ be  i.i.d.~with elliptical distribution~${\rm P}_{{\boldsymbol\mu},\boldsymbol{\Sigma},f}$, and assume that   $\hat{\boldsymbol\mu}\n$ and $\hat{\boldsymbol\Sigma}^{(n)}$ are strongly consistent  estimators of~${\boldsymbol\mu}$ and~${\boldsymbol\Sigma}$, respectively. Then,~${\bf F}_{\scriptscriptstyle\text{ell}} $ and~${\bf F}_{\pms}$ coincide, and\vspace{-1mm} 
$$   \max_{1\leq i\leq n}\Vert {\bf F}_{\scriptscriptstyle\text{ell}}\n({\bf Z}\n_i) - {\bf F}_{\pms}\n({\bf Z}\n_i)
 \Vert , \quad\text{hence also}\quad     \max_{1\leq i\leq n}\Vert 
 {\bf F}_{\scriptscriptstyle\text{ell}}\n({\bf Z}\n_i) - {\bf F}_{\pms}({\bf Z}\n_i)
  \Vert\vspace{-1mm}$$
%$$(iii)\quad   \max_{1\leq i\leq n}\Vert \Vert$$
tend to zero a.s., as $n\to\infty$, where % ${\bf F}_{\scriptscriptstyle\text{ell}}$ is given in \eqref{eq:ell} and
  ${\bf F}_{\pms}$ denotes the  center-outward distribution function of ${\rm P}_{{\boldsymbol 0},\boldsymbol{I},f}$.\vspace{-1mm}
    \end{Prop}\label{ellprop}

    This result connects the center-outward ranks and signs with the well-studied   elliptical ones.  The consistency of $ {\bf F}_{\scriptscriptstyle\text{ell}}\n$, however, requires %the assumption of
     ellipticity, whereas~${\bf F}_{\pms}\n$ remains consistent under any   ${\rm P}\in{\mathcal P}_d^{\pms}$. Note also that $ {\bf F}_{\scriptscriptstyle\text{ell}}\n$ determines~$n$ ellipsoidal contours,
%      ($n$ distinct values for the $ {\bf F}_{\scriptscriptstyle\text{ell}}\n({\bf Z}\n_i)$'s),
       while ${\bf F}_{\pms}\n$ only determines~$n_R$ of them (which, moreover,  for finite $n$ do not define an ellipsoid). \vspace{-1mm}

 \section{Proofs for Section~\ref{MKsec}% and~\ref{sec3}
 }\label{Proofsec2}
  \subsection{Proofs of Propositions~\ref{Fpmprop} and \ref{Qpmprop}%, and~\ref{Figalliprop}
  }
  
     {\color{black} \noindent \textbf{Proof of Proposition~\ref{Fpmprop}}. 
     Part (i) is satisfied by construction  (see the conclusion following~\eqref{inverse}). For Part (ii), since ${\bf F}_\pms\#{\rm P}={\rm U}_d$, the joint distribution of $\Vert{\bf F}_\pms({\bf Z}) \Vert$ and ${\bf S}({\bf Z})={\bf F}_\pms({\bf Z})/\Vert{\bf F}_\pms({\bf Z}) \Vert$ are those of $\Vert{\bf U} \Vert$ and ${\bf U}/\Vert{\bf U} \Vert$, where~${\bf U}\sim{\rm U}_d$; the claim follows.   Turning to Part~(iii), for any Borel set~$C$ of~$\mathbb{R}^d$, we have ${\rm P}\big(C  \big) = {\rm P}\big(C\cap\text{spt}({\rm P})
     \big)$. Now, the fact that the restrictions of~${\bf F}_\pms$ and ${\bf Q}_\pms$   to~\text{spt}$({\rm P})$ and~$\mathbb{S}_d$, respectively, are the inverse of each other, pushing $\rm P$ forward to ${\rm U}_d$ and ${\rm U}_d$ back to $\rm P$, entails 
     \begin{align*} {\rm P}\big(C\cap\text{spt}({\rm P})
     \big) &=  {\rm P}\big({\bf Q}_\pms\circ{\bf F}_\pms (C\cap\text{spt}({\rm P}))
     \big)={\rm U}_d\big({\bf F}_\pms(C\cap\text{spt}({\rm P}))
     \big)\\
     &={\rm U}_d\big({\bf F}_\pms(C)\cap{\bf F}_\pms(\text{spt}({\rm P})) 
     = {\rm U}_d\big({\bf F}_\pms(C)\cap{\mathbb S}_d);
     \end{align*}  
     the claim follows. Finally, Part~(iv) readily follows from the fact that, in dimension $d=1$, $2F-1$ is the only monotone  mapping from   $\mathbb{R}$ to $\mathbb{S}_1=(-1,1)$ pushing ${\rm P}\in{\mathcal P}_1$ forward to ${\rm U}_1={\rm U}_{(-1,1)}$: See Appendix~\ref{pmranksec}.\hfill $\Box$
}
     \medskip
     
     {\color{black} \noindent \textbf{Proof of Proposition~\ref{Qpmprop}}. 
     Parts (i) and (ii) are direct consequences of the definition of ${\bf Q}_\pms$. As for Part~(iii), it follows Proposition~\ref{Fpmprop}(iv) by adapting the traditional definition of a quantile function as a general inverse. \hfill $\Box$

%          \medskip
  \subsection{Proofs of Proposition~\ref{Figalliprop},  Proposition~\ref{DFProp}, and Corollary~\ref{gridcor}}\label{DFPropsec}

    \color{black} \noindent \textbf{Proof of Proposition~\ref{Figalliprop}}. Parts (i), (ii) and (iii) of the proposition are proved in del Barrio et al.~(2019). Hence, we only have to prove the claims about~$\mathbf{F}_\pm(\mathbf{x})$ for $\mathbf{x}\notin{\text{\rm spt}}({\rm P})$. Since $\phi$ is a finite convex function on $\mathbb{R}^d$,  it has a nonempty subgradient at every point. Let $\mathbf{x}\notin{\text{\rm spt}}({\rm P})$ and consider $\mathbf{u}\in\partial \phi(\mathbf{x})$. Since $\phi$ is~1-Lipschitz, we have $\|\mathbf{u}\|\leq 1$. We claim that $\|\mathbf{u}\|=1$. To show 
  this, assume that, on the contrary,   $\|\mathbf{u}\|<1$. Then, from part (i) of the proposition, we have that $\mathbf{u}=\nabla \phi(\mathbf{x}_0)$ for some $\mathbf{x}_0\in{\text{\rm spt}}({\rm P})$. But this means that both $\mathbf{x}$ and $\mathbf{x}_0$ are in $\partial\phi^*(\mathbf{u})$. Then, by convexity, $(1-t)\mathbf{x}+
t \mathbf{x}_0\in  \partial\phi^*(\mathbf{u})$ for every $t\in [0,1]$. Equivalently, %this means that 
 $\mathbf{u}\in\partial\phi((1-t)\mathbf{x}+
t \mathbf{x}_0)$ for every $t\in [0,1]$. Since ${\text{\rm spt}}({\rm P})$ is open, this means that different points in ${\text{\rm spt}}({\rm P})$ are mapped through $\nabla \phi$ to $\mathbf{u}$, contradicting the injectivity of $\nabla \phi$ in ${\text{\rm spt}}({\rm P})\!\setminus\! K$. We conclude that, necessarily, $\|\mathbf{u}\|=1$. 
  
Next, let us assume that $\mathbf{x}\notin{\text{\rm spt}}({\rm P})$ is such that $\mathbf{u_1}\neq\mathbf{u_2}\in\partial \phi(\mathbf{x})$. Then,  for every $t\in[0,1]$, $(1-t)\mathbf{u}_1+t\mathbf{u}_2\in \partial \phi(\mathbf{x})$. But $\|(1-t)\mathbf{u}_1+t\mathbf{u}_2\|<1$ for~$t\in~\!\!(0,1)$ unless
$\mathbf{u}_1=\mathbf{u}_2$. This proves that $\partial \phi(\mathbf{x})$ is a singleton, hence that $\phi$ is differentiable at $\mathbf{x}$. The fact that the gradient of a convex function is continuous in the differentiability set completes the proof. \hfill $\Box$

    \medskip

\noindent \textbf{Proof of Proposition~\ref{DFProp}}. Part\,{\it (i)}. Sufficiency of ${\bf Z}\n_{{\scriptscriptstyle{(\, .\, )}}}\vspace{-0.5mm}$---equivalently, sufficiency of  the sub-$\sigma$-field ${\mathcal B}\n_{{\scriptscriptstyle{(\, .\, )}}}$ of permutationally invariant\footnote{Permutation here means permutation among the $n$ $d$-dimensional subspaces of ${\mathcal B}^n_d$.} events of ${\mathcal B}^n_d$---follows from a trivial application of the classical Fisher-Neyman factorization criterion for dominated families (Corollary 2.6.1 in Lehmann and Romano (2005)). Completeness  is established  (under the name of {\it symmetric completeness}) in Lemma~3 of Bell et al.\ (1960)\footnote{That lemma establishes completeness of the order statistic for   nonparametric families of the form $\{{\rm P}^n\vert {\rm P}\ll {\rm P}_1 \}$ where ${\rm P}_1$ is non-atomic. In that notation, ${\cal P}_d= \{{\rm P}^n\vert {\rm P}\ll {\cal N}({\bf 0}, {\bf I}) \}$ where ${\cal N}({\bf 0}, {\bf I})$ indeed  is non-atomic; the result thus applies to ${\cal P}\n_d\!$.} and minimal sufficiency follows from the fact (see, e.g., Proposition~1.4.8 in Pfanzagl~(2011)) that  complete sufficient $\sigma$-fields are automatically minimal sufficient.

Part~{\it (ii)}. %Let us start with the distribution of the $n$-tuple~${\bf F}\n_\pms({\bf Z}\n)$. 
 Assume~$n_0=0$ or 1.    Conditionally on ${\bf Z}\n_{{\scriptscriptstyle{(\, .\, )}}}\vspace{-0mm}$,~${\bf F}\n_\pms({\bf Z}\n)$ takes values in the set of the $n!$ permutations of the 
 $n$ gridpoints. Because of the permutational symmetry of the~${\bf Z}\n\!$ likelihood, all those values are equally likely, hence have conditional probability  $1/n!\vspace{-0.5mm}$. Since that (uniform) conditional distribution does not depend on~${\bf Z}\n_{{\scriptscriptstyle{(\, .\, )}}}\vspace{-0mm}$, it is also unconditional. If $n_0>1$, the situation is exactly the same, except that the $n_0!$  permutations involving the $n_0$ copies of the origin are undistinguishable, so that the $n!$ permutations of the grid reduce to~$n!/n_0!$ permutations with repetitions.  This, however, can be avoided by breaking the~$n_0$ ties at the origin.\footnote{For instance, one may replace the $n_0$ copies of $\bf 0$ with an i.i.d.\ $n$-tuple of gridpoints (distinct with probability one) simulated from a uniform over~$[1/2(n_R+1)]{\mathbb{S}}_d$ or a uniform over $[1/2(n_R+1)]{\mathcal{S}}_{d-1}$. Uniformity (conditional on the simulation results) over the~$n!$ permutations of the resulting~$n$ gridpoints, hence distribution-freeness,  is recovered. Some of the resulting ranks, however, are losing their nature as integers---much in the same way as the {\it midranks} resulting from traditional univariate tie-breaking (see, e.g., Section III.8 in H\' ajek and \v Sid\' ak~(1967)).}

 Part~{\it (iii)}:  Assume $n_0=0$. The result  readily follows from the uniformity of~${\bf F}\n_\pms({\bf Z}\n)$ over the $n!$ permutations of the  gridpoints,  which are indexed by a product of the set of~$n_R$ integers $\{1,\ldots n_R \}$ and the set of $n_S$ unit vectors. It no longer holds for $n_0\geq 1$, even after performing the tie-break prodedure just described: indeed, $0\leq R\n_{\pms,i}<1$ tells something about ${\bf S}\n_{\pms,i}$. However, the proportion $n_0/n$ of ties tends to zero as $n\to\infty$; moreover, the non-independence between  ranks and (multivariate) signs has no decision-theoretic consequences as long as joint distribution-freeness   holds. 
 
  Part~{\it (iv)}, in view of  {\it (i)} and {\it (ii)}, is an immediate consequence  of the classical Basu theorem---Basu's Second Theorem in  Appendix~\ref{Basuapp} below.  
   
    Turning to Part~{\it (v)}, either assume that   $n_0\leq 1$ or, for $n_0>1$, assume that  the previously described tie-breaking grid randomization device has been performed, so that the grid does not exhibit any multiplicity at the origin. Then, the mapping~${\bf z}\mapsto \big({\bf z}_{{\scriptscriptstyle{(\, .\, )}}},{\bf F}\n_\pms({\bf z})\big)$  is injective for ${\bf z}\in\mathbb{R}^{nd}\!\setminus\! N$.  Distribution-freeness  (Part {\it (ii)} of the proposition) entails the ancillarity of the $\sigma$-field ${\mathcal B}\n_\pms$ generated by ${\bf F}\n_\pms({\bf Z}\n)$. In view of Corollary~\ref{Basu3strong}, the completeness of the sufficient \linebreak $\sigma$-field~${\mathcal B}\n_{{\scriptscriptstyle{(\, .\, )}}}$ generated by the order statistic~${\bf Z}\n_{{\scriptscriptstyle{(\, .\, )}}}\vspace{-0mm}$, and the ancillarity of   ${\mathcal B}\n_\pms\!$, we only have to show that  the $\sigma$-field $\sigma\big({\bf Z}\n_{{\scriptscriptstyle{(\, .\, )}}}, {\bf F}\n_\pms({\bf Z}\n)
\big)$ is strongly ${\mathcal P}\n_d$-essentially equivalent to the Borel $\sigma$-field ${\mathcal B}^{nd}$.  This readily follows, however, from the injectivity, over $\mathbb{R}^{nd}\!\setminus\! N$, of ${\bf z}\mapsto \big({\bf z}_{{\scriptscriptstyle{(\, .\, )}}},{\bf F}\n_\pms({\bf z})\big)$. The claim follows. 
%The same injectivity also entails that   $\big({\bf Z}\n_{{\scriptscriptstyle{(\, .\, )}}},{\bf F}\n_\pms({\bf Z}\n)\big)$ and ${\bf Z}\n$ are strongly ${\mathcal P}\n_d\!$-essentially equivalent. The Basu factorization property thus holds in view of     {\it (i)},  {\it (ii)}, and {\it (iv)}.
 \hfill $\Box$
    
     \medskip

 \noindent \textbf{Proof of Corollary~\ref{gridcor}}. It follows from the injectivity of the restriction\linebreak to $\mathbb{R}^{nd}\!\setminus\!\!N$ of ${\bf z}\mapsto\big({\bf z}_{{\scriptscriptstyle{(\, .\, )}}}, {\bf F}\n_\pms({\bf z}) \big)$
%    Irrespective of the choice of a grid, we have
     that the Borel $\sigma$-field ${\mathcal B}^{nd}$ is strongly~${\mathcal P}\n_d\!$-essentially equivalent (for some null set $N$) to $\sigma\big({\bf Z}\n_{{\scriptscriptstyle{(\, .\, )}}}, {\bf F}\n_\pms({\bf Z}\n) \big)$. For the same reason, ${\mathcal B}^{nd}$ is strongly~${\mathcal P}\n_d\!$-essentially equivalent (for some null set $\tilde{N}$) to~$\sigma\big({\bf Z}\n_{{\scriptscriptstyle{(\, .\, )}}}, \tilde{\bf F}\n_\pms({\bf Z}\n) \big)$ . These strong essential equivalences still hold true with~$N$ and $\tilde{N}$ replaced with~$M:=N\cup\tilde{N}$. It follows that, for ${\bf z}\in \mathbb{R}^{nd}\!\setminus\!M$,  a bijection exists between $\big({\bf z}_{{\scriptscriptstyle{(\, .\, )}}}, {\bf F}\n_\pms({\bf z}) \big)$ and $\big({\bf z}_{{\scriptscriptstyle{(\, .\, )}}}, \tilde{\bf F}\n_\pms({\bf z}) \big)$, hence between the~permutations~${\bf F}\n_\pms({\bf z})$ and $\tilde{\bf F}\n_\pms({\bf z})$ of the two $n$-points grids. The result follows.\hfill$\Box$
%    \color{blue} \noindent \textbf{Proof of Corollary~\ref{gridcor}}. Irrespective of the choice of a grid, we have that~${\bf Z}\n$ is~${\mathcal P}\n_d\!$-essentially equivalent to $\big({\bf Z}\n_{{\scriptscriptstyle{(\, .\, )}}}, {\bf F}\n_\pms({\bf Z}\n) \big)$ (for some null set $N$) and~${\mathcal P}\n_d\!$-essentially equivalent to $\big({\bf Z}\n_{{\scriptscriptstyle{(\, .\, )}}}, \tilde{\bf F}\n_\pms({\bf Z}\n) \big)$ (for some null set $\tilde{N}$). This still holds true with $N$ and $\tilde{N}$ replaced with $M=N\cup\tilde{N}$. It follows that, for ${\bf z}\in \mathbb{R}^{nd}\!\setminus\!M$, \linebreak a bijection exists between $\big({\bf z}_{{\scriptscriptstyle{(\, .\, )}}}, {\bf F}\n_\pms({\bf z}) \big)$ and $\big({\bf z}_{{\scriptscriptstyle{(\, .\, )}}}, \tilde{\bf F}\n_\pms({\bf z}) \big)$, hence between the~permutations~${\bf F}\n_\pms({\bf z})$ and $\tilde{\bf F}\n_\pms({\bf z})$ of the two $n$-points grids. The result follows.\hfill$\Box$

  \section{Minimal sufficiency and maximal ancillarity} }\label{Basuapp}
%  \subsection{Basu's theorems}
  This appendix collects, for ease of reference,  some classical and less classical definitions and results about sufficiency and ancillarity which are scattered  across Basu's papers; some of them (such as the concept of {\it strong essential equivalence})  are slightly modified to adapt our needs.   

The celebrated result commonly known as    Basu's Theorem was first established as Theorem~2 in Basu~(1955).    The same paper also contains a Theorem~1, of which Theorem~2 can be considered a partial converse. Call them    Basu's First and Second Theorems, respectively.

  \begin{prop}[Basu's First Theorem]\label{Basu1}  Let $S$ be   sufficient for a family~$\cal P$  of distributions over some abstract space $({\mathcal X},\mathcal A)$. Then, if a statistic $W$ is $\rm P$-independent of $S$ for all ${\rm P}\in{\mathcal P}$, it is    distribution-free over~${\mathcal P}$.
  \end{prop}

  \begin{prop}[Basu's Second Theorem]\label{Basu2}  Let $T$ be (boundedly) complete and sufficient for a family~$\cal P$  of distributions over some abstract space $({\mathcal X},\mathcal A)$. Then, if a statistic $W$ is distribution-free over~${\mathcal P}$, it is $\rm P$-independent of $T$ for all ${\rm P}\in{\mathcal P}$. 
    \end{prop}

Basu's original proof of Proposition~\ref{Basu1} was flawed, however, and Basu's First Theorem does not hold with full generality.   Basu~(1958) realized that problem and fixed it by imposing on $\mathcal P$ a sufficient additional  condition of {\it connectedness}. Some twenty years later, that condition has been replaced (Koehn and Thomas~1975) with a considerably weaker necessary and sufficient one (same notation as in Proposition~\ref{Basu1}). 

   \begin{prop}[Koehn and Thomas 1975]\label{Koehn} Basu's First Theorem holds true if and only if ${\mathcal P}$ does not admit a measurable {\rm splitting set}, namely, a set~$A\in~\!{\mathcal A}$ along with a partition   ${\mathcal P}= {\mathcal P}_0\oplus {\mathcal P}_1$ of ${\mathcal P}$ into two nonempty subsets~${\mathcal P}_0$ and~${\mathcal P}_1$ such that ${\rm P}(A)=0$ for all ${\rm P}\in {\mathcal P}_0$ and ${\rm P}(A)=1$ for all ${\rm P}\in {\mathcal P}_0$. 
   \end{prop}
   
   Recall that a sub-$\sigma$-field ${\mathcal A}_{0}$ of $\mathcal A$ such that ${\rm P}_1(A)= {\rm P}_2(A)$ for all $A\in{\mathcal A}_0$ and all ${\rm P}_1, {\rm P}_2$ in $\mathcal P$ is called {\it ancillary}. Clearly, the $\sigma$-field ${\mathcal A}_{V}$ generated by a distribution-free statistic $V$ is ancillary. Contrary to sufficient $\sigma$-fields (the smaller, the better), it is desirable for ancillary $\sigma$-field to be a large as possible. While minimal sufficient $\sigma$-fields, when they exist, are unique,  maximal ancillary $\sigma$-fields typically exist, but are neither unique nor easily characterized---due, mainly, to null-sets issues. 
   
   Basu~(1959) therefore introduced the notions of {$\mathcal P$-essentially  equivalent}\linebreak and~{$\mathcal P$-essentially  maximal} sub-$\sigma$-fields.
   
   \begin{defin}\label{esseqdef} Two sub-$\sigma$-fields ${\mathcal A}_1$ and ${\mathcal A}_2$ of ${\mathcal A}$ are said to be $\mathcal P$-{\em essentially  equi\-valent} if, for any $A_1\!\in\!{\mathcal A}_1$, there exists an $A_2\!\in\!{\mathcal A}_2$ and,  for any $A_3~\!\!\in~\!\!{\mathcal A}_2$, an  $A_4\!\in\!{\mathcal A}_1$ such that ${\rm P}(A_1\Delta A_2) = 0={\rm P}(A_3\Delta A_4) $ for any ${\rm P}\in{\mathcal P}$.   An ancillary sub-$\sigma$-field essentially  equivalent to a maximal ancillary sub-$\sigma$-field is called {\em essentially  maximal}. 
      \end{defin}
      
      The same reference then establishes the following sufficient condition for an ancillary statistic to be essentially  maximal.
      
    \begin{prop}[Basu's Third Theorem]\label{Basu3}  Denote by ${\mathcal A}_{\text{\rm suff}}$   a  (boundedly) complete and sufficient (for a family~$\cal P$  of distributions over   $({\mathcal X},\mathcal A)$)  sub-$\sigma$-field of $\mathcal A$.  Then, any ancillary   sub-$\sigma$-field ${\mathcal A}_{\text{\rm anc}}$ such that $\sigma\big({\mathcal A}_{\text{\rm suff}}\cup {\mathcal A}_{\text{\rm anc}}\big)=\mathcal A$  is essentially maximal ancillary. 
      \end{prop}
      
      Let us slightly reinforce Definition~\ref{esseqdef} and the concepts of {essentially  equivalent} and {essentially  maximal} sub-$\sigma$-fields.
        \begin{defin}\label{esseqdefstrong} Two sub-$\sigma$-fields ${\mathcal A}_1$ and ${\mathcal A}_2$ of ${\mathcal A}$ are said to be {\em strongly} $\mathcal P$-{\em essentially  equi\-valent} if there exists $N\in{\cal A}$ such that ${\rm P}(N)=0$ for all ${\rm P}\in{\mathcal P}$ and~${\mathcal A}_1\cap({\mathcal X}\!\setminus\!N)={\mathcal A}_1\cap({\mathcal X}\!\setminus\!N)$. 
          An ancillary sub-$\sigma$-field strongly $\mathcal P$-essentially  equivalent to a maximal ancillary sub-$\sigma$-field is called {\em strongly $\mathcal P$-essentially  maximal}. 
      \end{defin}
      
Clearly, strong essential equivalence and maximal ancillarity imply essential equivalence and maximal ancillarity, respectively. The following slightly modified version of Basu's Third Theorem then readily follows.

    \begin{cor}\label{Basu3strong}  Denote by ${\mathcal A}_{\text{\rm suff}}$   a  (boundedly) complete and sufficient, for a family~$\cal P$  of distributions over   $({\mathcal X},\mathcal A)$,  sub-$\sigma$-field of $\mathcal A$.  Then, any ancillary   sub-$\sigma$-field ${\mathcal A}_{\text{\rm anc}}$ such that $\sigma\big({\mathcal A}_{\text{\rm suff}}\cup {\mathcal A}_{\text{\rm anc}}\big)$ is strongly $\mathcal P$-essentially  equivalent to~$\mathcal A$  is strongly $\mathcal P$-essentially maximal ancillary. 
      \end{cor}

       \color{black}
       
        \section{Proofs for Section~\ref{sec3}
 }\label{Proofsec3}

 \subsection{Proof of Proposition~\ref{step1}}\label{Proofsec31} Duality  yields, for the linear program~\eqref{assumption3}, 
\begin{equation}\label{dual0}
\begin{aligned}
\min_{\pi}   & \sum_{i=1}^n \sum_{j=1}^n c_{i,j}\pi_{i,j} &=  & & \max_{a,b}   & \frac 1 n \sum_{i=1}^n a_i + \frac 1 n \sum_{j=1}^n b_j     \\
\mbox{s.t. } & \sum_{i=1}^n \pi_{i,j}=\sum_{j=1}^n \pi_{i,j}=\frac 1 n, & &&  \mbox{s.t. } & a_i+b_j\leq c_{i,j},\, i,j=1,\ldots,n.\\
& \pi_{i,j}\geq 0,\  i,j=1,\ldots,n & & & &
 \end{aligned}
 \end{equation}
Moreover,  $\pi=\{\pi_{i,j}\vert  \, i,j=1,\ldots,n\}$ is a minimizer for the left-hand side program, and~$(a,b)=(a_1,\ldots ,a_n,b_1,\ldots ,b_n)$ a maximizer for the right-hand side  one, if and only if they satisfy the corresponding constraints and
$$\sum_{i=1}^n \sum_{j=1}^n c_{i,j}\pi_{i,j}=\frac 1 n \sum_{i=1}^n a_i + \frac 1 n \sum_{j=1}^n b_j.$$
 With the change of variables
$a_i=:\|{\bf x}_i\|^2-2 \varphi_i$, $b_j=:\|{\bf y}_j\|^2-2\psi_j$, the dual programs~\eqref{dual0} take the form
\begin{equation}
\begin{aligned}\label{dual1}
\max_{\pi}   & \sum_{i=1}^n \sum_{j=1}^n \pi_{i,j} \langle {\bf x}_i, {\bf y}_j \rangle &=  & & \min_{\varphi,\psi}   & \frac 1 n \sum_{i=1}^n \varphi_i + \frac 1 n \sum_{j=1}^n \psi_j     \\
\mbox{s.t. } & \sum_{i=1}^n \pi_{i,j}=\sum_{j=1}^n \pi_{i,j}=\frac 1 n, & &&  \mbox{s.t. } & \varphi_i+\psi_j\geq \langle {\bf x}_i, {\bf y}_j \rangle,\, i,j=1,\ldots,n\\
& \pi_{i,j}\geq 0,\, i,j=1,\ldots,n & & & &
\end{aligned}
\end{equation}
where $\pi$ is a maximizer for the left-hand side program and  $(\varphi,\psi)$ a minimizer for the right-hand side one if  and only if they satisfy the corresponding constraints and
$$\sum_{i=1}^n \sum_{j=1}^n \pi_{i,j}\langle {\bf x}_i, {\bf y}_j \rangle=\frac 1 n \sum_{i=1}^n \varphi_i + \frac 1 n \sum_{j=1}^n \psi_j.$$
Let  
$(\varphi,\psi)$ be a minimizer for the right-hand side program in (\ref{dual1}). Then, replacing~$\varphi_i$ with~$\tilde{\varphi}_i:=\max_{j=1,\ldots,n} (\langle {\bf x}_i, {\bf y}_j \rangle-\psi_j)$ yields  a new feasible solution~$(\tilde{\varphi}, \psi)$ satisfying $\varphi_i\geq \tilde{\varphi}_i$. Optimality of $(\varphi,\psi)$ thus implies that $\varphi_i=\tilde{\varphi_i}$, so that,   at optimality,
\begin{equation}\label{optimality1}
\varphi_i=\max_{j=1,\ldots,n} (\langle {\bf x}_i, {\bf y}_j \rangle-\psi_j),\quad i=1,\ldots,n.
\end{equation}
Now, if~\eqref{assumption3} is minimal, then $\pi_{i,i}=  1/n$, $\pi_{i,j}=0$, $j\ne i$ is the unique maximizer in the left-hand side  linear program in
(\ref{dual1}). Therefore, $(\varphi,\psi)$ is a minimizer for the right-hand side program  if and only if 
$$\frac 1 n \sum_{i=1}^n (\varphi_i+\psi_i-\langle {\bf x}_i, {\bf y}_i\rangle)=0.$$
In view of (\ref{optimality1}) this implies that 
\begin{equation}\label{eqcondition1}
\langle {\bf x}_i, {\bf y}_i \rangle-\psi_i=\max_{j=1,\ldots,n} (\langle {\bf x}_i, {\bf y}_j \rangle-\psi_j),\quad i=1,\ldots,n.
\end{equation}

Conversely, assume that the weights $\psi_1,\ldots,\psi_n$ are such that (\ref{eqcondition1}) holds. Then, letting~$\varphi_i=\max_{j=1,\ldots,n} (\langle {\bf x}_i, {\bf y}_j \rangle-\psi_j)$, we have that
 $(\varphi,\psi)$ is a feasible solution for which
 $$\frac 1 n \sum_{i=1}^n (\varphi_i+\psi_i-\langle {\bf x}_i, {\bf y}_i \rangle)=0,$$ which, in view of the discussion above, implies that the map $$T: {\bf x}_i\mapsto T({\bf x}_i)={\bf y}_i$$
is cyclically monotone. This completes the proof of Part {\it (i)} of the proposition. 

As for Part {\it (ii)}, %note that 
$T$ is the unique cyclically monotone map from~$\{{\bf x}_1,\ldots,{\bf x}_n\}$ to~$\{{\bf y}_1,\ldots,{\bf y}_n\}$
if and only if, for any choice of indices $\{i_0,i_1, \ldots,i_m\}$ in~$\{1,\ldots,n\}$, we have
\begin{equation}\label{keyconditioneq}
\langle {\bf x}_{i_0}, {\bf y}_{i_0}-{\bf y}_{i_1}\rangle+\langle {\bf x}_{i_1}, {\bf y}_{i_0}-{\bf y}_{i_2}\rangle+\cdots+\langle {\bf x}_{i_m}, {\bf y}_{i_m}-{\bf y}_{i_0}\rangle >0,
\end{equation}
while \eqref{keycondition} holds if and only if there exist real numbers $\psi_1,\ldots,\psi_n$ such that
$$\langle {\bf x}_i,{\bf y}_i-{\bf y}_j\rangle> \psi_i-\psi_j\quad \mbox{ for all } i\ne j.$$
On the other hand, defining $f_{i,j}(\psi):=\psi_i-\psi_j-\langle {\bf x}_i,{\bf y}_i-{\bf y}_j\rangle$ for $i\ne j$, we can apply Farkas' Lemma (see, e.g., Theorem 21.1. in Rockafellar~(1970))
to see that either there exists $\psi\in\mathbb{R}^n$ such that $f_{i,j}(\psi)<0 $ for all $i\ne j$ (equivalently,~\eqref{keycondition}  holds), 
or there exist nonnegative weights  $\lambda_{i,j}$, not all zero, such that
$$\sum_{i\ne j} \lambda_{i,j} f_{i,j}(\psi)\geq 0 \quad  \mbox{ for all } \psi \in\mathbb{R}^n.$$
Consider the graph with vertices $\{1,\ldots,n\}$ and (directed) edges corresponding to those pairs~$(i,j)$ for which~$\lambda_{i,j}>0$. 
There cannot be a vertex of degree one in the graph since, in that case, $\sum_{i\ne j} \lambda_{i,j} f_{i,j}(\psi)$ could not be bounded from below.
Hence, the graph contains at least one cycle, that is, there exist $i_0,i_1,\ldots,i_m$ such that $\lambda_{i_0,i_1}$, $\lambda_{i_1,i_2},\ldots$, and 
$\lambda_{i_m,i_0}$ all are strictly positive. Part {\it (i)} of the lemma then implies the existence of $\bar{\psi}_1,\ldots,\bar{\psi}_n$ such that~$f_{i,j}(\bar{\psi})\leq 0$ for all $i\ne j$. But then  
$0\leq \sum_{i\ne j} \lambda_{i,j} f_{i,j}(\bar{\psi})\leq 0,$ 
which implies that $f_{i,j}(\bar{\psi})=0$ for each pair $i,j$ with $\lambda_{i,j}>0$, so that
$$f_{i_0,i_1}(\bar{\psi})+f_{i_1,i_2}(\bar{\psi})+\cdots+f_{i_m,i_0}(\bar{\psi})=0.$$ This in turn entails (observe that the sum $\bar{\psi_i}-\bar{\psi_j}$
along a cycle $i_0,i_1,\ldots,i_m,i_0$ vanishes) 
\begin{equation}\label{Ah}\langle {\bf x}_{i_0},{\bf y}_{i_0}-{\bf y}_{i_1}\rangle+\langle {\bf x}_{i_1},{\bf y}_{i_1}-{\bf y}_{i_2}\rangle +\cdots+\langle {\bf x}_{i_m},{\bf y}_{i_m}-{\bf y}_{i_0}\rangle=0.
\end{equation}
But \eqref{Ah} contradicts (\ref{keyconditioneq}), which implies that if $T$ is the unique cyclically monototone map from $\{{\bf x}_1,\ldots,{\bf x}_n\}$ to $\{{\bf y}_1,\ldots,{\bf y}_n\}$, 
then \eqref{keycondition} holds. Conversely, if \eqref{keycondition} holds, then, for every cycle $i_0,i_1,\ldots,i_m,i_0$, we have
$$\langle {\bf x}_{i_0}, {\bf y}_{i_0}-{\bf y}_{i_1}\rangle +\langle {\bf x}_{i_1}, {\bf y}_{i_0}-{\bf y}_{i_2}\rangle +\cdots+\langle {\bf x}_{i_m}, {\bf y}_{i_m}-{\bf y}_{i_0}\rangle \hspace{30mm}$$
$$\hspace{30mm}>(\psi_{i_0}-\psi_{i_1})+(\psi_{i_1}-\psi_{i_2})+\cdots +(\psi_{i_m}-\psi_{i_0})=0,$$
and  $T$ is the unique cyclically monotone map from $\{{\bf x}_1,\ldots,{\bf x}_n\}$ to~$\{{\bf y}_1,\ldots,{\bf y}_n\}$. This completes the proof.\hfill $\Box$

\subsection{%Appendix II.2. 
Proof of Proposition~\ref{InterpolationTheorem}}\label{InterpolationTheoremProofSec} The map $\varphi_\varepsilon$ is convex and continuously differentiable since~$\varphi$ is convex (see, e.g., Theorem 2.26 in Rockafellar and Wets~(1998)).
Hence $T_\varepsilon :=\nabla \varphi_\varepsilon$ is a cyclically monotone, continuous map for every~$\varepsilon>0$. Setting  
$$\tilde{\varepsilon}_0=\min_{1\leq i\leq n} \Big( (\langle {\bf x}_i, {\bf y}_i \rangle-\psi_i)-\max_{j\ne i} (\langle {\bf x}_i, {\bf y}_j \rangle-\psi_j)\Big),$$
let    $\varepsilon_0=\frac{1}{2} {\tilde{\varepsilon}_0}  \min(1,1/\max_{1\leq i\leq n}\|{\bf y}_i\|)$. Note that   $\tilde{\varepsilon}_0$, by~\eqref{assumption3},  is strictly positive; hence,  so is~$\varepsilon_0$.  
If ${\bf x}$ lies in the $\varepsilon_0$-ball $B({\bf x}_i,\varepsilon_0)$ centered at ${\bf x}_i$,  then, if~$j \neq i$,
\begin{eqnarray*}
\langle {\bf x}, {\bf y}_i \rangle-\psi_i&=& \langle {\bf x}_i, {\bf y}_i \rangle-\psi_i+\langle {\bf x}-{\bf x}_i, {\bf y}_i\rangle >\langle {\bf x}_i, {\bf y}_j \rangle-\psi_j+\tilde{\varepsilon}_0-\varepsilon_0 \|{\bf y}_i\|\\
&\geq & \langle {\bf x}_i, {\bf y}_j \rangle-\psi_j+\frac{1}{2}{\tilde{\varepsilon}_0}\geq \langle {\bf x}, {\bf y}_j\rangle -\psi_j.
\end{eqnarray*}
This shows that $B({\bf x}_i,\varepsilon_0)\subset C_i$ and 
$$\varphi({\bf x})=\langle {\bf x}, {\bf y}_i \rangle-\psi_i,\quad {\bf x} \in B({\bf x}_i,\varepsilon_0).$$

Assume now that 
$$0< \varepsilon\leq 
\frac{1}{2}{{\varepsilon}_0}  \min\left(1,\frac{1}{\max_{1\leq i\leq n}\|{\bf y}_i\|}\right)
,$$
 and let~${\bf x}\!\in~\!\!B({\bf x}_i,\varepsilon)$. The map ${\bf y} \mapsto  \langle {\bf y}, {\bf y}_i\rangle-\psi_i+\frac 1 {2\varepsilon}\|{\bf y}-{\bf x}\|^2$
attains its global minimum at ${\bf y}={\bf x}-\varepsilon {\bf y}_i\in B({\bf x}_i,\varepsilon_0)$. For any $\bf y$, we have 
\begin{eqnarray*}
\varphi({\bf y})+\frac{1}{2\varepsilon} \|{\bf y}-{\bf x}\|^2&\geq & \langle {\bf y}, {\bf y}_i \rangle -\psi_i +\frac{1}{2\varepsilon} \|{\bf y}-{\bf x}\|^2\\
&\geq&\varphi({\bf x}-\varepsilon {\bf y}_i)+\frac{1}{2\varepsilon} \|{\bf x}-\varepsilon {\bf y}_i-{\bf x}\|^2\\ 
&=&\langle {\bf x}, {\bf y}_i \rangle-\psi_i-\frac \varepsilon 2 \|{\bf y}_i\|^2.
\end{eqnarray*}
This proves that 
$$\varphi_\varepsilon({\bf x})=\langle {\bf x}, {\bf y}_i \rangle-\psi_i-\frac \varepsilon 2 \|{\bf y}_i\|^2, \quad {\bf x}\in B({\bf x}_i,\varepsilon);$$
in particular, we conclude that  $T_\varepsilon({\bf x}_i)={\bf y}_i$. 
{%\color{red} 

Turning to the last claim,   note that $$T_\varepsilon({\bf x})=\frac 1 \varepsilon ({\bf x}-{\bf y}_0),$$ where  ${\bf y}_0$ is the unique minimizer of ${\bf y}\mapsto \varphi({\bf y})+{\|{\bf y}-{\bf x}\|^2} /{2\varepsilon}$
(again by Theorem~2.26 in  Rockafellar and Wets~(1998)). But ${\bf y}_0$ is such a minimizer if and only if~$0\in \partial \varphi({\bf y}_0)+\frac 1 \varepsilon ({\bf y}_0-{\bf x})$, that is, if and only if~$T_\varepsilon({\bf x})%=\frac 1 \varepsilon (x-y_0)
\in \partial \varphi({\bf y}_0)$, where~$\partial \varphi({\bf y}_0)$ denotes the subdifferential of $\varphi$ at ${\bf y}_0$. Now  (this is Theorem~25.6 in Rockafellar~(1970)), for every~${\bf x}\in~\!\mathbb{R}^d$, 
$\partial \varphi({\bf x})$ is the closure of the convex hull of the set of limit points of sequences of the type $\nabla \varphi({\bf x}_n)$ with ${\bf x}_n\to {\bf x}$.~The map~$\varphi$ is differentiable in the regions $C_i$, with gradient~${\bf y}_i$. Hence, for every~${\bf x}$,~$T_\varepsilon({\bf x})$ belongs to the convex hull of $\{{\bf y}_1,\ldots,{\bf y}_n\}$. This completes the proof,
} \hfill $\Box$\medskip

%{\sc Remark A.1.} 
\begin{rem}\label{notaA1} 
{\rm (Remark~\ref{rem1} continued)  
It is important to note that, in spite of what intuition may suggest, and except for  the one-dimensional case ($d=1$), 
linear interpolation does not work in this problem. Assume that~$n~\!\geq~\!d+1$ and  that  $\{{\bf x}_1, \ldots,{\bf x}_n\}$ are in general position. Denoting by~$\cal C$   the convex hull of $\{{\bf x}_1, \ldots,{\bf x}_n\}$,   there exists a partition of $\cal C$  into $d$-di\-mensional simplices determined by points in  $\{{\bf x}_1, \ldots,{\bf x}_n\}$: every point in~$\cal C$ thus can be written in a unique way as a linear convex combination of the points determining the simplex it belongs to (with   obvious modification for   boundary points).  Therefore, for all~${\bf x} \in {\cal C}$, there exist uniquely defined coefficients~$\lambda_i^{\bf x} \in [0,1]$, $i=1,\ldots, n$, with $\sum_i \lambda_i^{\bf x}=1$ and~$\#\{i | \lambda _i^{\bf x} \neq 0\} \leq d+1$, such that~${\bf x} = \sum_{i=1}^k \lambda_i^{\bf x} {\bf x}_i$. 
A ``natural" linear interpolation of~$T$ on~$\cal C$ would be~$
{\bf x} \mapsto \sum_{i=1}^k \lambda_i^{\bf x} {\bf y}_i, \ x \in \cal C
$. 
For $d=1$, this map is trivially monotone  increasing, hence cyclically monotone. Starting with~$d=2$, however, this is no longer true, as the following counterexample shows. Let (for $d=2$)
$$
{\bf x}_1 = \left(\begin{array}{c}0\\ 0\end{array}\right) ,
\ {\bf x}_2 = \left(\begin{array}{c}0\\ 1\end{array}\right) , \  
{\bf x}_3 = \left(\begin{array}{c}1\\ 1\end{array}\right) , $$ % \quad\text{and}
\ %quad 
$$ {\bf y}_1 =  \left(\begin{array}{c}-5\\ -.01\end{array}\right)  ,\  
 {\bf y}_2 =  \left(\begin{array}{c}.5\\ .01\end{array}\right)   ,\ 
 {\bf y}_3 =   \left(\begin{array}{c}1\\ 0\end{array}\right)  .
$$
It is easily checked that the map ${\bf x}_i \mapsto {\bf y}_i$,  $i=1,\, 2,\, 3$ is the only cyclically monotone one pairing those points. Now, let us consider the points 
$${\bf x}_0 = .8{\bf x}_1 + .1{\bf x}_2 + .1{\bf x}_3  \quad\text{and}\quad  {\bf y}_0 = .8{\bf y}_1 + .1{\bf y}_2 + .1{\bf y}_3.$$
 The computation of all possible 24 pairings  shows that the only cyclically monotone mapping between the sets $\{{\bf x}_0,\ldots, {\bf x}_3\}$ and  $\{{\bf y}_0,\ldots, {\bf y}_3\}$ is 
%\[
%{\bf x}_i \mapsto \left\{\begin{array}{ll}
%{\bf y}_i & \mbox{ if } i=1,3
%\\
%{\bf y}_0 & \mbox{ if } i=2
%\\
%{\bf y}_2 & \mbox{ if } i=0
%\end{array}
%\right.
%\]
 $${\bf x}_0 \mapsto {\bf y}_2, \ \  
{\bf x}_1 \mapsto {\bf y}_1, \ \  
{\bf x}_2 \mapsto {\bf y}_0, \ \  
{\bf x}_3 \mapsto {\bf y}_3$$
where, obviously, ${\bf x}_0$ is not paired with ${\bf y}_0$ (nor ${\bf x}_2$ with ${\bf y}_2$).
}
\end{rem}

\subsection{%Appendix II.3. 
Proof of Proposition~\ref{PropGC}}\label{GCproofsec}  

The proof of Proposition~\ref{PropGC} relies on  the following two preliminary propositions.

\begin{prop}\label{sumlem} Let ${\bf Z}\n_1,\ldots , {\bf Z}\n_n$ be i.i.d.~with   distribution ${\rm P}\in{\cal P}_d$ and denote by $\mu\n$  the corresponding empirical distribution. Then, 
$$\gamma\n:= (\text{\,\rm identity}\times {\bf F}\n_{\pms})\#\mu\n \text{ converges weakly to } \gamma =  (\text{\rm identity}\times {\bf F}_{\pms})\#{\rm P}
$$
as~$n\to\infty$,  $ {\rm P}-\text{a.s.}$,
 where ${\bf F}_{\!\pms}$ is $\rm P$'s center-outward distribution.
%where $\nabla\psi$, by definition of ${\bf F}_{\pms}$ and the unicity (up to a set of $\rm P$-probability zero) part of McCann's main result, is precisely ${\bf F}_{\pms}$.  
\end{prop}

%\begin{prop}\label{monotonicity}
%%\hspace{-1.5mm}
%Assume $\varphi\!: \mathbb{R}^d\to \mathbb{R}$ is a differentiable convex function such that~$\nabla \varphi$ is a homeomorphism from $\mathbb{R}^d$
%to the open unit ball. %If we take 
%Let ${\bf x}_n=\lambda_n {\bf u}_n$ with $0<\lambda_n\to \infty$, $\|{\bf u}_n\|=1$ and ${\bf u}_n\to {\bf u}$: then,  $\nabla \varphi ({\bf x}_n)\to~\!{\bf u}$.
%\end{prop}

{\color{black}
\begin{prop}\label{monotonicity}
%\hspace{-1.5mm}
Let ${\rm P}\in {\cal P}_d^\pms%^{\text{\em conv}}
$ have center-outward distribution function~$\mathbf{F}_\pms$ and 
let ${\bf x}_n=\lambda_n {\bf u}_n$ with $0<\lambda_n\to \infty$, $\|{\bf u}_n\|=1$, and ${\bf u}_n\to {\bf u}$ as $n\to\infty$: then,~$\mathbf{F}_\pm ({\bf x}_n)\to~\!{\bf u}$.
\end{prop}
}

%\color{magenta}

%\subsection{%Appendix 1.2. 
%Proof of Propositions~5.1 and~5.2  (Glivenko-Cantelli)}\label{GCproofsec} 
  
%  $\;$\vspace{-6mm}
%  
%{\bf Proof of Proposition~5.1.} 
  The proof of Proposition~\ref{sumlem} involves  four    lemmas, three  from McCann~(1995) and one from Rockafellar (1966), which we reproduce here for the sake of completeness.   Throughout this section,~$\mu$ and~$\nu$ denote elements of the    set ${\cal P}( \mathbb{R}^d)$ of all probability distributions on $ \mathbb{R}^d$,   ${\cal P}( \mathbb{R}^d\times  \mathbb{R}^d)$  the set of all probability distributions on $\mathbb{R}^d\times  \mathbb{R}^d$, 
 and~$\Gamma (\mu, \nu)$ the set of probability distributions in~${\cal P}( \mathbb{R}^d\times  \mathbb{R}^d)$ with given marginals $\mu$ and~$\nu$ in ${\cal P}( \mathbb{R}^d)$. A measure~$\gamma$ in~${\cal P}( \mathbb{R}^d\times  \mathbb{R}^d)$ is said to have {\it cyclically  monotone support} if there exists a cyclically monotone closed Borel set $S$ in $\mathbb{R}^d\times  \mathbb{R}^d$ such that~$\gamma(S)=~\!1$.

\begin{lem}\label{McCprop1} {\em (McCann 1995, Corollary~14).} 
Let $\mu$, $\nu\in{\cal P}( \mathbb{R}^d)$, and suppose that one of those two measures vanishes on all sets of Hausdorff dimension~$d-~\!1$. Then, there exists one and only one measure $\gamma\in \Gamma (\mu, \nu)$ having cyclically  monotone support.
\end{lem}

\begin{lem}\label{McCprop2} {\em (McCann 1995, Lemma~9). } 
Let $\gamma\n \in {\cal P}( \mathbb{R}^d \times  \mathbb{R}^d)$ converge weakly as $n\to\infty$ to~$\gamma \in {\cal P}( \mathbb{R}^d \times  \mathbb{R}^d)$. Then, 
\begin{enumerate}
\item[(i)] if $\gamma\n$ has cyclically monotone support for all $n$, so does $\gamma$;
\item[(ii)] if $\gamma\n\in\Gamma(\mu\n, \nu\n)$ where $\mu\n$ and $\nu\n$ converge weakly, as $n\to\infty$,  to $\mu$ and $\nu$, respectively, then $\gamma\in\Gamma(\mu, \nu)$.
\end{enumerate}
\end{lem}

%Next, recall that the subdifferential $\partial\psi$ of a convex function $\psi: \ \mathbb{R}^d\to\mathbb{R}$ 
%is the collection of pairs $({\bf x},{\bf y})\in\mathbb{R}^d\times\mathbb{R}^d$ such that 
% $$
%\psi({\bf{z}})\geq \psi({\bf x}) +  \langle {\bf y}, {\bf z}-{\bf x}
%\rangle \qquad {\bf z}\in\mathbb{R}^d,
%$$ 
%that is, such that~$\psi({\bf z})$ lies entirely ``above" the (supporting) hyper\-plane $\{{\bf z} :\ {\bf y}\pr({\bf z}-{\bf x}) = 0\}$; ${\bf y}$ is called a {\it subgradient} of $\psi$ at~${\bf x}$.   A convex function being  Lebesgue-a.e.\ differentiable, the subdifferential of a convex function $\psi$
%  coincides Lebesgue-a.e.\ with  the collection $\{({\bf x}, \nabla\psi({\bf x}))\}$.

\begin{lem}\label{McCprop3} {\em (McCann 1995, Proposition~10).}    Suppose that $\gamma \in \Gamma(\mu, \nu)$ is supported on the subdifferential $\partial\psi$ of some convex function $\psi$ on $\mathbb{R}^d$ (meaning that the support  of $\gamma $ is a subset of~$ \partial\psi$). Assume that $\mu$ vanishes on Borel  sets of Hausdorff dimension $d-1$. Then, $\nabla\psi\# \mu =\nu$, that is,  $\gamma = (\text{\,\rm identity}\times \nabla\psi)\#\mu $, 
where $ (\text{\,\rm identity}\times \nabla\psi){\bf x} := ({\bf x},\nabla\psi({\bf x}))$. 
\end{lem}

%Finally, the following lemma by Rockafellar (1966) establishes a strong relation between cyclical monotonicity and convex functions (Rockafellar's statement actually holds for more general topological vector space).
%
%

\begin{lem}\label{Rockfprop} {\em (Rockafellar 1966, Theorem~1).} The subdifferential $\partial\psi$ of a convex function $\psi$ on $\mathbb{R}^d$ enjoys cyclical monotonicity. Conversely, any cyclically monotone set $S$ of $\mathbb{R}^d\times \mathbb{R}^d$ is contained in the  subdifferential $\partial\psi$ of some convex function $\psi$ on $\mathbb{R}^d$. %\vspace{2mm}
 \end{lem}

This implies the existence of a gradient of convex function running through any $n$-tuple of cyclically monotone couples $(({\bf x}_1, {\bf y}_1),\ldots , ({\bf x}_n, {\bf y}_n))\in\mathbb{R}^d\times\mathbb{R}^d$.   \smallskip

We now turn to the proof  of Propositions~\ref{sumlem} and~\ref{monotonicity}. \smallskip
 
 {\bf Proof of Proposition~\ref{sumlem}.}  Denote by~$(\Omega, {\cal A}, {\rm P})$ the (unimportant) probability space underlying the observation of  the sequence of~${\bf Z}\n_i$'s, $n\in{\mathbb{N}}$,   by~$\gamma\n =(\text{identity}\times {\bf F}\n_{\pms})\#\mu\n$ the empirical distribution, with marginals~$\mu\n$ and~$U\n$,  of the couples $({\bf Z}\n_i, {\bf F}\n_{\pms}({\bf Z}\n_i))$,  and 
 by~$\gamma=(\text{identity}\times {\bf F}_{\pms})\#{\rm P}$ (with marginals ${\rm P}, {\rm U}_d$) the joint distribution of $({\bf Z}, {\bf F}_{\pms}({\bf Z}))$. Here, $\mu\n$, hence also $\gamma\n$, are  random measures, with realizations $\mu\n_\omega$ and  $\gamma\n_\omega$.

 A sequence $\gamma\n_\omega$, $n\in{\mathbb{N}}$, is $\rm P$-a.s.\ asymptotically tight since $\mu\n_\omega$ converges weakly to $\rm P$ with probability one and ${\rm U}\n$ has uniformly bounded support. By Prohorov's theorem,  subsequences~$\gamma^{(n_k)}_\omega$ can be extracted that converge weakly (to some~$\gamma^{\infty}_\omega$'s).  Those $\gamma^{(n_k)}_\omega$'s by construction have cyclically monotone supports, and their marginals  $\mu^{(n_k)}_\omega$ and~${\rm U}^{(n_k)}$  converge weakly to $\rm P$ and ${\rm U}_d$. Hence, by Lemma~\ref{McCprop2}, all limiting $\gamma^{\infty}_\omega$'s have cyclically monotone supports and margi\-nals $\rm P$ and ${\rm U}_d$, respectively. 

In view of Lemma~\ref{McCprop1}, there exists only one $\gamma$ with cyclically monotone support and mar\-ginals~$\rm P$ and ${\rm U}_d$. Hence, irrespective of the choice of the weakly converging subsequence~$\gamma^{(n_k)}_\omega$, all limiting $\gamma^\infty_\omega$'s coincide with $\gamma$, which implies that the original sequence is converging weakly to~$\gamma$. Moreover, that limit is the same for any $\omega$ in some $\Omega_1\subseteq\Omega$ such that $\rm P (\Omega_1)=1$.

Rockafellar's Theorem (Lemma~\ref{Rockfprop}) provides a convex function $\psi$ the subgradient of which contains the support of $\gamma$. Lemma~\ref{McCprop3} and the definition of~${\bf F} _{\pms}$ conclude that 
 $\gamma = (\text{identity}\times \nabla\psi)\#{\rm P} = (\text{identity}\times {\bf F}_{\pms}
)\#{\rm P}.$ \cqfd
\medskip 

{\color{black}
{\bf Proof of Proposition~\ref{monotonicity}.}  Being the gradient $\nabla \phi$ of a convex func-\linebreak tion,~$\mathbf{F}_\pm=\nabla \phi$ is a monotone function (see, e.g., Rockafellar and Wets (1998)). The sequence $\nabla \phi(\mathbf{x}_n)$ is bounded. Taking subsequences if necessary, we can assume that 
$\nabla \phi(\mathbf{x}_n)\to\mathbf{y}$ for some $\mathbf{y}$  with $\|\mathbf{y}\|\leq 1$. By monotonicity, we have that 
$$\langle {\bf x}_n-{\bf x},\nabla \phi ({\bf x}_n)-\nabla \phi({\bf x})\rangle\geq 0$$ 
for every ${\bf x}\in\mathbb{R}^d$. In particular, 
$$\langle {\bf x}_n-(\nabla \phi)^{-1}({\bf w})), \nabla \phi ({\bf x}_n)-{\bf w}\rangle \geq 0$$ 
for every $\bf w$ with $0<\|{\bf w}\|<1$. But this means that
$$\langle {\bf u}_n-{\textstyle \frac 1 {\lambda_n}}(\nabla \phi)^{-1}({\bf w}), \nabla \phi ({\bf x}_n)-{\bf w}\rangle\geq 0$$ 
and, taking limits, that $\langle {\bf u},{\bf y}-{\bf w}\rangle\geq 0$
for every $\bf w$ with $\|{\bf w}\|\leq 1$. From this we conclude that $\langle {\bf u}, {\bf y}\rangle\geq \|{\bf u}\|$. But, since $\|{\bf y}\|\leq 1$, this only  can happen if~${\bf y}={\bf u}$.~\hfill$\Box$%\vspace{20mm}
\medskip }

We now can proceed with the proof of Proposition~\ref{PropGC}.\medskip

 {\bf Proof of Proposition~\ref{PropGC}.}  Denote by $\mathrm{U}^{(n)}_{d}$ the discrete probability measure assigning mass ${n_0}/n$ to the origin and 
mass $ 1 /n$ to the remaining points in the regular grid used for the definition of ${\bf F}_\pms^{(n)}$, and note that $\mathrm{U}^{(n)}_{d}$ converges weakly to   
${\rm U}_d$. Also write ${\rm P}^{(n)}$ for the empirical measure on~$Z\n_1,\ldots ,Z\n_n$. Over a probability one set $\Omega_0$, say, of the underlying probability space $\Omega$, the sequence~${\rm P}^{(n)}$ converges weakly to $\rm P$.
In the remainder of this proof, we tacitly \linebreak assume   that $\omega\in\Omega_0$. 
Note that~$\overline{\bf F}\n_\pms=\nabla \phi_n$ for some convex~$\phi_n$, and\linebreak that~${\bf F}_{\pms}=\nabla \phi$ with  $\phi$ convex and continuously differentiable over~$\mathbb{R}^d$. 
Recall, moreover, that~$\nabla \varphi_n$, by construction, maps~${\rm P}^{(n)}$ to $\mathrm{U}^{(n)}_{d}$. By Theorem~2.8  in del Barrio and Loubes~(2019), after subtracting centering constants if necessary, we can assume that $\phi_n({\bf x})\to \phi({\bf x})$ for  every~${\bf x}\in %\mathcal{X}=
\text{spt}({\rm P})$. Actually,  the statement of that  result assumes 
convergence in transportation cost metric rather than weak convergence; the proof, however, only depends  on the fact that, in that case, the sequence~$\pi_n=(\text{identity}\times \nabla \varphi_n)\#  {\rm P}^{(n)}$
converges weakly to~$\pi=(\text{identity}\times \nabla \varphi)\#  {\rm P}$, which, in view of Proposition~\ref{sumlem}, holds  here. 
%{\color{blue} 
We claim that, in fact, ${\phi}_n({\bf x})\to \phi({\bf x})$ for every~${\bf x}\in \mathbb{R}^d$. 
To see this,    first  note that, by Proposition~\ref{InterpolationTheorem}, ~$\|\overline{\bf F}\n_\pms(\mathbf{x})\|=\|\nabla \phi_n(\mathbf{x})\|\leq 1$
for every $\mathbf{x}\in\mathbf{R}^d$, which implies that the sequence $\{\phi_n\}$ is uniformly 1-Lipschitz, hence uniformly equicontinuous on $\mathbb{R}^d$. Also, since $\phi_n$ is pointwise convergent in spt$({\rm P})$, we can apply the Arzel\`a-Ascoli Theorem to conclude that we can extract a uniformly convergent subsequence over any compact subset of $\mathbb{R}^d$. By extracting a further subsequence, we can assume that $\phi_n\to \rho$ pointwise on all of $\mathbf{R}^d$ for some function~$\rho$. This function must be convex and 1-Lipschitz (in particular, finite over all $\mathbb{R}^d$)
and, obviously, $\rho(\mathbf{x})=\phi(\mathbf{x})$ for every $\mathbf{x}\in\text{spt}({\rm P})$. We note also that for every~$\mathbf{u}\in \mathbb{S}_d$ there exists some $\mathbf{x}\in\text{spt}({\rm P})$ with $\mathbf{u}=\nabla \phi(\mathbf{x})= \nabla \rho(\mathbf{x})$. Hence, for every~$\mathbf{z}$,
 $\rho(\mathbf{z})-\rho(\mathbf{x})\geq\langle\mathbf{u},\mathbf{z}-\mathbf{x} \rangle$.  
By duality, $$\rho(\mathbf{x})=\phi(\mathbf{x})=\langle \mathbf{u},\mathbf{x}\rangle-\phi^*(\mathbf{u})=\langle \mathbf{u},\mathbf{x}\rangle-\psi(\mathbf{u}).$$ 
This shows that
\begin{equation}\label{upboundGC}
 \rho(\mathbf{z})\geq \sup_{u\in\mathbb{S}_d}(\langle \mathbf{u}, \mathbf{z}\rangle-\Psi(\mathbf{u}))=\phi(\mathbf{z}),\quad \mathbf{z}\in\mathbb{R}^d.
\end{equation}

To get an upper bound we note that, for every $\mathbf{x}$, $\mathbf{u}_n:=\nabla \phi_n(\mathbf{x})\in\mathbb{S}_d$. Since~$\langle \mathbf{x}, \mathbf{u}_n\rangle=\phi_n(\mathbf{x})+\phi_n^*(\mathbf{u}_n)$ we obtain that
\begin{equation}\label{lowerGC}
\phi_n(\mathbf{x})=\langle \mathbf{x}, \mathbf{u}_n\rangle-\phi_n^*(\mathbf{u}_n)\leq
\sup_{\mathbf{u}\in\mathbb{S}_d}(\langle \mathbf{x}, \mathbf{u}\rangle-\phi_n^*(\mathbf{u}))=:\tilde{\phi}_n(\mathbf{x}).
\end{equation}
Now, $\tilde{\phi}_n$ is the convex conjugate of the function 
$$\mathbf{u}\mapsto \tilde{\psi}_n(\mathbf{u})=\left\{
\begin{array}{cl}
{\phi}_n^*(\mathbf{u})\quad &\mathbf{u}\in\bar{\mathbb{S}}_d\\ 
 \infty\quad & \mathbf{u}\notin\bar{\mathbb{S}}_d .
 \end{array}
 \right. 
 $$
Using Theorem 2.8 in del Barrio and Loubes~(2019) again, we obtain that,  for every
$\mathbf{u}\in\bar{\mathbb{S}}_d$, ${\phi}_n^*(\mathbf{u})\to \psi(\mathbf{u})$ and, consequently,  for every $\mathbf{u}\in\mathbb{R}^d\setminus \mathcal{S}_{d-1}$, that~$\tilde{\psi}_n(\mathbf{u})\to \psi(\mathbf{u})$. Combining this with Theorems 7.17 and 11.34 in Rockafellar and Wets~(1998), we conclude that 
$\tilde{\phi}_n(\varphi(\mathbf{x})\to \psi^*(\mathbf{x})=\phi(\mathbf{x})$ for every~$\mathbf{x}\in\mathbf{R}^d$.
Combined with (\ref{lowerGC}), this shows that $\rho(\mathbf{x})\leq \phi(\mathbf{x})$ which, along with~(\ref{upboundGC}),  yields 
$\phi_n(\mathbf{x})\to \phi(\mathbf{x})$. But then, Theorem~25.7 in Rockafellar~(1970) implies that
$$\overline{\bf F}\n_\pms({\bf x})=\nabla \phi_n({\bf x})\to \nabla \phi({\bf x})={\bf F}_{\pms}({\bf x}),\quad \mathbf{x}\in\mathbf{R}^d,$$ 
uniformly over compact sets. 

It only remains
to show that uniform convergence holds over~$\mathbb{R}^d$. %}
For this, it suffices to show that, for every ${\bf w}\in\mathbb{R}^d$,
\begin{equation}\label{unifnormproj}
\sup_{{\bf x}\in\mathbb{R}^d}\big|\langle \big(\overline{\bf F}\n_{\pms}({\bf x}) -{\bf F}_{\pms}({\bf x})\big) , {\bf w}\rangle\big| \to 0.
\end{equation}
Let us assume that, on the contrary, there exist $\varepsilon >0$, ${\bf w}\in \mathbb{R}^d\setminus \{\bf 0\}$, and~${\bf x}_n\in ~\!\mathbb{R}^d$ such that %{\color{blue}
\begin{equation}\label{contradiction3}
\big|\langle \big(\overline{\bf F}\n_{\pms}({\bf x}_n) -{\bf F}_{\pms}({\bf x}_n)\ \big), {\bf w}\rangle \big|>\varepsilon
\end{equation}
for all $n$. The sequence ${\bf x}_n$ must be unbounded (otherwise (\ref{contradiction3}) cannot hold). Hence, using compactness of the unit sphere and taking subsequences if necessary,
we can assume that~${\bf x}_n=\lambda_n {\bf u}_n$ with $0<\lambda_n\to \infty$, $\|{\bf u}_n\|=1$, and~${\bf u}_n\to {\bf u}$ for some $\bf u$ with $\|{\bf u}\|=1$. Again by compactness, we can assume that~${\bf F}_{\pms}({\bf x}_n)\to~\!{\bf y}$ and~$\overline{\bf F}\n_{\pms}({\bf x}_n)\to~\!{\bf z}$. By Proposition~\ref{monotonicity}, we have that~${\bf y}={\bf u}$. On the other hand, by monotonicity, for every ${\bf x}\in\mathbb{R}^d$,
$$\langle \overline{\bf F}\n_{\pms}({\bf x}_n)-\overline{\bf F}\n_{\pms}({\bf x}), {\bf x}_n-{\bf x}\rangle\geq 0.$$
Taking $\tau>0$ and ${\bf x}=\tau {\bf u}_n$, we obtain that, if $n$ is large enough (to ensure~$\lambda_n>~\!\tau$), then
$$\langle \overline{\bf F}\n_{\pms}({\bf x}_n)-\overline{\bf F}\n_{\pms}(\tau{\bf u}_n), {\bf u}_n\rangle \geq 0.$$
We conclude that, for every $\tau>0$
$$\langle {\bf z}-{\bf F}_{\pms}(\tau {\bf u}), {\bf u}\rangle\geq 0.$$
Now, we can take $\tau_n\to\infty$ and use Proposition~\ref{monotonicity}  to obtain that ~$\langle {\bf z}-{\bf u}, {\bf u}\rangle\geq~\!0$, that is,~$\langle {\bf z}, {\bf u}\rangle \geq \|{\bf u}\|^2=1$. This, however,  implies that~${\bf z}={\bf u}={\bf y}$, which  contradicts~(\ref{contradiction3}), hence completes the proof. %}
\hfill $\Box$

 \color{black}

%\subsection{%Appendix II.4. 
%Lemma~\ref{monotonicity}} We now state and prove the Lemma that has been used in the Proof of Proposition~\ref{PropGC}.
%

\section{A ``multivariate step function" version  of~${\bf F}_\pms\n\!\!$}\label{stepF}
 Although, for~$d=~\!1$,  a smooth monotone increasing interpolation of the $n$-tuple %of  points
 $(X\n_i,F\n(X\n_i))$ in general provides a better approximation of $F$, empirical distribution functions 
%, in dimension~$d=1$,
 are traditionally defined as right-continuous step functions---the exact opposite of smooth functions. Such step function interpolation yields some interpretational advantages in terms of the empirical measure of  regions of the form $(-\infty, x]$, $x\in\mathbb{R}$. Still for $d=1$, an outward-continuous center-outward counterpart can be defined in a very natural way, with %an 
 interpretation in terms of  the empirical measure of central regions of the form $[x-, x^+]$\vspace{-0mm} where~$[x^-,x\n_{1/2})$ and $(x\n_{1/2}, x^+]$ ($x\n_{1/2}$ an empirical median) contain the same number of observations: see Figure~\ref{Ffig} in~Appendix~\ref{1dimsec}.  

A similar solution can be constructed for $d\geq 2$. 
% as follows. 
 Let $\overline{\bf F}\n_{\!{\pms}}$ be some smooth interpolation of ${\bf F}\n_{\!{\pms}}$. For any $r\in[0,1]$ and ${\bf u}$ on the unit sphere~${\cal S}_{d-1}$, define~$\lfloor r{\bf u} \rfloor _{n_R}:=  {\lfloor (n_R +1)r \rfloor {\bf u} }/{n_R +1} $. Then,  
 $r{\bf u}\mapsto\lfloor r{\bf u} \rfloor _{n_R}$ maps an outward-open, inward-closed spherical annulus comprised in between two hyperspheres of the grid onto its inner boundary  sphere while preserving directions. A ``multivariate step function" version of the empirical center-outward distribution function ${\bf F}_{\pms}^{(n)}$, continuous from outward, can be defined~as 
 $%\begin{equation*}
\overline{\bf F}_{\pms}^{(n)*} := \lfloor \overline{\bf F}_{\pms}^{(n)}  \rfloor _{n_R}
$.%\end{equation*}%\label{stepF}

Instead of {\it steps}, those functions yield  {\it plateaux} or {\it hyperplateaux}, the boundaries (equivalently, the discontinuity points) of which are the continuous {\it quantile contours} or {\it hypersurfaces} characterized by $\overline{\bf F}\n_{\!{\pms}}$. Those ``quantile contours" present an obvious statistical interest.  In contrast with the univariate case, this ``step function version"  %\eqref{stepF} 
of the empirical center-outward distribution function~${\bf F}_{\pms}^{(n)}$, for $d>1$, is not unique, and depends on the  smooth interpolation~$\overline{\bf F}_{\pms}^{(n)}$ adopted.  However, all its versions enjoy cyclical monotonicity and obviously satisfy the sup form of Glivenko-Cantelli: % with probability one, 
$\sup_{{\bf x}\in\mathbb{R}^d}\|\overline{\bf F}_\pms^{(n)*}({\bf x})-{\bf F}_{\pms}({\bf x})\|\to 0$ a.s.  as~$n\to\infty$.

\section{Further numerical results}\label{NUMAPP}

\subsection{Center-outward quantiles and Tukey depth}\label{secTukey} 
Statistical depth and our measure transportation approach are sharing the same ultimate objective of defining a concept of  multivariate quantile. Some compa\-risons   thus are quite natural---although not entirely straightforward, as we shall see, as the two concepts are of a different nature. The discussion below is restricted to Tukey depth, but similar conclusions   hold for other depth concepts. %It is a continuation of Section~\ref{secShape1}. 
 \begin{figure}[htbp] 
\begin{center} 
\arraycolsep = 5pt \renewcommand{\arraystretch}{1}
$\noindent\begin{array}{cc}\framebox{\includegraphics*[width=5.5cm,height=5.6cm,trim=30 30 20 30,clip]
%, trim= 40 180 20 180, clip]
{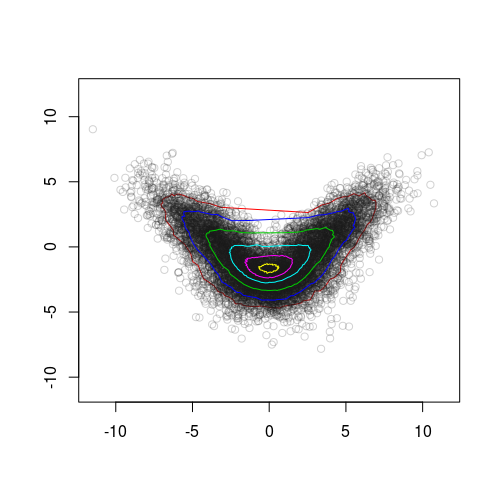}}
&
\framebox{\includegraphics*[width=5.5cm,height=5.6cm,trim=30 30 20 30,clip]%, trim= 40 180 20 180, clip]
{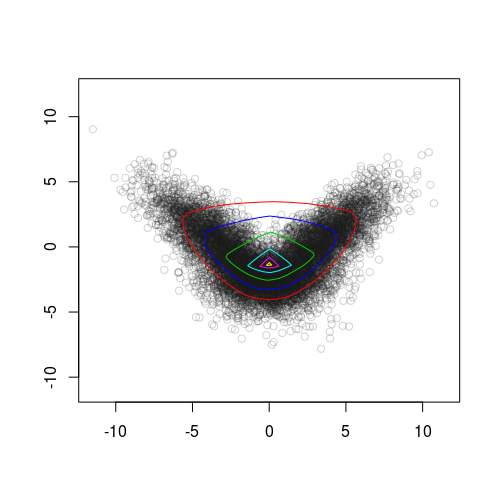}}
\end{array}$
\caption{Center-outward quantile contours (left)  and  Tukey contours (right) for the same Gaussian mixture as in the middle panel of Figure~\ref{FigGM},  with $n=10000$.}\label{Figbanana}  %\vspace{-3mm}
\end{center}
\end{figure}

Whether theoretical or empirical, center-outward quantile functions and Tukey depth produce contours---in short, quantile contours and Tukey contours. 
 For a spherical distribution with center $\boldsymbol\mu$, the family of population  quantiles  (indexed by their probability contents) and the family  of population  Tukey  contours (indexed by depth)  coincide (see Section~2.4 of Chernozhukov et al~(2017)) with the family of (hyper)spheres centered at $\boldsymbol\mu$. Empirical quantile  and empirical Tukey contours   both consistently reconstruct those (hyper)spheres. As a rule, the empirical  Tukey contours are smoother than the empirical quantile ones---although pairwise comparisons are difficult (or meaningless), as the probability contents of  a Tukey contour  (indexed by depth), unlike that of a quantile contour, depends on the underlying distribution. If smooth estimation of the {\it family} of population quantile contours were the objective, Tukey contours thus are doing a better job here. This is somewhat misleading, though---the (deterministic) family of (hyper)spheres centered at $\boldsymbol\mu$ is doing  even better!  
 %Spherical distributions, thus, are perhaps not    the best context for a comparison. 
 But the objective here is not the smooth reconstruction of the {\it family} of quantile contours: we want something consistent that for finite $n$  has the nature of an empirical  quantile function, which requires  (cyclical) monotonocity  properties that Tukey depth, even under spherical distributions, does not satisfy. And, of course, things  only get worse under non-spherical densities. 
 
\begin{figure}[htbp]
\begin{center}
$\noindent\begin{array}{c}
\framebox{\includegraphics*[width=9.6cm,height=9.4cm,trim=30 30 20 30,clip]
{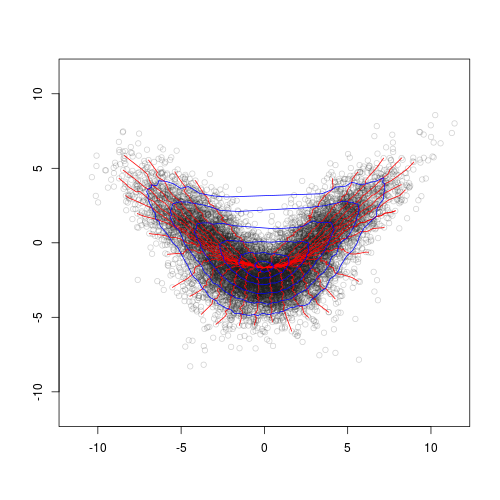}}
%&
%\framebox{\includegraphics*[width=5.5cm,height=5.4cm,trim=30 30 20 30,clip]
%{Fig6b.png}}
\end{array}$
\caption{Center-outward quantile contours and  sign curves for the same Gaussian mixture as in  Figure~\ref{Figbanana},  with $n=20000$. \vspace{-4mm}
% (left) and
% $n=20000$ (right).
 }\label{BananaRays'}   
\end{center}
\end{figure}

 Now, let us have a closer look at the banana-shaped Gaussian mixtures of Figure~\ref{FigGM}. Figure~\ref{Figbanana} is providing, side by side, a plot of some quantile and Tukey contours for $n=10000$. The concave shape of the distribution is only partially picked up by 
%  hardly visible in
  the outer quantile contours (left-hand panel)---despite of proven asymptotic concavity. The same concavity is not picked up at all (not even asymptotically so) by the Tukey contours (right-hand panel), which are inherently convex. 
 Even worse,  the inner Tukey contours display a misleading spike pointing upwards to the empty region. 
 %But at first sight, the global pictures are not that different. 
 Despite of this, and although  the theoretical weakness (lack of cyclical monotonicity) of Tukey contours as multivariate quantiles remain the same as previously discussed, one may feel that   Tukey depth, as a descriptive tool, is doing almost as well, with less computational efforts, as  empirical center-outward quantiles. As mentioned in Section~\ref{numsec}, this  is neglecting   directional information  contained in the empirical sign curves. Tukey depth, which is scalar-valued, has nothing equivalent to offer.  
% In the spherical case, those sign curves are, more or less, straight halflines running through the center-outward median and  uniformly distributed among   all directions: they are quite uninformative, thus, and, therefore, we did not plot them in Figure~2. In the highly non-spherical case of the banana-shaped Gaussian mixture considered here, those sign curves are conveying an essential information. 

 Figure~\ref{BananaRays'} is providing the full picture for $n=20000$. % and $n=20000$.
  The sign curves to the left and to the right of the vertical direction are very neatly combed to the left and to the right parts of the contours. Since each curvilinear sector comprised between two consecutive sign curves roughly has the same probability contents, Figure~\ref{BananaRays'}   provides evidence of a very low density in the central concavity bridged by the contours, thus producing a clear  visualization of the banana shape of the dataset. Such figures, rather than contours alone, are the descriptive plots associated  with  empirical center-outward quantile functions.

 Irrespective of the point of view adopted---be it inferential or data-analytical---center-outward quantile plots, thus,     are carrying an  information that Tukey depth plots cannot provide, which is well worth the additional computational   effort.

\subsection{Compact convex supports}\label{secShape2}
All simulations in Section~\ref{numsec} have been conducted under $\mathbb{R}^d$-supported distributions. In this section, we consider two simple  compactly supported cases. 

\begin{figure}[h!] 
\begin{center} 
\arraycolsep = 5pt \renewcommand{\arraystretch}{1}
$\noindent\begin{array}{cc}
\framebox{\includegraphics*[width=5cm, height =5cm, trim= 40 180 20 180, clip]{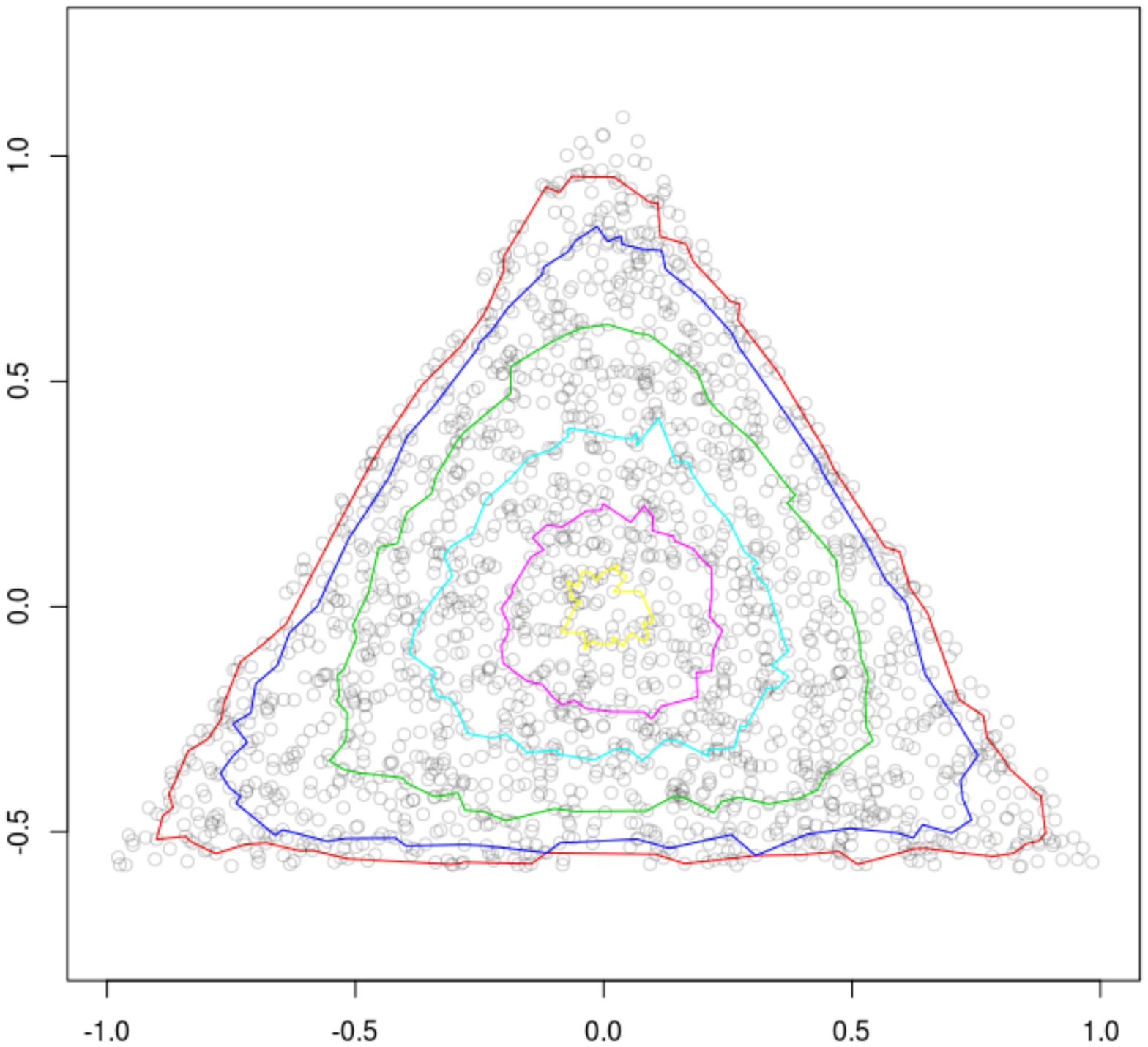}}
&
\framebox{\includegraphics*[width=5cm, height =5cm, trim= 40 180 20 180, clip]{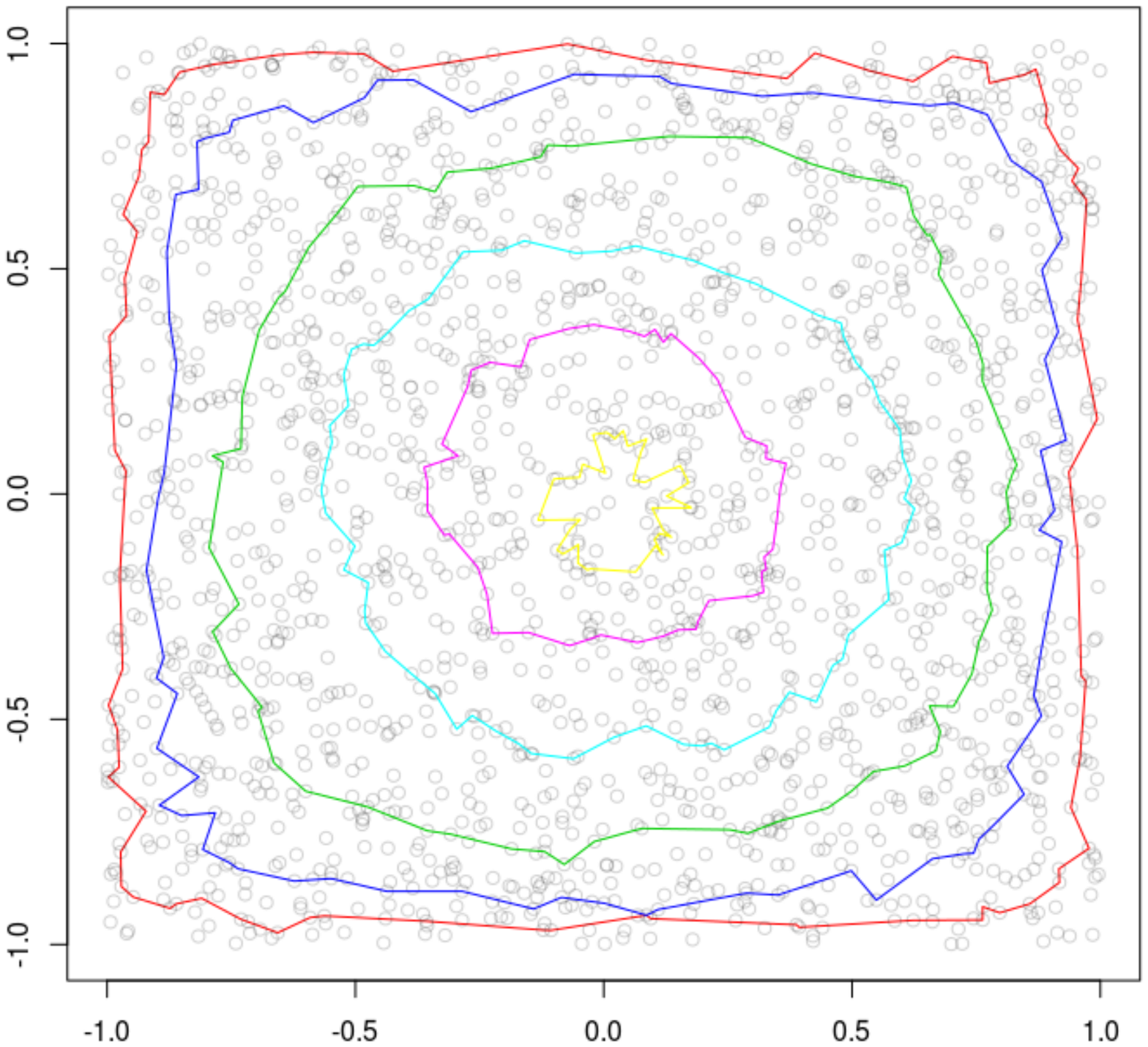}}\vspace{-1mm} 
\end{array}$
\caption{Smoothed empirical center-outward quantile contours (probability contents   .50   (green),   .75  (red), .90 (black)) computed from $n=2000$ i.i.d.\ observations from  Lebesgue-uniform distributions over the triangle and the square, respectively.\vspace{-0mm}}\label{Figtriang}   
\end{center}
\end{figure}

Figure~\ref{Figtriang}  provides  simulations for Lebesgue-uniforms with triangular and squared supports (sample size  $n=2000$, with~$n_R=~\!50$ and~$n_S=40$), and shows how the contours evolve from nested circles in the center, where boundary effects are weak or absent,  to nested triangles and squares as the boundary  effects become dominant.

\subsection{Disconnected supports}\label{Secmoons}

\begin{figure}[h!] 
\begin{center} 
\arraycolsep = 5pt \renewcommand{\arraystretch}{1}
$\noindent\begin{array}{cc}
\framebox{\includegraphics*[width=5cm, height =5.5cm%, trim= 40 180 20 180, clip
]{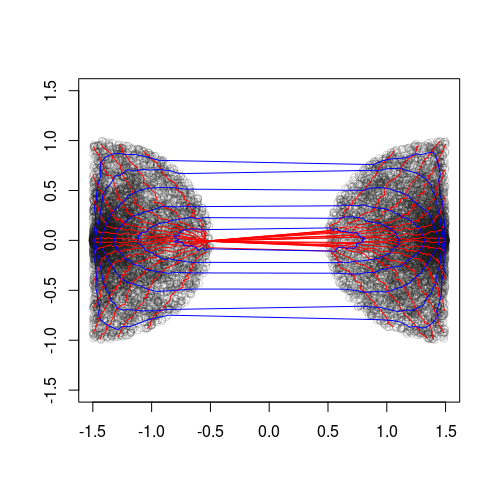}}
&
\framebox{\includegraphics*[width=5cm, height =5.5cm%, trim= 40 180 20 180, clip
]{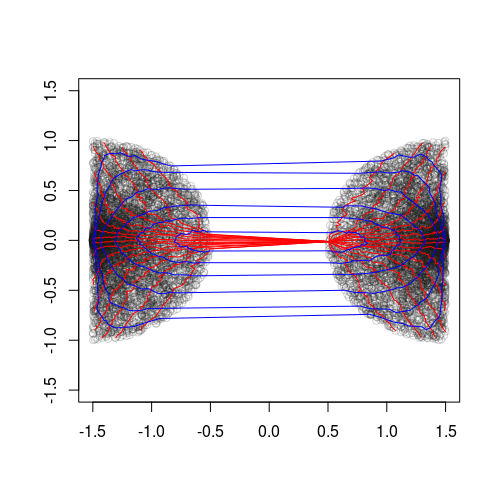}}\vspace{-1mm} 
\end{array}$
\caption{Smoothed empirical center-outward quantile contours and sign curves computed from $n=10000$ i.i.d.\ observations from  a Lebesgue-uniform distribution over two half balls (two independent simulations, showing the instability of the median set).\vspace{-0mm}}\label{Figmoons}   
\end{center}
\end{figure}

Figure~\ref{Figmoons}   provides  two independent simulations (sample size  $n=10000$) from Lebesgue-uniforms supported on two disconnected half-balls. Although the assumptions for consistency are not satisfied, the contours and sign curves provide a very good description of the dataset, demonstrating, as in Figure~\ref{BananaRays'},  their complementarity: while the contours alone   fail to disclose disconnectedness, that crucial feature of the dataset is  fully revealed by the sign curves---an information that no depth concept can provide. Note also the  (unsurprising) instability   of the median set in such case; the same instability would occur in dimension one with a density supported on two disjoint intervals of equal probability 1/2, due to the lack of injectivity of the distribution function.

%The quantile contours are similar in both examples,
%apparently capturing at least part of the shape of the moons. One thing that
%we can see is that the sample median is unstable in this setup (similar to
%what we would have in the univariate case if the underlying distribution was a
%mixture of two distributions with a density in disjoint intervals and giving
%mass 1/2 to each interval. One run has produced a sample median on the left
%moon, and you see the quantile rays emanating from that point. The other run
%has a sample median on the right moon. I have included also (Moons2b.png) a
%picture without the quantile rays (it is the same simulation as in Moons2.png,
%but without the rays).

\subsection{A non-connected $\alpha$-hull contour}\label{Secalphahull}

In this section, we provide an example of the dangers attached with the so-called $\alpha$-hull interpolation, as considered in Chernozhukov et al.~(2017).

 Consider the six points 
$${\bf x}_1=\left(\begin{array}{c}-2 \\ -1/2\end{array}\right), \ \ {\bf x}_2=\left(\begin{array}{c}-1 \\ -1/2\end{array}\right), \ \ 
{\bf x}_3=\left(\begin{array}{c}-3/2 \\ \frac{\sqrt{3}}2   -1/2\end{array}\right),$$
$${\bf x}_4=\left(\begin{array}{c}1\\ -1/2\end{array}\right), \ \ {\bf x}_5=\left(\begin{array}{c} 2\\ -1/2\end{array}\right), \ \  {\bf x}_6=\left(\begin{array}{c}3/2 \\ \frac{\sqrt{3}}2-1/2\end{array}\right).$$
 Note that ${\bf x}_1$, ${\bf x}_2$, ${\bf x}_3$ and ${\bf x}_4$, ${\bf x}_5$, ${\bf x}_6$
are the vertices of two equilateral triangles with sides of   length one; denote them as~$\cal A$ and~$\cal B$, respectively.

The complement of the $\alpha$-hull of the set
$\bf{\mathcal{X}}:=\{{\bf x}_1,\ldots ,{\bf x}_6\}$ is defined as the union of all   open balls
of radius $\alpha$ that have empty intersection with~$\mathcal{X}$. Put~$\alpha= 3/2$. In order for its intersection   with $\mathcal{X}$ to be empty, a ball of radius~$\alpha$ 
  must be centered at distance at
least $\alpha$ from each point in $\mathcal{X}$. 
%Let us denote by $A$ and $B$ the closed triangles with vertices ${\bf x}_1,{\bf x}_2,{\bf x}_3$
%and ${\bf x}_4,{\bf x}_5,{\bf x}_6$, respectively.
Clearly, any point outside the triangles $\cal A$ and $\cal B$ belongs to some ball of
radius $\alpha$ that does not intersect $\mathcal{X}$; hence, the  $\alpha$-hull of 
$\bf{\mathcal{X}}$ is contained in~${\cal A}\cup{\cal B}$. %Some points in $\cal A$ and $\cal B$

\vspace{2mm}

\begin{figure}[htbp] 
\begin{center} 
%\arraycolsep = 5pt \renewcommand{\arraystretch}{1}
%$\noindent\begin{array}{cc}
\framebox{\includegraphics[width=4cm, height =4cm, trim= 40 180 20 180, clip]{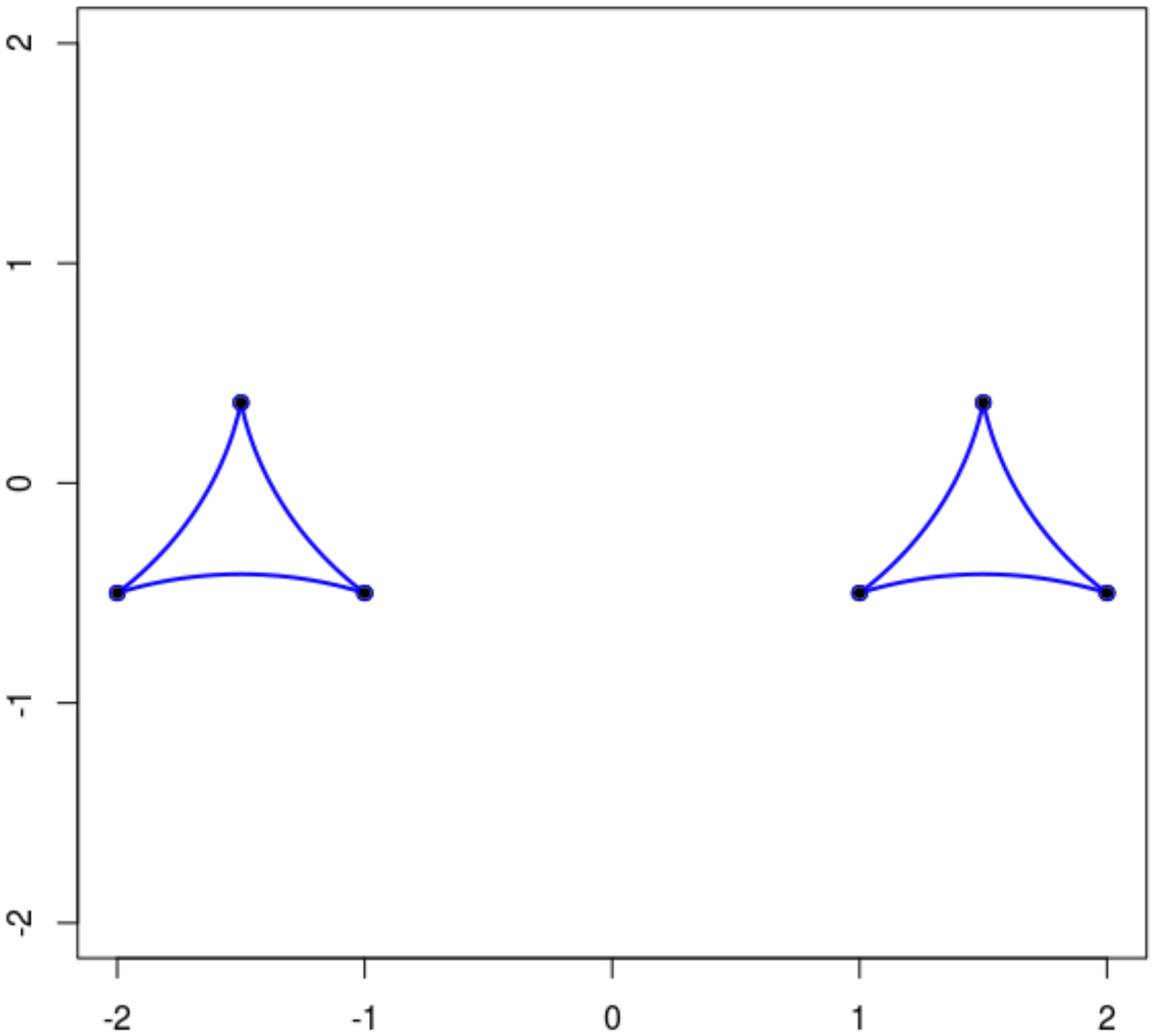}}
%&
%\framebox{\includegraphics*[width=4cm, height =4cm, trim= 40 180 20 180, clip]{Square.pdf}}\vspace{-1mm} 
%\end{array}$
\begin{caption}{A disconnected $\alpha$-hull contour. The picture has been produced with the 
\texttt{alphahull} R-package.}\label{Figalphahull}\end{caption}
\end{center}
\end{figure}

Some balls of radius $\alpha$ that do not intersect with $\mathcal{X}$ nevertheless intersect with~$\cal A$ or $\cal B$. The ``worst" case, that is, the balls of radius $\alpha$ that do not
intersect with $\mathcal{X}$ while having 
 largest intersection with $\cal A$ and $\cal B$ are those  centered at~${\bf c}_1,\ldots,{\bf c}_6$ where ${\bf c}_1$, for instance, is maximizing, among all points  
at distance~$\alpha$ from ${\bf x}_1$ and ${\bf x}_2$, the distance from 
${\bf x}_3$;  similarly,  ${\bf c}_2$, say, is maximizing, among all points at distance~$\alpha$ from ${\bf x}_2$ and ${\bf x}_3$, the distance from 
${\bf x}_1$, etc. %  ,\ldots,y_6$.
As a consequence,  the $\alpha$-hull  of $\mathcal{X}$, for $\alpha= 3/2$,  is the
union of the two curvilinear triangles shown in Figure~\ref{Figalphahull}---obviously not a connected contour.

\bigskip

\noindent For  convenience and ease of use,  this list  regroups the references  for both  the Appendix and the main text.

\end{document}